\documentclass[lineno]{jfm}

\usepackage{graphicx}
\usepackage{newtxtext}
\usepackage{newtxmath}
\usepackage{natbib}
\usepackage{hyperref}
\usepackage{soul}
\usepackage{xcolor}
\hypersetup{
    colorlinks = true,
    urlcolor   = blue,
    citecolor  = black,
}

\newcommand{\RomanNumeralCaps}[1]

\usepackage{amsmath}

\usepackage[caption=false]{subfig}
\usepackage{bbm}
\usepackage{tabularx}

\newcommand{\reviewerone}[1]{\color{black} #1 \color{black}}
\newcommand{\reviewertwo}[1]{\color{black} #1 \color{black}}
\newcommand{\reviewerthree}[1]{\color{black} #1 \color{black}}
\newcommand{\reviewers}[1]{\color{black} #1 \color{black}}

\DeclareRobustCommand{\widefrac}[3][5pt]{%
  \frac{\hspace{#1}#2\hspace{#1}}{\hspace{#1}#3\hspace{#1}}}

\shorttitle{Wavelet-based resolvent analysis}
\shortauthor{E.~Ballouz, B.~Lopez-Doriga, S.~T.~M.~Dawson and H.~J.~Bae}

\title{Wavelet-based resolvent analysis of non-stationary flows}

\author{Eric~Ballouz\aff{1}
  \corresp{\email{eballouz@caltech.edu}},
  Barbara~Lopez-Doriga\aff{2},
  Scott~T.~M.~Dawson\aff{2},
 \and H.~Jane~Bae\aff{3}}

\affiliation{\aff{1} Mechanical and Civil Engineering, California Institute of Technology,
Pasadena, CA 91125, USA
\aff{2} Mechanical, Materials, and Aerospace Engineering, Illinois Institute of Technology, Chicago, IL 60616, USA
\aff{3} Graduate Aerospace Laboratories, California Institute of Technology,
Pasadena, CA 91125, USA}

\begin{document}
\nolinenumbers
\maketitle
\nolinenumbers
\begin{abstract}
This work introduces a formulation of resolvent analysis that uses wavelet transforms rather than Fourier transforms in time. Under this formulation, resolvent analysis may extend to turbulent flows with non-stationary mean states. The optimal resolvent modes are augmented with a temporal dimension and are able to encode the time-transient trajectories that are most amplified by the linearised Navier-Stokes equations. 
We first show that the wavelet- and Fourier-based resolvent analyses give equivalent results for statistically stationary  flow by applying them to turbulent channel flow. 
We then use wavelet-based resolvent analysis to study the transient growth mechanism in \reviewerone{the near-wall region of a turbulent channel flow} by windowing the resolvent operator in time and frequency. The computed principal resolvent response mode, \emph{i.e.} the velocity field optimally amplified by the linearised dynamics of the flow, exhibits characteristics of the Orr mechanism, which supports the claim that this mechanism is key to linear transient energy growth.
We also apply this method to non-stationary parallel shear flows such as an oscillating boundary layer, and three-dimensional channel flow in which a sudden spanwise pressure gradient perturbs a fully developed turbulent channel flow. 
In both cases, wavelet-based resolvent analysis yields modes that are sensitive to the changing mean profile of the flow.
For the oscillating boundary layer, wavelet-based resolvent analysis produces oscillating principal forcing and response modes that peak at times and wall-normal locations associated with high turbulent activity. 
For the turbulent channel flow under a sudden spanwise pressure gradient, the resolvent modes gradually realign themselves with the mean flow as the latter deviates.
Wavelet-based resolvent analysis thus captures the changes in the transient linear growth mechanisms caused by a time-varying turbulent mean profile.
\end{abstract}


\section{Introduction}

Though turbulent flows are highly chaotic systems, they are very often organised into large-scale energetic structures \citep{jimenez2018coherent}. These coherent structures have been observed for wall-bounded flows, jet flows, and flows over wings or other bodies. Since coherent structures are important vehicles of mass and energy, they constitute a popular research topic in a variety of fields \reviewerthree{including} climate sciences and aerodynamics. 

In this paper, we focus on the coherent structures present in near-wall turbulence. We note the ubiquity of streamwise streaks near the wall, \emph{i.e.} regions of low and high velocity elongated in the streamwise direction whose shape, life-cycle and interactions with the outer flow are studied extensively through experiments and numerical simulations \citep{klebanoff1962three, kline1967structure, bakewell1967viscous, kim1971production, blackwelder1979streamwise,smithmetzler1983characteristics, johansson1987generation, robinson1991coherent, adrian2007hairpin, smits2011high}. 
These streaks are often described as undergoing a quasiperiodic cycle of formation and breakdown, 
the drivers of which many works are dedicated to understanding \citep{landahl1980note,  butler1993optimal, hamilton1995regeneration, panton2001overview, chernyshenko2005mechanism, delalamo2006linear,  jimenez2018coherent}. The structure of near-wall turbulence has inspired the pursuit of lower-dimensional models, wherein high-dimensional flows are described by the dynamical evolution of large spatial structures. Often, these structures are extracted from spatiotemporal correlations exhibited in experimental or numerical data \citep{lumley1967structure, lumley2007stochastic,berkooz1993proper, boree2003extended, mezic2013analysis, abreu2020spectral, tissot2021stochastic}.

In contrast to data-driven approaches, many works seek to understand the generation and sustenance of coherent structures through the equations of motion. In the context of wall-bounded turbulence, despite the central role of nonlinearities, linear mechanisms have been proposed as sources of highly energetic large scale coherent structures \citep{panton2001overview, chernyshenko2005mechanism, delalamo2006linear, jimenez2013linear,lozano2021cause}. 
One example is the Orr mechanism \citep{orr1907, jimenez2013linear}, in which the mean shear profile near the wall rotates wall-normal velocity perturbations forward in the streamwise direction and stretches vertical scales; to preserve continuity, wall-normal fluxes and velocity perturbations are intensified. 
Another linear mechanism that has been studied as a possible energy source for coherent velocity perturbations is lift-up \citep{hwang2010self}, which occurs when wall-normal velocity perturbations transport fluid 
initially near the wall to regions farther away from the wall, 
allowing it to be accelerated by the faster mean flow away from the wall.
The key role of linear mechanisms in near-wall turbulence has been emphasised in works like \citet{delalamo2006linear} and \citet{pujals2009note}, which show that, even after removing the nonlinear term from the perturbation equations, linear transient growth via the mean shear generates the dominant (streaky) structures in wall-bounded turbulence. 
The numerical experiments in \citet{lozano2021cause} show that turbulence can be sustained in the minimal flow unit even without the nonlinear feedback between the velocity fluctuations and the mean velocity profile. The only exception is when the authors suppress either the aforementioned Orr-mechanism or the push-over mechanism, \emph{i.e.}  the momentum transfer from the spanwise mean shear into the streamwise velocity perturbation, suggesting the prominence of linear transient growth in energising near-wall streaks.

Given these results, it is not entirely surprising that resolvent analysis has been fruitful in the analysis and modelling of near-wall turbulence, despite relying \reviewerthree{on} a linearisation of the Navier-Stokes equations \citep{butler1993optimal, farrell1998perturbation,Jovanovic2005,mckeon2010critical}.
In resolvent analysis, the Navier-Stokes equations are written as a linear dynamical system for velocity and pressure fluctuations about a mean profile. The nonlinear term, along with any additional exogenous force on the system, is represented as a forcing term acting on this system. 
The resolvent operator refers to the linear map between the forcing inputs and the flow states. In this linearised setting, without computing the nonlinear terms, we can solve for the input (or forcing) terms that would generate the output trajectories (or responses) with the largest kinetic energy \citep{Jovanovic2005}. This is done in practice by taking a singular value decomposition (SVD) of the discretised resolvent operator:
the first right singular mode reveals the inputs to which the linearised equations of motion are most sensitive; the first left singular mode reveals the most amplified outputs, and the first singular value squared yields the kinetic energy amplification. 
The assumption underpinning this approach is that the optimal structures computed by resolvent analysis will be preferentially amplified by the linear dynamics of the flow, believed to be prominent in near-wall turbulence as discussed previously, and will thus manifest as sustained coherent structures. In the context of wall-bounded turbulent flows, resolvent analysis is successful at identifying streamwise rolls as the most perturbing structures, and streamwise streaks as the most amplified structures \citep{mckeon2010critical, bae2021nonlinear}.

Since resolvent response modes are expected to figure prominently in the flow, a linear combination of the leading response modes have been used to construct low-dimensional approximations of turbulent flows, including channel and pipe flow \citep{moarref2014foundation, gomez2016reduced, beneddine2017unsteady, illingworth2018estimating, bae2020resolvent,bae2020studying,ahmed2021resolvent,arun2023towards}. This is especially tractable when the singular values decay quickly, and the resolvent operator can be represented by a heavily truncated SVD. Other works have also explored the use of resolvent modes in estimating and predicting flows with sparse measurements. Specifically, a low-rank approximation of the resolvent operator can be used to model correlations between different spatial locations of the flow \citep{martini2020resolvent,towne2020resolvent}. Moreover, the dynamical relevance of resolvent modes in controlling the fully turbulent flow has been probed \citep{luhar2014opposition,yeh2019resolvent,bae2021nonlinear}. We highlight the work of \citet{bae2021nonlinear}, who demonstrate the effectiveness of resolvent modes in transferring energy to coherent near-wall turbulent perturbations within a turbulent minimal channel: by subtracting out the contribution of the leading resolvent forcing mode from the nonlinear term at every time step, the streak-regeneration process is interrupted and buffer layer turbulence is suppressed.  


\reviewertwo{Traditionally, resolvent analysis has been applied to systems that are homogeneous in two-spatial dimensions and time, but the framework has been extended to explore the influence of spatial variations in at least one additional dimension. This
``global" resolvent analysis has produced promising results in the study of steady flows over airfoils \citep{kojima2020, ribeiro2020randomized, yeh2019resolvent, ribeiro2023triglobal} and jets \citep{pickering2020orr, towne2022efficient}. The main obstacle that this method faces is its large memory and computation costs, which can be mitigated with the use of randomised methods \citep{ribeiro2020randomized} or matrix-free time stepping algorithms \citep{martini2021efficient, towne2022efficient, farghadan2024efficient}.
}

Despite these extensions, the formulation of the resolvent operator relies on Fourier-transforming the linearised Navier-Stokes equations in time. This restricts its formulation to statistically steady and quasiperiodic flows \citep{Padovan2020}. Moreover, the resulting SVD modes will be Fourier modes in time, and cannot represent temporally local effects. 
However, the linear energy amplification mechanisms that are important to near-wall turbulence, namely the Orr-mechanism, are transient processes. Accounting for transient effects is also important in estimation and control problems. In \citet{martini2020resolvent}, time-colouring is employed to improve their estimates, and in \citet{yeh2019resolvent}, which studies flow separation over an airfoil, the resolvent operator is modified to select forcing and responses modes acting on a time scale of interest using the exponential discounting method introduced in \citet{jovanovic2004modelling}.
Be it for analysis, estimation or control, resolvent modes capable of encoding time are a potentially valuable extension.

In this work, we propose using a wavelet transform \citep{Meyer1992} in time to construct the resolvent operator so that the SVD modes for the newly formulated resolvent operator are localised in time. 
Wavelets are indeed functions (in time, for this application) whose mass is concentrated in a subset of their domain. This allows a projection onto wavelets to preferentially capture information centred  in a time interval. Each wavelet onto which a function is projected also captures a subset of the Fourier spectrum. 
Due to their properties, wavelets \reviewerthree{have} been used extensively in fluid mechanics research, particularly spatial wavelets which allow for the analysis of select length scales concentrated in a region of interest \citep{meneveau1991analysis,lewalle1993wavelet}. Temporal wavelet transforms have also been used to decompose turbulent flows. In \citet{barthel2023wavelet}, the authors show that high-frequency phenomena upstream over an airfoil are correlated with low-frequency extreme events downstream, and exploit the time-frequency localisation in wavelet space to build more robust predictors of these extreme events.
Other work has focused on constructing an orthogonal wavelet basis from simulation data to best capture self-similarity in the data \citep{ren2021image,floryan2021discovering}.
An operator-based approach is given in \cite{lopez2023sparsity, lopez2024spacetime}, in which the authors use a time-resolved resolvent analysis to extract transient structures that are preferentially amplified by the linearised flow; these modes notably exhibit a wavelet-like profile in time.
In the context of resolvent analysis, the additional time and frequency localisation provided by the wavelet transform will allow us to formulate the flow states and forcing around non-stationary mean profiles. The resolvent modes would thus reflect time-localised changes due to transient events in the mean profile. Moreover, resolvent modes that encode both time and frequency information could help analyse  linear amplification phenomena that occurs transiently and that separates forcing and response events in time and/or frequency.

The present work is organised as follows. In \S\ref{sec:formulation}, we describe traditional Fourier-based resolvent analysis and introduce a wavelet-based formulation.
We highlight the properties of the wavelet transform and discuss the choice of wavelet basis; we also discuss the efficiency and robustness of the numerical methods to compute the resolvent modes.
In \S\ref{sec:application}, we develop and validate wavelet-based resolvent analysis for a variety of systems, ranging from quasiparallel wall-bounded turbulent flows to spatiotemporally evolving systems.  
In \S\ref{sec:app:channel}, we establish the equivalence of Fourier- and wavelet-based resolvent analyses for the statistically stationary turbulent channel flow and, in \S\ref{sec:transient}, we showcase the additional capacity of the wavelet-based resolvent to capture linear transient growth under transient forcing. Specifically, we use the time- and frequency-augmented system to capture the Orr-mechanism in turbulent channel flow.
Then we apply wavelet-based resolvent analysis to statistically non-stationary flows in \S\ref{sec:app_time}, notably the turbulent Stokes boundary layer flow in which the mean oscillates periodically in time (\S\ref{sec:stokes}), as well as a turbulent channel flow subjected to a sudden transverse pressure gradient (\S\ref{sec:3DChannel}). A preliminary version of this work is published in \citet{ballouz2023wavelet}; \reviewertwo{novel contributions include an application of the method to the study of two additional phenomena: the Orr mechanism in turbulent channel flow, and the transient reduction of Reynolds stresses in the non-stationary channel flow under a sudden transverse pressure gradient. This work also includes a deeper discussion on the computational costs of the method.}
Conclusions and a discussion of the results are given in \S\ref{sec:conclusion}.

\section{Mathematical formulation}\label{sec:formulation}
\subsection{Fourier-based resolvent analysis}

The non-dimensional incompressible Navier-Stokes equations are given by 
\begin{equation}
    \frac{\partial\bar{u}_i}{\partial{t}} + \bar{u}_j\frac{\partial\bar{u}_i}{\partial{x_j}} = -\frac{\partial\bar{p}}{\partial x_i} + \frac{1}{\Rey}\frac{\partial^2\bar{u}_i}{\partial{x}_j\partial{x}_j},\quad \frac{\partial\bar{u}_i}{\partial{x_i}} = 0,
    \label{eq:NS}
\end{equation}
where $\bar{u}_i$ is the total velocity (including the mean and the fluctuating component) in the $x_i$ direction and $\bar{p}$ is the total pressure. The Reynolds number is given by $\Rey = u^* L^*/\nu$, where $\nu$ is the kinematic viscosity, and $u^*$ and $L^*$ are respectively a reference velocity and lengthscale used to non-dimensionalise $\bar u_i$, $x_i$, and $t$. Likewise, $\bar p$ is non-dimensionalised by a reference density $\rho^*$ and $u^*$. The non-dimensionalisations for each of the cases studied in this work are \reviewerthree{given} in table 1.
The total velocity can be split into $\bar{u}_i = U_i + u_i$. Here, $U_i := \langle \bar u_i \rangle$ represents the average over ensembles and homogeneous directions, with $\langle \cdot \rangle$ denoting the averaging operation, and $u_i$ is the fluctuating component. Similarly, pressure can be decomposed as $\bar{p} = P + p := \langle \bar p \rangle + p$.

We can split equations ~\eqref{eq:NS} into equations for the mean and the fluctuating components of the flow
\begin{gather}
    \frac{\partial{U_i}}{\partial t} + \left\langle \bar{u}_j\frac{\partial\bar{u}_i}{\partial{x_j}}\right\rangle = -\frac{\partial{P}}{\partial{x_i}} + \frac{1}{\Rey} \frac{\partial^2U_i}{\partial{x_j}\partial{x_j}},\quad \frac{\partial{U_i}}{\partial{x_i}}= 0,\\ \label{eq:NS_fluc}
    \frac{\partial{u_i}}{\partial t} + U_j\frac{\partial{u_i}}{\partial{x_j}} + u_j\frac{\partial{U_i}}{\partial{x_j}} = -\frac{\partial{p}}{\partial{x_i}} + \frac{1}{\Rey} \frac{\partial^2u_i}{\partial{x_j}\partial{x_j}} + f_i,\quad \frac{\partial{u_i}}{\partial{x_i}}= 0, 
\end{gather}
where $f_i$ is the remaining nonlinear terms in the fluctuating equations. The equations above do not have an analytic solution unless in very particular situations and are most commonly solved numerically. Note that some of the terms in the fluctuating equations may be zero depending on the flow configuration.
\reviewerthree{In traditional resolvent analysis and in this work, equations \eqref{eq:NS_fluc} are Fourier-transformed in the homogeneous spatial directions. For the cases considered in this work, these are the streamwise and spanwise directions. Homogeneity in these directions also implies $\partial U_i /\partial x_1 = \partial U_i /\partial x_3 = 0$. 
Further simplifications of equations \eqref{eq:NS_fluc} are also possible in some cases, for example when certain components of the mean flow are zero.
Moreover, in traditional resolvent analysis, the  mean flow is statistically stationary (and therefore independent of time), and the fluctuation equations are assumed to be periodic in time and further Fourier transformed in time. 

After applying a Fourier transformation in time and in one or more spatial dimensions in which the system is homogeneous, and discretising the equations over the remaining spatial dimensions of the grid, we obtain the following
\begin{equation} \label{eq:Disc_NS}
        \widehat{\mathsfbi{D}_t} \widehat{\mathsfbi u}_i +  \widehat{\mathsfbi u_j} \mathsfbi{dU}_{i,j} + \mathsfbi U_j \widehat{\mathsfbi D_j}  \widehat{\mathsfbi u}_i = -\widehat{\mathsfbi D}_i \widehat{\mathsfbi p} + \frac{1}{\Rey} \widehat{\mathsfbi L} \widehat{\mathsfbi u}_i + \widehat{\mathsfbi f}_i, \quad \widehat{\mathsfbi D}_i \widehat{\mathsfbi u}_i = 0,
\end{equation}
where $\widehat{\mathsfbi u_i}$, $\widehat{\mathsfbi f_i}$ and $\widehat{\mathsfbi p}$ respectively denote the transformed and discretised $u_i$, $f_i$ and $p$, 
}
$\widehat{\mathsfbi{D}_t}$ and $\widehat{\mathsfbi{D}_i}$ are the transformed discrete differentiation operators in time and the $x_i$-- directions respectively, and $ \widehat{\mathsfbi L}$ is the transformed Laplacian. 
For systems that are homogeneous in the streamwise and spanwise directions and periodic in time, we can write $\widehat{\mathsfbi{D}_t} = -\mathrm{i} \omega$, $\widehat{\mathsfbi{D}_1} = \mathrm{i}k_1$, $\widehat{\mathsfbi{D}_3} = \mathrm{i}k_3$, $\widehat{\mathsfbi{D}_2} = \mathsfbi{D}_2$, and $ \widehat{\mathsfbi L} = -k_1^2 + \mathsfbi D_2^2 -k_3^2$, where $\omega$ is the chosen frequency for the temporal Fourier transform,  $(k_1, k_3)$ are the streamwise and spanwise wavenumber pair, and
$\mathsfbi D_2$ is the discrete derivative in the $x_2$ direction.
We use $\mathsfbi{U}_{i}$ and $\mathsfbi{dU}_{i,j}$ to denote the diagonal matrices whose diagonal terms are respectively $U_i$ and $\partial U_i/\partial x_j$ evaluated at the grid points, and 
we note that since the system is homogeneous in $x_1$ and $x_3$, $\mathsfbi U_2 = \mathsfbi{dU}_{i,1} = \mathsfbi{dU}_{i,3} = \boldsymbol{0}$. Each of the discretised momentum and continuity equations is an $ N_{2}$-dimensional system, where $N_2$ is the spatial resolution in the $x_2$ direction. 
\reviewerthree{These linearised equations can then be cast as} 
\begin{equation}
\left[\begin{array}{c}\hat{\mathsfbi u}_1(x_2)\\ \hat{\mathsfbi u}_2(x_2)\\ \hat{\mathsfbi u}_3(x_2)\\ \hat{\mathsfbi p}(x_2)\end{array}\right] = \hat{\mathsfbi {H}}^{(k_1,k_3,\omega)} \left[\begin{array}{c}\hat{\mathsfbi f}_1(x_2)\\ \hat{\mathsfbi f}_2(x_2)\\ \hat{\mathsfbi f}_3(x_2)\\ \boldsymbol{0} \end{array}\right],
\label{eq:in_out_fourier}
\end{equation}
\reviewerthree{where $\widehat{\mathsfbi {H}}^{(k_1,k_3,\omega)}$ denotes the traditional Fourier-based resolvent operator
\begin{multline}\label{eq:def_fourier_based_resolvent}
    \widehat{\mathsfbi{H}}^{(k_1, k_3, \omega)} = \\
    \left[\left( -\mathrm{i} \omega \mathsfbi I - \frac{1}{\Rey} \mathsfbi L + \mathrm{i} k_1 \mathsfbi U_1 + \mathsfbi U_2 \mathsfbi D_2 + \mathrm{i} k_3 \mathsfbi U_3  \right ) 
    \left(
    \begin{array}{cccc}
    \mathsfbi I & \boldsymbol 0 & \boldsymbol 0 & \boldsymbol 0 \\
    \boldsymbol 0 & \mathsfbi I & \boldsymbol  0 & \boldsymbol 0 \\
    \boldsymbol 0 & \boldsymbol 0 & \mathsfbi I & \boldsymbol 0 \\
    \boldsymbol 0 & \boldsymbol 0 & \boldsymbol 0 & \boldsymbol 0 
    \end{array}
    \right)\right.
    +
    \left.\left(
    \begin{array}{cccc} 
    \boldsymbol{0} & \widehat{\mathsfbi{dU}}_{1,2} & \boldsymbol{0} & \mathrm{i} k_1 \mathsfbi I \\
    \boldsymbol{0} & \widehat{\mathsfbi{dU}}_{2,2} & \boldsymbol{0} & \mathsfbi D_2 \\
    \boldsymbol{0} & \widehat{\mathsfbi{dU}}_{3,2} & \boldsymbol{0} & \mathrm{i} k_3 \mathsfbi I \\
    \mathrm{i} k_1 \mathsfbi I        & \mathsfbi D_2        & \mathrm{i} k_3 \mathsfbi I       & \boldsymbol 0 
    \end{array}
    \right)\right]^{-1},
\end{multline}}
and the superscript $(k_1,k_3,\omega)$ indicates the choice of streamwise and spanwise wavenumbers $k_1$ and $k_3$, and frequency $\omega$ used in the Fourier transforms. \reviewerthree{The functional dependence on $x_2$ indicates the discretisation over the wall-normal spatial dimension}. 
Typically, the SVD of the linear resolvent operator $\widehat{\mathsfbi{H}} ^{(k_1,k_3,\omega)} \in \mathbb{C}^{4N_2} \times \mathbb{C}^{4N_2}$ is taken to study the left and right singular vectors as response and forcing modes, and the singular values as amplification factors or gains. 
We denote the principal forcing and response modes by $\boldsymbol{\hat \phi}(x_2) = [
    \boldsymbol{\hat \phi}_1^T, \boldsymbol{\hat \phi}_2^T, \boldsymbol{\hat \phi}_3^T, \boldsymbol{0}^T]^T$
and $\boldsymbol{\hat \psi}(x_2) = [
    \boldsymbol{\hat \psi}_1^T, \boldsymbol{\hat \psi}_2^T, \boldsymbol{\hat \psi}_3^T, \boldsymbol{\hat \psi}_p^T]^T
$ respectively. For a wall-normal spatial domain $[0,L_2]$, the modes are normalised \reviewerthree{such that their integrated kinetic energy satisfies
\begin{align}
    [\hat \phi_1] + [\hat \phi_2] + [\hat \phi_3] = 1, \\
    [\hat \psi_1] + [\hat \psi_2] + [\hat \psi_3] = 1,
\end{align}
}
where we use $[\cdot] =\frac{1}{L_2} \int_0^{L_2} |\cdot|^2dx_2$ to denote the  $x_2-$integrated energy.

\subsection{Wavelet-based resolvent analysis}

\subsubsection{Formulation}

To extend resolvent analysis to statistically non-stationary flows by accounting for transient behaviour in the mean flow or the fluctuations, we introduce the wavelet-based resolvent analysis. The benefit of the wavelet transform in time is that it preserves both time and frequency information. The wavelet transform projects a function onto a wavelet basis composed of scaled and shifted versions of a mother function $\eta(t)$. The transformed function depends on the scale $\alpha$ and shift $\beta$ parameters respectively linked to frequency and time information, whereas the Fourier transform is a function of only frequency. We propose using a wavelet transform in time to the left of the spatially Fourier-transformed equations \eqref{eq:NS_fluc} while keeping the Fourier transform in homogeneous spatial directions. 
\reviewerthree{In practice, we discretise the spatially Fourier-transformed equations in both time and the wall-normal dimension, and apply a discrete wavelet transform $\mathsfbi W$ to the left of the discrete equations. For a system that is homogeneous in the streamwise and spanwise directions, and that has been wavelet-transformed in time and Fourier-transformed in $x_1$ and $x_3$ for a streamwise and spanwise wavenumber pair $(k_1, k_3)$, we obtain}
\begin{equation} 
\label{eq:Disc_Wav_NS}
        \widetilde{\mathsfbi{D}_t} \widetilde{\mathsfbi u}_i +  \widetilde{\mathsfbi u_j} \widetilde{\mathsfbi{dU}_{i,j}} + \widetilde{\mathsfbi U_j} \widetilde{\mathsfbi D_j}  \widetilde{\mathsfbi u}_i = -\widetilde{\mathsfbi D}_i \widetilde{\mathsfbi p} + \frac{1}{\Rey} \widetilde{\mathsfbi L} \widetilde{\mathsfbi u}_i + \widetilde{\mathsfbi f}_i, \quad \widetilde{\mathsfbi D}_i \widetilde{\mathsfbi u}_i = 0.
\end{equation}
Here, $\widetilde{\mathsfbi u_i}$, $\widetilde{\mathsfbi f_i}$ and $\widetilde{\mathsfbi p}$ respectively denote the transformed $u_i$, $f_i$ and $p$ discretised over $x_2$, $\alpha$ and $\beta$. 
We introduce $\widetilde{\mathsfbi U_i} := \mathsfbi W \mathsfbi U_i \mathsfbi W^{-1}$ and $\widetilde{\mathsfbi{dU}_{i, j}} := \mathsfbi W \mathsfbi{dU}_{i, j} \mathsfbi W^{-1}$, where $\mathsfbi U_i$ and $\mathsfbi{dU}_{i, j}$ are defined in \S2.1. As before, $\mathsfbi U_2 = \mathsfbi{dU}_{i,1} = \mathsfbi{dU}_{i,3} = \boldsymbol{0}$. 
Moreover, $\widetilde{\mathsfbi D}_1 = \widehat{\mathsfbi D}_1 = \mathrm i k_1$, $\widetilde{\mathsfbi D}_2 = \widehat{\mathsfbi D}_2 = \mathsfbi D_2$, $\widetilde{\mathsfbi D}_3 = \widehat{\mathsfbi D}_1 = \mathrm i k_3$, $\widetilde{\mathsfbi L} = \widehat{\mathsfbi L} = -k_1^2 \mathsfbi D_2^2 -k_3^2$, and $\widetilde{\mathsfbi{D}_t} := \mathsfbi W \mathsfbi{D}_t \mathsfbi W^{-1}$, where $\mathsfbi D_t$ refers to the discrete time differentiation matrix. 
\reviewertwo{We denote the temporal resolution of our discretised system by $N_t$; the wavelet-transformed vectors will also have a resolution of $N_t$ in the space of wavelet parameters $\alpha$ and $\beta$.}
Similarly to the Fourier-based resolvent analysis, the equations can be written in matrix form as
\begin{equation}
\left[\begin{array}{c}\tilde{\mathsfbi u}_1(x_2,\alpha, \beta)\\ \tilde{\mathsfbi u}_2(x_2,\alpha, \beta)\\ \tilde{\mathsfbi u}_3(x_2,\alpha, \beta)\\ \tilde{\mathsfbi p}(x_2,\alpha, \beta)\end{array}\right] = \tilde{\mathsfbi{\mathsfbi H}}^{(k_1,k_3)} \left[\begin{array}{c}\tilde{\mathsfbi f}_1(x_2,\alpha, \beta)\\ \tilde{\mathsfbi f}_2(x_2,\alpha, \beta)\\ \tilde{\mathsfbi f}_3(x_2,\alpha, \beta)\\ \boldsymbol 0 \end{array}\right],
\label{eq:in_out_w}
\end{equation}
where \reviewerthree{the functional dependence on $x_2$, $\alpha$ and $\beta$ represents the discretisation over the wall-normal spatial dimension, and the wavelet shifts and scales,} and the wavelet-based resolvent operator $\tilde{\mathsfbi{H}} \in \mathbb{C}^{4N_t\times N_2} \times \mathbb{C}^{4N_t\times N_2}$ is defined as
\begin{multline}\label{eq:def_wavelet_based_H}
    \widetilde{\mathsfbi{H}}^{(k_1, k_3)} =\\ \left[\left(\widetilde{\mathsfbi D_t} - \frac{1}{\Rey} \mathsfbi L + \mathrm{i} k_1 \widetilde{\mathsfbi U_1} + \widetilde{\mathsfbi U_2} \mathsfbi D_2 + \mathrm{i} k_3 \widetilde{\mathsfbi U_3} \right) 
    \left(
    \begin{array}{cccc}
    \mathsfbi I & \boldsymbol 0 & \boldsymbol 0 & \boldsymbol 0 \\
    \boldsymbol 0 & \mathsfbi I & \boldsymbol  0 & \boldsymbol 0 \\
    \boldsymbol 0 & \boldsymbol 0 & \mathsfbi I & \boldsymbol 0 \\
    \boldsymbol 0 & \boldsymbol 0 & \boldsymbol 0 & \boldsymbol 0 
    \end{array}
    \right)\right.
    +
    \left.\left(
    \begin{array}{cccc} 
    \boldsymbol{0} & \widetilde{\mathsfbi{dU}}_{1,2} & \boldsymbol{0} & \mathrm{i} k_1 \mathsfbi I \\
    \boldsymbol{0} &
    \widetilde{\mathsfbi{dU}}_{2,2} & 
    \boldsymbol{0} & \mathsfbi D_2 \\
    \boldsymbol{0} & 
    \widetilde{\mathsfbi{dU}}_{3,2} & 
    \boldsymbol{0}& 
    \mathrm{i} k_3 \mathsfbi I\\
    \mathrm{i} k_1 \mathsfbi I & 
    \mathsfbi D_2 & 
    \mathrm{i} k_3 \mathsfbi I & 
    \boldsymbol 0 
    \end{array}
    \right)\right]^{-1}.
\end{multline}
This formulation allows us to study transient flows using resolvent analysis. We denote the principal forcing and response modes obtained under this formulation by $\boldsymbol{\tilde \phi}(x_2, \alpha, \beta) = [
    \boldsymbol{\tilde \phi}_1^T, \boldsymbol{\tilde \phi}_2^T, \boldsymbol{\tilde \phi}_3^T, \boldsymbol 0^T]^T$
and $\boldsymbol{\tilde \psi}(x_2, \alpha, \beta) = [
    \boldsymbol{\tilde \psi}_1^T, \boldsymbol{\tilde \psi}_2^T, \boldsymbol{\tilde \psi}_3^T, \boldsymbol{\tilde \psi}_p^T]^T
$ respectively. 
\reviewerthree{We denote their respective inverse wavelet-transforms by $\boldsymbol{\breve \phi}(x_2, t) := \mathsfbi W^{-1} \boldsymbol{\tilde \phi}$ and $\boldsymbol{\breve \psi}(x_2, t):= \mathsfbi W^{-1} \boldsymbol{\tilde \psi}$.
These are normalised such that integrated kinetic energy satisfies
\begin{align}
    \frac{1}{T}\frac{1}{L_2} \int_0^T \int_0^{L_2} |\breve \phi_1 |^2 + |\breve \phi_1|^2 + |\breve \phi_3|^2 dx_2 dt =1, \\
    \frac{1}{T} \frac{1}{L_2}\int_0^T \int_0^{L_2} |\breve \psi_1|^2 + |\breve \psi_2|^2 + |\breve \psi_3|^2 dx_2 dt = 1,
\end{align}
where $[0, T)$ represents the temporal domain. We denote the inverse Fourier transforms of the modes to the physical domain by $\boldsymbol{\phi}(x_1, x_2, x_3, t)$ and $\boldsymbol{\psi}(x_1, x_2, x_3, t)$, repectively.}

\subsubsection{Wavelet-based resolvent analysis with windowing} \label{sec:windowing}

We can formulate a resolvent map between forcing and response at specific time shifts and scales by defining a windowed resolvent operator
\begin{equation}
\left[\begin{array}{c}\tilde{\mathsfbi u}_1(x_2,\alpha, \beta)\\ \tilde{\mathsfbi u}_2(x_2,\alpha, \beta)\\ \tilde{\mathsfbi u}_3(x_2,\alpha, \beta)\\ \tilde{\mathsfbi p}(x_2,\alpha, \beta)\end{array}\right] = \mathsfbi{C}\tilde{\mathsfbi{H}}^{(k_1,k_3)} \mathsfbi{B} \left[\begin{array}{c}\tilde{\mathsfbi f}_1(x_2,\alpha, \beta)\\ \tilde{\mathsfbi f}_2(x_2,\alpha, \beta)\\ \tilde{\mathsfbi f}_3(x_2,\alpha, \beta)\\ 0 \end{array}\right],
\label{eq:windowed_res}
\end{equation}
where $\mathsfbi{B}$ and $\mathsfbi{C}$ are windowing matrices on the forcing and response modes, respectively \citep{jeun2016, kojima2020}. The windowing matrices select a subset of the full forcing and response states. For example, to select a particular scale and shift parameter $(\alpha_s,\beta_s)$ for the forcing mode, we set
\begin{equation}
    \mathsfbi{B} = \text{diag}\big(\mathbbm{1}(\alpha = \alpha_s)\mathbbm{1}(\beta = \beta_s)\big),
\end{equation}
where $\mathbbm{1}(\cdot)$ is an indicator function. Here, 
\reviewerthree{$\mathsfbi B$ selects the relevant columns of $\mathsfbi W^{-1}$ corresponding to the wavelet scales and shifts to which we wish to restrict our forcing. Analogously, $\mathsfbi C$ selects the rows of $\mathsfbi W$ corresponding to the wavelet scales and shifts to which we restrict our response modes.}
The SVD of the windowed resolvent operator, $\mathsfbi{C}\tilde{\mathsfbi{H}}^{(k_1,k_3)} \mathsfbi{B}$, allows us to identify forcing and response modes restricted to a limited frequency and time interval \reviewerthree{determined by the selected wavelets}. 


\subsubsection{Choice of wavelet basis}

\begin{figure}
\begin{center}
\vspace{0.1cm}
\subfloat[]{\includegraphics[width=0.44\linewidth]{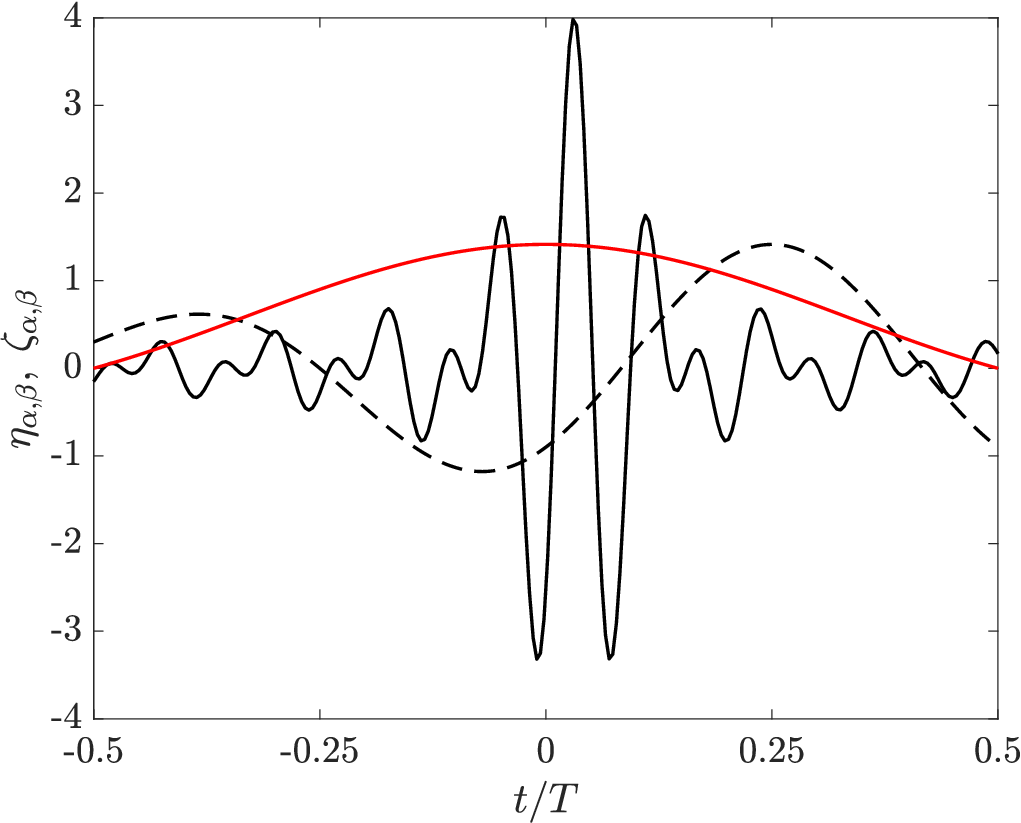}}
\hspace{0.2cm}
\subfloat[]
{\includegraphics[width=0.465\linewidth]{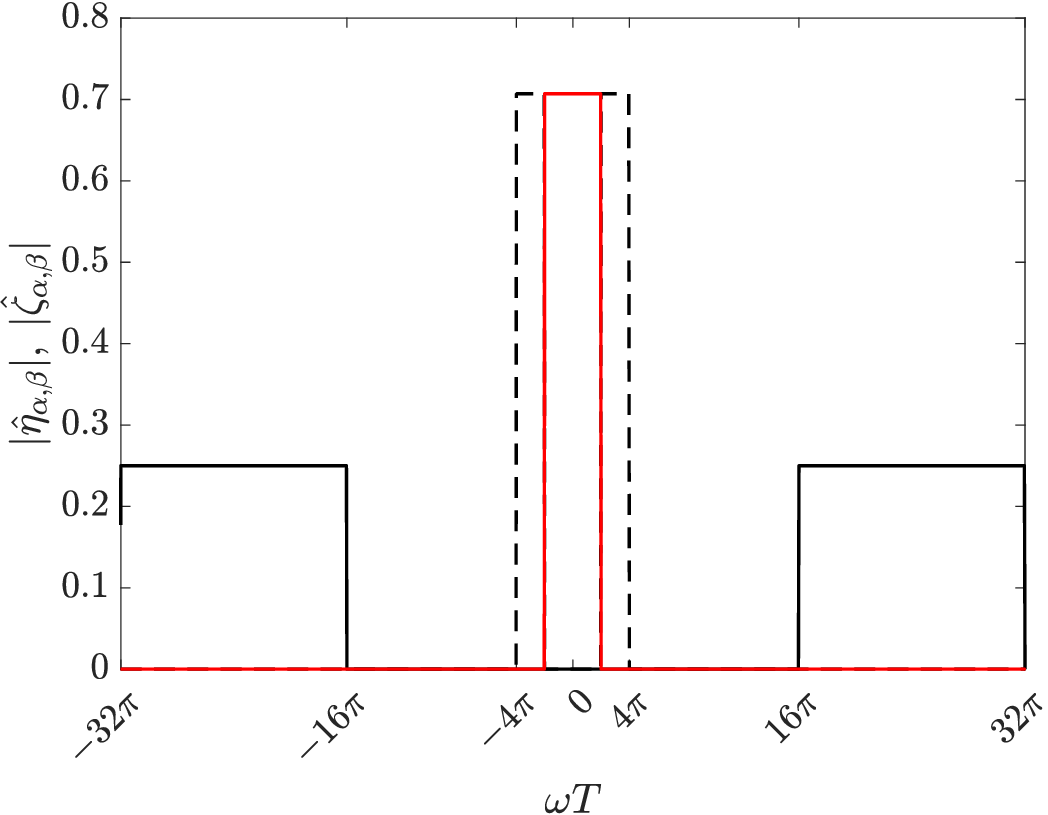}}
\\
\subfloat[]{\includegraphics[width=0.44\linewidth]{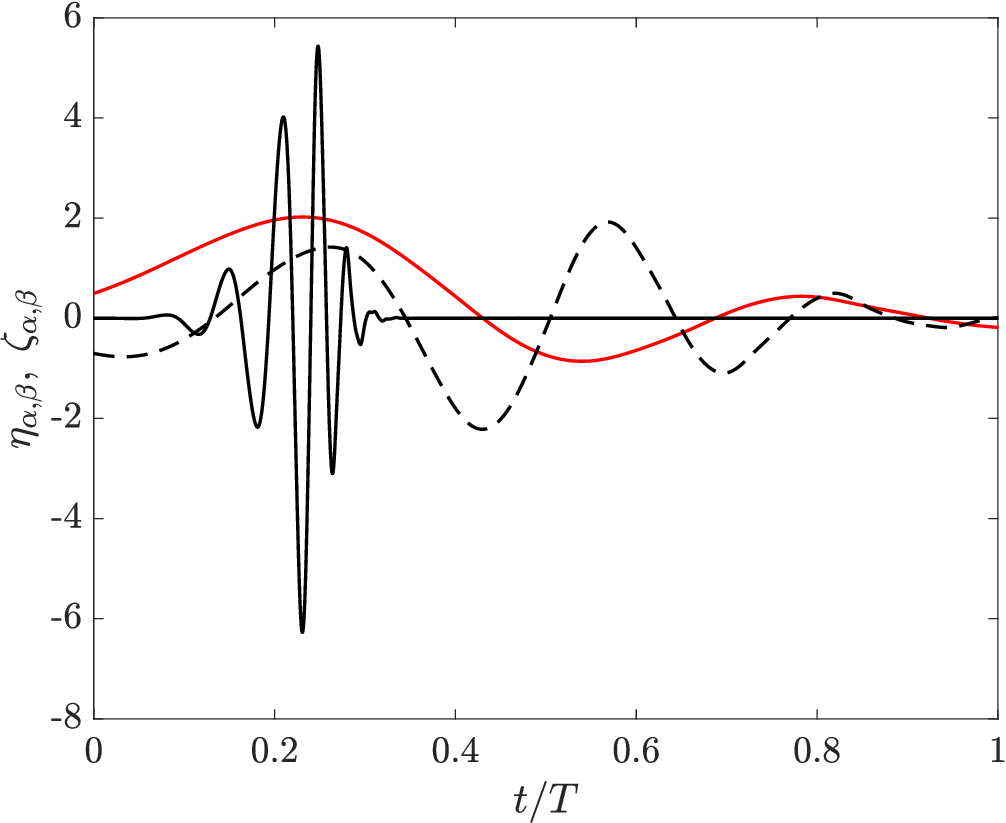}}
\hspace{0.2cm}
\subfloat[]{\includegraphics[width=0.465\linewidth]{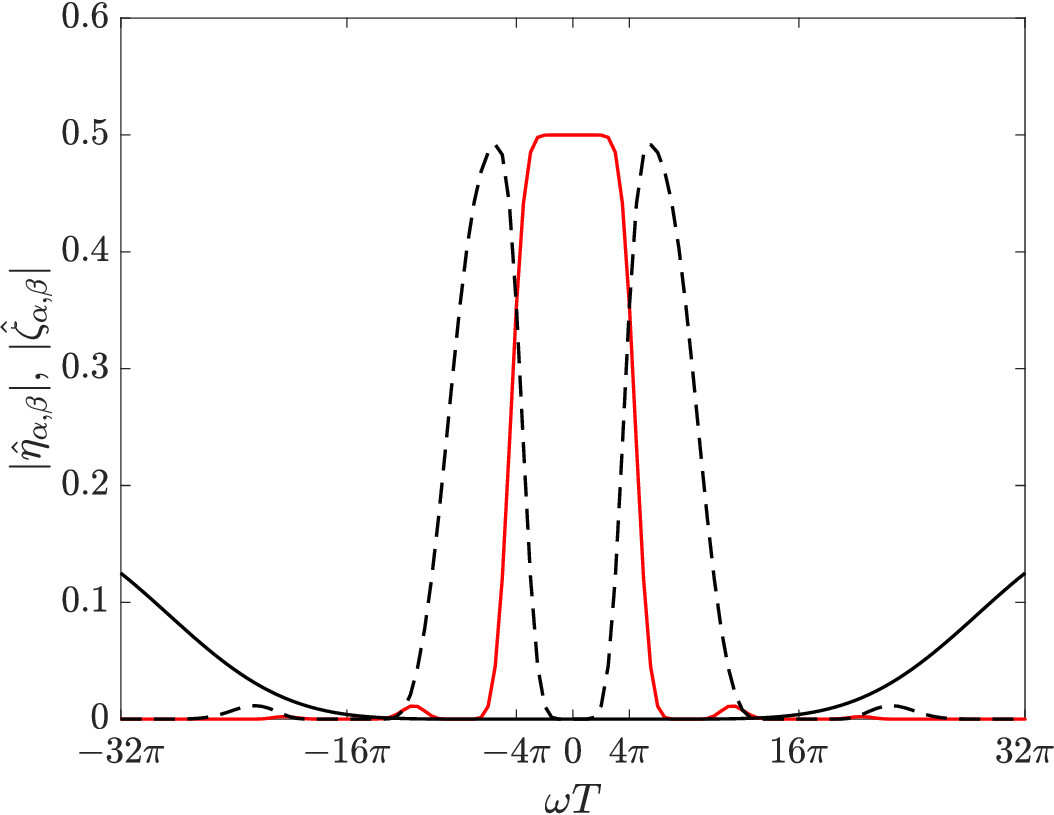}}
\end{center}
\caption{Shannon wavelets and scaling function in time (a) and frequency (b) domains. Daubechies-16 wavelets and scaling function in time (c) and frequency (d) domains. The functions shown are two wavelets for $\alpha = 2^L = N_t$ (black ---) and $\alpha = 2^{L-3}$ (black $--$), and a scaling function for $\alpha = 2^L$ (red ---) \reviewerthree{for arbitrary shift parameters.}}
\label{fig:shannon}
\end{figure}

Wavelet transforms are not unique and are determined by the choice of the mother wavelet $\eta(t)$.
The translations and dilations of a real mother wavelet are given by
\begin{equation}
    \eta_{\alpha,\beta}(t) \equiv \frac{1}{\sqrt{\alpha}}\eta \left(\frac{t-\beta}{\alpha} \right),
\end{equation}
where $\alpha$ and $\beta$ correspond respectively to the scale and shift parameters, and respectively represent location in the frequency and time domains. The dilations of the wavelet capture information at varying scales, and its translations capture information at different time intervals. 

Consider an arbitrary \reviewerthree{square-integrable} function $f(t)$. Its Fourier and wavelet transform are
\begin{equation}
\hat f(\omega) = \int_{-\infty}^{+\infty}f(t)e^{- \mathrm{i} \omega t} dt,
\end{equation}
\begin{equation}
\tilde f^{(w)} (\alpha, \beta)= \int_{-\infty}^{+\infty}f(t)\eta_{\alpha, \beta}(t) dt,
\end{equation}
where $\mathrm{i} = \sqrt{-1}$. In practice, we define the wavelet transform on a dyadic grid, \emph{i.e.} $\alpha = 2^\ell$, $\beta = k2^\ell$ for $ k, \; \ell \in \mathbb Z$. 
However, we usually do not dilate the mother wavelet indefinitely; the dilations and translations of a scaling function $\zeta(t)$ are used to capture the residual left-over from a scale-truncated wavelet expansion. We define the projection onto these functions as
\begin{equation}
\tilde f^{(s)} (\alpha, \beta)= \int_{-\infty}^{+\infty}f(t)\zeta_{\alpha, \beta}(t) dt,\quad\zeta_{\alpha, \beta}(t) \equiv \frac{1}{\sqrt{\alpha}} \zeta \left(\frac{t-\beta}{\alpha} \right).
\end{equation}
The wavelet expansion of an arbitrary function $f(t)$ at dyadic scales is thus given by

\begin{equation}\label{eq:waveletExpansion}
    f(t) = \sum_{\ell = -\infty}^L\sum_{k = -\infty}^{+\infty} \tilde f^{(w)} (2^\ell, 2^k  ) \eta \left (\frac{t}{2^\ell}-k \right) + \sum_{k = -\infty}^{+\infty} \tilde f^{(s)}(2^L, 2^k) \zeta \left (\frac{t}{2^L} - k \right ),
\end{equation}
where $L \in \mathbb Z$ represents the largest scale captured by the wavelet expansion, the $\tilde f^{(w)} (2^\ell, 2^k  )$ terms approximate $f(t)$ at scales $-\infty < 2^\ell \leq 2^L $, and the $\tilde f^{(s)}(2^L, 2^k)$ terms capture the residual at scales $2^\ell > 2^L$. 
\reviewerthree{For each $\ell$, the projection onto $\eta ( t/2^\ell -k)$ will roughly capture a portion of the frequency content of $f$, centred in a time interval determined by $k$. Larger $\ell$ corresponds to a narrower band of frequencies closer to zero, while larger $k$ corresponds to later times.}
In a discretised setting, we use the finite resolution wavelet expansion, which approximates equation \eqref{eq:waveletExpansion} for a discrete signal and where $2 \leq 2^\ell \leq 2^L \leq N_t$. 
\reviewerthree{The choice of the largest scale $L$ will depend on the band of frequencies we wish to isolate with our wavelet-based resolvent operator: narrower bands closer to zero will require larger $L$.}

The wavelet and scaling function coefficients are produced by a premultiplication by the discrete wavelet transform
$\mathsfbi W$, which approximates the convolution against the wavelets and scaling functions. The choice of the wavelet and scaling function pair determines the properties of 
$\mathsfbi W$. Wavelets/scaling functions of compact support in time result in banded 
$\mathsfbi W$, since the convolution with these functions will also have compact support. Orthonormal wavelets/scaling functions result in a unitary
$\mathsfbi W$.
Each wavelet or scaling function captures a portion of the temporal and frequency domains. There is a trade-off between precision in frequency and precision in time, i.e., one cannot find a function $\eta(t)$ that is well localised in both time and frequency \citep{Mallat2001}. As two extreme examples, consider the Dirac delta centred  at $t=1$, which is perfectly localised in \reviewertwo{time} but with an infinite spread in frequency space, and the Fourier mode $e^{\mathrm{i}t}$, which is perfectly localised in frequency space at $\omega = 1$ but has infinite spread in time. 
In the context of windowed resolvent analysis, we may wish to highlight specific bands of the frequency spectrum, or conversely, narrow bands in time, which will inform the choice of wavelet transform.

For this study, we work with the Shannon and Daubechies-16 wavelets. The Shannon wavelet is notable because it acts as a perfect bandpass filter \citep{Mallat2001, Najmi2012} and covers a frequency band 
$N_t / 2^{\ell}([-2\pi, -\pi] \cup [\pi, 2\pi])$ 
(figure \ref{fig:shannon}). Though the Shannon wavelet does not have the perfect frequency localisation provided by the Fourier transform, it allows the separation of the frequency content into distinct non-overlapping bands for different scales. One disadvantage of the Shannon wavelet is that it does not have a compact support in time and its corresponding discrete wavelet transform is dense, thus increasing the computational cost of the inversion of $\tilde{\mathsfbi{H}}$ and SVD of its inverse. For problems where the sparsity of the wavelet transform is important, we use the Daubechies wavelets, which trade the perfect bandpass property in frequency domain for a compact support in time. Higher index Daubechies wavelets will have larger temporal supports and will behave closer to perfect bandpass filters. For both the wavelets described, the discrete transform matrix is unitary \citep{Najmi2012, Mallat2001}. 

\reviewerone{In addition to numerical cost, we also consider the physical interpretation implied by the wavelet transform to inform our choice of wavelets. For example, in the channel flow case where critical layer dynamics are important, the Shannon wavelet would allow us to study wave speeds of interest with more precision due to its being a perfect bandpass filter \citep{ballouz2023wavelet}. 
Constraining the resolvent modes to a particular dilation of the Shannon wavelet would highlight waves of frequencies contained in the frequency band of the chosen wavelet.
If more precision is required in the time domain to study time-localised phenomena, a wavelet that is compactly supported in time is more pertinent.  
This is the case, for example, in \cite{ballouz2023transient}, which studies how a compactly supported resolvent forcing function affects a turbulent channel beyond its time support.
In some cases, both properties can be combined through the use of wavelets that are compactly supported in time but whose frequency Fourier spectra decay quickly outside a certain frequency band, making them quasibandpass filters. The Daubechies or the Fej{\'e}r-Korovkin wavelets satisfy these properties, and in the case of turbulent channel flow, produce similar results as when a Shannon wavelet-transform is used.}

\subsubsection{Computational cost}

The construction of $\tilde{\mathsfbi{H}}$ (equation \eqref{eq:def_wavelet_based_H}) requires the inversion of a $4N_y N_t \times 4N_y N_t$ matrix, a computation that costs $O(N_y^3 N_t^3)$ operations when solved directly. The full SVD of $\tilde{\mathsfbi{H}}$ would also require $O(N_y^3 N_t^3)$ operations. With a direct solve, the wavelet-based resolvent analysis would cost $O(N_t^2)$-times more than performing $N_t$ separate Fourier-based resolvent for each temporal scale, though the latter would fail to capture the interactions between the different time scales. This penalty of $O(N_t^2)$ is the nominal cost of constructing time-localised resolvent modes. Below, we discuss some methods to reduce the cost and memory storage requirements of such a computation.

One method for reducing the memory and computational cost of wavelet-based resolvent analysis is to use sparse finite difference operators and wavelet transforms when constructing $\tilde{\mathsfbi{H}}^{-1}$. We then factor the resulting sparse matrix using specialised packages such as MATLAB's 'decomposition' function,
and save the factors in order to later solve linear equations of the form $\tilde{\mathsfbi{H}}^{-1} \mathsfbi v = \mathsfbi w$, where $\mathsfbi v$ and $\mathsfbi w$ are arbitrary vectors, without having to invert $\tilde{\mathsfbi{H}}$ again. This is useful in the context of iterative methods for computing the SVD of $\tilde{\mathsfbi{H}}^{-1}$. Though much of the sparsity of $\tilde{\mathsfbi{H}}$ is lost by the factorisation process, we note that the factors still exhibit significant sparsity. In this work, to take advantage of the sparse precomputed factors of $\tilde{\mathsfbi{H}}$, we opt for an iterative method to perform the SVD. We use a one-sided Lanczos bidiagonalisation \citep{simon2000low}, which additionally allows us to compute a truncated SVD and accurately estimate a number $q < 4N_yN_t$ of the most significant singular input and output modes. 

Other efficient SVD algorithms rely on randomised approaches, in particular by subsampling the high-dimensional matrix and performing the SVD on the lower-dimensional approximation \citep{halkotropp, drineas2016randnla, tropp2017}.  
Modifications of randomised SVD algorithms, notably randomised block Krylov methods \citep{musco2015randomized}, have been additionally developed for matrices with slow-decaying singular values, a property exhibited by the resolvent operator in \S\ref{sec:stokes}. 
A randomised SVD of a high-dimensional discrete resolvent operator was used in \citet{ribeiro2020randomized} and \citet{yeh2020resolvent}.

Another option that would avoid the direct inversion of $\mathsfbi{ \tilde H}^{-1}$ involves taking the SVD of $\mathsfbi{\tilde H}^{-1}$ first. The left and right singular vectors of $\mathsfbi {\tilde H}^{-1}$ are respectively the right and left singular vector of $\mathsfbi{\tilde H}$. However, since we are looking for the largest singular values of $\mathsfbi{\tilde H}$ and their corresponding singular vectors, we would have to compute the full SVD of $\mathsfbi {\tilde H}^{-1}$ to find its smallest singular values and corresponding singular vectors. Though this method avoids the inversion of $\mathsfbi {\tilde H}^{-1}$, it does not preserve the efficiency gains of a (heavily) truncated SVD, and should only be used if the sparse factorisation of $\mathsfbi {\tilde H}^{-1}$ remains the costliest operation. Suppose for example that $\mathsfbi {\tilde H}^{-1}$ has $n_{nz} \leq (4 N_y N_t)^2$ non-zero elements, and that the LU-factorisation of $\mathsfbi {\tilde H}$ has at most $m_{nz} \leq (4 N_y N_t)^2$ non-zero elements. Suppose that $n_{nz}$ is small enough that the cost of the LU-factorisation is small. A full iterative SVD of $\mathsfbi {\tilde H}^{-1}$ has complexity $O(4N_yN_tn_{nz})$, whereas a $q$-truncated SVD of $\mathsfbi {\tilde H}$ has complexity $O(m_{nz} q)$. Thus, if $n_{nz}/m_{nz} < q/(4N_yN_t)$, it is more efficient to compute an SVD of $\mathsfbi {\tilde H}^{-1}$ without computing an LU-factorisation. For the turbulent Stokes boundary layer problem considered in \S\ref{sec:application}, $q = 400$ modes are calculated and $q/(4N_yN_t) \approx 0.001$. Using a second-order finite difference operator in time and Daubechies-8 wavelet transform, $n_{nz}/m_{nz} \approx 0.24$, making the factorisation and truncated SVD method more efficient. In general, since we compute a heavily truncated SVD ($q/(4N_yN_t)$ is very small), we find that a factorisation of the sparse system prior to the SVD is advantageous.

Resolvent analysis can also be performed more efficiently for the windowed systems described in \S\ref{sec:windowing}. Indeed, $\mathsfbi{B}\tilde{\mathsfbi{H}}\mathsfbi{C} = (\mathsfbi{B}^\dagger\tilde{\mathsfbi{H}}^{-1}\mathsfbi{C}^\dagger)^{\dagger}$, where the superscript $\dagger$ indicates the Moore-Penrose pseudoinverse. 
Rather than form the resolvent operator $\tilde{\mathsfbi{H}}$ first through an inversion, we can reduce the dimension of the system by windowing the linearised Navier-Stokes operator \emph{prior} to taking the pseudoinverse of the windowed system. The matrix pseudoinversion and SVD will be applied to a lower-dimensional matrix of size defined by the non-zero block of $\mathsfbi{B}\mathsfbi{C}$.

\subsubsection{Choice of time differentiation matrix} \label{sec:timediff}
The choice of the discrete time differentiation operator $\mathsfbi{D}_t$ has a significant impact on the computation of resolvent modes. The sparsity of $\mathsfbi{D}_t$ controls the sparsity of the resolvent operator, which heavily affects the memory and complexity requirements of the computation of the resolvent modes. However, though a sparse $\mathsfbi{D}_t$ seems beneficial, it also distorts the time differentiation for high-frequency waves and can falsify the results of the SVD. 

To illustrate this, we study the spectra of two time-derivative matrices, $\mathsfbi D_{t,2}$, a second-order centred  finite difference matrix, and $\mathsfbi D_{t,F}$, a Fourier derivative matrix. The eigenvectors of both operators are the discrete Fourier modes. The eigenvalues of the Fourier derivative matrix $\mathsfbi D_{t,F}$ are simply the wavenumbers $\mathrm{i} 2\pi k /T$ for $k = 1,\,2\, ... \lceil{N_t/2}\rceil-1$, while those of the second-order centred  difference matrix $\mathsfbi D_{t, 2}$ are the modified wavenumbers $\mathrm{i} \, \mathrm{sin}(2\pi k / N_t) N_t/T$. For a fixed wavenumber $k$, we note that

\begin{equation}
    \lim_{N_t \to + \infty} \mathrm{sin}\left (\frac{2\pi k}{N_t} \right ) \frac{N_t}{T} = \frac{2\pi k} {T},
\end{equation}
and our modified wavenumber converges to the correct value. Now consider the maximum wavenumer $k = \lceil{N_t/2}\rceil - 1$. Suppose without loss of generality that $N_t$ is even so that $\lceil{N_t/2}\rceil -1  = N_t/2 -1$. The limit as the resolution in time increases becomes
\begin{equation}
    \lim_{N_t \to +\infty} \mathrm{sin} \left ( \frac{2\pi}{N_t} \left ( \frac{N_t}{2}-1 \right )  \right ) \frac{N_t}{T} = \frac{2\pi}{T},
\end{equation}
which does not converge to the correct wavenumber $2\pi(N_t/2-1)/T$. Indeed, the error grows as $O(N_t)$. 

\reviewerthree{To study how the numerical properties of the resolvent operators introduced in sections \S2.1 and \S2.2 vary with the choice of discrete time derivative, we consider}
an arbitrary 
matrix $\mathsfbi A$, \reviewerthree{which can represent the spatial derivatives of the linearised Navier-Stokes equations, and} write the following approximation
\begin{equation}
\begin{split}
    (\mathsfbi D_{t,2} + \mathsfbi A)^{-1} & = (\mathsfbi D_{t, F} + \mathsfbi A + \mathsfbi D_{t,2} - \mathsfbi D_{t,F})^{-1}  \\
    &\approx (\mathsfbi D_{t, F} + \mathsfbi A)^{-1} - (\mathsfbi D_{t, F} + \mathsfbi A)^{-1} (\mathsfbi D_{t,2} - \mathsfbi D_{t,F}) (\mathsfbi D_{t, F} + \mathsfbi A)^{-1} + ... \; .
\end{split}
\end{equation}
Thus, 
\begin{equation}
\Vert (\mathsfbi D_{t,2} + \mathsfbi A)^{-1} - (\mathsfbi D_{t, F} + \mathsfbi A)^{-1}\Vert_2 \leq  \mathcal O (\sqrt{N_t}) \; \Vert (\mathsfbi D_{t, F} + \mathsfbi A)^{-1} \Vert_2^2 + \mathcal O (N_t).
\end{equation}
The lack of convergence as $N_t$ increases suggests that the use of a finite difference operator rather than a Fourier derivative can significantly distort the SVD of the resolvent operator. To benefit from the advantages of a sparse temporal finite difference operator while avoiding spurious SVD modes, 
we propose using the windowing procedure described in \S\ref{sec:windowing} to filter-out the wavelet scales associated with the high-frequency wavenumbers more susceptible to distortion. 
Specifically, rather than choose the windowing matrices $\mathsfbi B$ and $\mathsfbi C$ to highlight a physically interesting range of the frequency spectrum, we use them to exclude the frequencies above a threshold $k_\mathrm{max} < N_t/2$. 
The maximum error between the eigenvalues of $\mathsfbi D_{t, 2}$ and $\mathsfbi D_{t, F}$ is given by the Taylor expansion 
\begin{equation}
\mathrm{sin} \left( \frac{2\pi k_\mathrm{max}}{N_t}\right) \frac{N_t}{T} - \frac{2\pi k_\mathrm{max}}{T} = \frac{4\pi^3 k_\mathrm{max}^3} {3 N_t^2 T} + \mathcal O \left ( \frac{k_\mathrm{max}^5}{N_t^4} \right ).
\end{equation}
Thus, assuming the chosen wavelet transform $\mathsfbi W$ is unitary, 
\begin{align}
\Vert \mathsfbi B \mathsfbi W \left( \mathsfbi D_{t,2} + \mathsfbi A)^{-1} - (\mathsfbi D_{t, F} + \mathsfbi A)^{-1} \right ) \mathsfbi W^{-1}\mathsfbi C \Vert_2 \leq  \mathcal O \left (\frac{k_\mathrm{max}^{3/2}}{N_t}\right ) \; \Vert (\mathsfbi D_{t, F} + \mathsfbi A)^{-1} \Vert_2^2 .
\end{align}
The error between the SVD of the two operators will decrease as $1/N_t$ provided $k_\mathrm{max}$ remains fixed. In \S\ref{sec:application}, we employ this filtering approach for cases using a finite difference time derivative operator.

\section{Application to statistically stationary  flow} \label{sec:application}

In this section, we first validate wavelet-based resolvent analysis on a statistically stationary  turbulent channel flow, for which the wavelet- and Fourier- approaches are equivalent provided that we use a unitary wavelet transform. Thus, for the channel flow case, we expect the two methods to produce identical resolvent modes. After confirming this result, we exploit the time-localisation property of wavelet-based resolvent analysis to study transient growth in turbulent channel flow.

\subsection{Turbulent channel flow} \label{sec:app:channel}

\begin{figure}
\begin{center}
\vspace{0.1cm}
\subfloat[]{\includegraphics[width=0.45\linewidth]{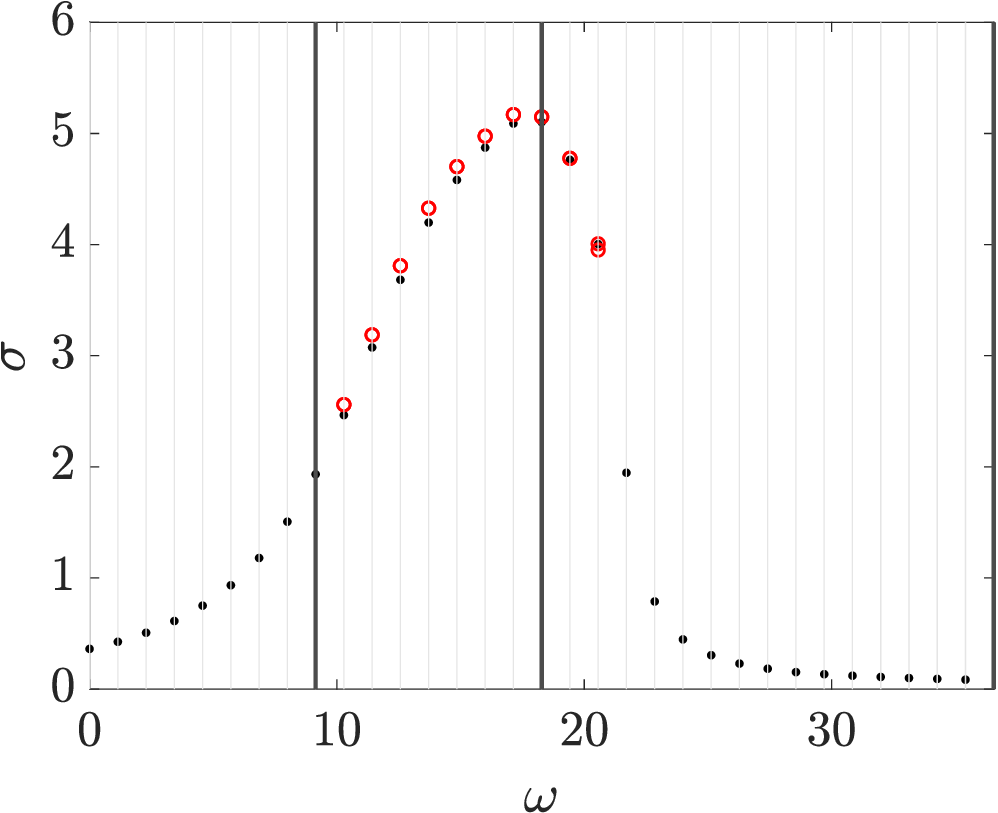}}
\hspace{0.3cm}
\subfloat[]{\includegraphics[width=0.47\linewidth]{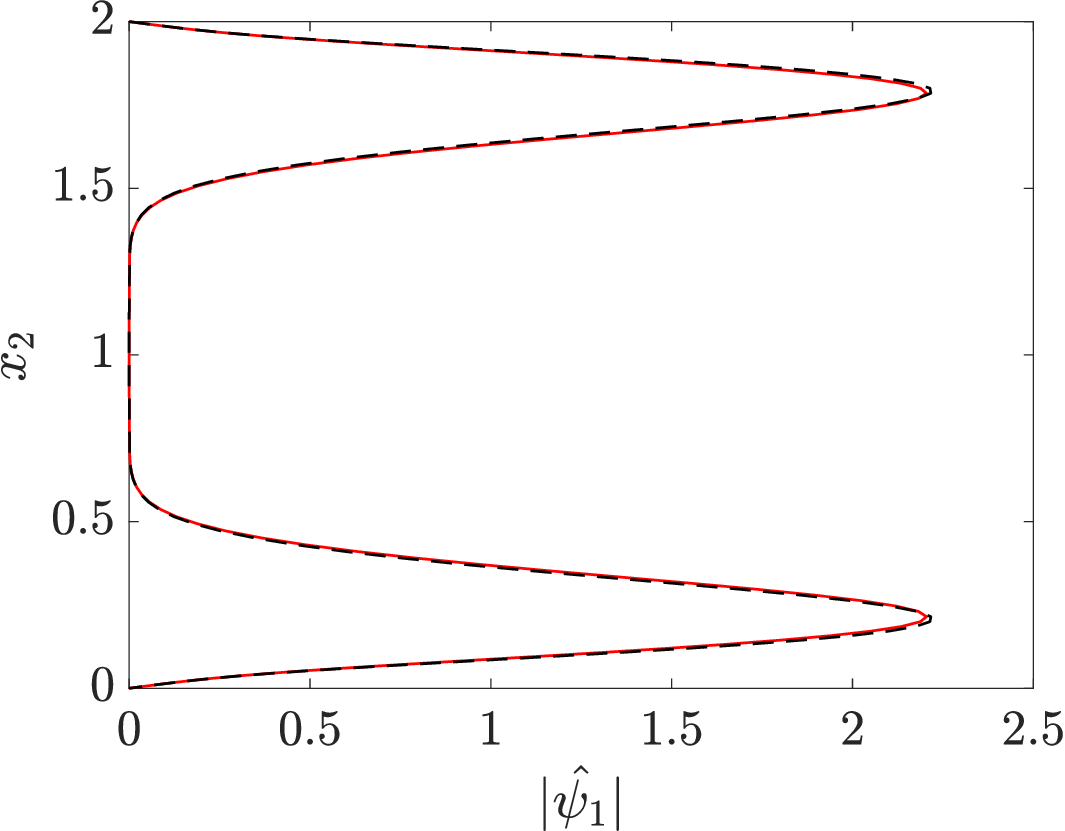}}  
\end{center}
\caption{\label{fig:compare}(a) First 10 singular values for the wavelet-based resolvent (red) and the largest singular value of the Fourier-based resolvent operator (black) computed for each $\omega_\ell$. The vertical gray lines indicate the frequencies resolved by the chosen temporal grid, and the vertical black lines delimit the frequency band covered by each of the wavelet scales used in the wavelet expansion. (b) Magnitude of the \reviewerthree{streamwise component} of the principal resolvent mode corresponding to $\omega\approx 17.14$. The red line represents $\mathsfbi F \boldsymbol{\breve \psi}_1$, the mode obtained from wavelet-based resolvent analysis after a Fourier transformation in time, and the black line the mode obtained from traditional Fourier-based resolvent analysis $\boldsymbol{\hat \psi}_1 $. In both cases, $\lambda_1^+ = 1000$ and $\lambda_3^+ = 100$.} 
\end{figure} 

\begin{table}
    
\begin{tabularx}{ \textwidth}  { >{\raggedright\arraybackslash} X  >{\raggedright\arraybackslash} X
   >{\raggedright\arraybackslash}X 
   >{\raggedright\arraybackslash}X  }

 Case & Section & $L^*$ & $u^*$ \\
Channel flow   & \S\ref{sec:application} &$\delta^*$ (channel half-height)    & $u_\tau^*$\\
\vspace{0.1cm} Turbulent Stokes boundary layer &\vspace{0.1cm}  \S\ref{sec:stokes} & \vspace{0.1cm} $\delta^*_\Omega$ (laminar boundary layer thickness)   & \vspace{0.1cm} $U_\mathrm{max}^*$ (max wall velocity) \\
\vspace{0.1cm}
 Channel flow with spanwise pressure gradient & \vspace{0.1cm} \S\ref{sec:3DChannel}& \vspace{0.1cm} $\delta^*$ (channel half-height)  & \vspace{0.1cm} $u_{\tau, 0}^*$ (wall shear velocity at $t=0$)\\
\end{tabularx}
\label{table:1}
\caption{Length and velocity used to non-dimensionalise the Navier-Stokes equations for each case considered in this paper.}
\end{table}

The mean profile of turbulent channel flow at friction Reynolds number $\Rey_\tau \approx 186$ is obtained from \cite{bae2021life}.
We non-dimensionalise using the channel half-height $\delta^*$ and the friction velocity $u_\tau^*$ as shown in table 1, so that $\Rey = \Rey _\tau$. We note that $\mathsfbi{U}_{3}$, $\mathsfbi{U}_{2}$, $\mathsfbi{dU}_{2,2}$, and $\mathsfbi{dU}_{3,2}$ are zero due to the absence of a spanwise and wall-normal contribution to the mean velocity profile.

For resolvent analysis, the wall-normal direction is discretised using a Chebyshev collocation method using $N_2 = 128$, and the mean streamwise velocity profile and its wall-normal derivative from the DNS are interpolated to the Chebyshev collocation points. We uniformly discretise the temporal domain, $[0, T)$, where $T = 5.5$ \reviewertwo{($T^+ = 1023$)}, with a temporal resolution of $N_t = 128$.
The superscript $+$ denotes wall units, which are defined to be $(\cdot)^+ := (\cdot) \Rey_\tau$ for length \reviewertwo{and time} scales and $(\cdot)^+ := (\cdot)$ for velocity scales.
We impose periodic boundary conditions at the edges of the time window. The temporal boundary conditions are encoded in the choice of time differentiation matrix $\mathsfbi D_t$, which we choose to be a Fourier differentiation matrix.
For spatial derivatives in the wall-normal direction, we use first- and second-order Chebyshev differentiation matrices, and impose no-slip and no-penetration boundary conditions at the wall. In choosing $k_1$ and $k_3$, we target spanwise and streamwise wavelengths of $\lambda_1^+ \approx 1000$ and $\lambda_3^+ \approx 100$ in wall units, \reviewertwo{which are the typical length scales for near-wall streaks and correspond to the peaks in the streamwise spectrum.}
\reviewertwo{For our wavelet transform, we choose a two-stage ($L = 2$) Shannon wavelet transform, \emph{i.e.} the state vectors in the resulting system described by equation \eqref{eq:in_out_w} will contain terms covering the following three intervals in frequency domain: $N_t/4 [-\pi, \pi]$ and $N_t/2^\ell [-2\pi, -\pi] \cup [-2\pi, -\pi]$ for $\ell = 1, \;2$.}
We note that, for this application, the resolvent modes converge despite the relatively low dimension of the resolvent operator. This permits us to use the aforementioned dense differentiation matrices. Sparse finite difference matrices may be used in higher-dimensional problems to improve efficiency.
\reviewerthree{
The wavelet- and Fourier-based cases, described by equations \eqref{eq:Disc_Wav_NS} and \eqref{eq:Disc_NS} respectively, only differ by their time differentiation matrices. 
These satisfy $\mathsfbi W^{-1}\widetilde{\mathsfbi D_t} \mathsfbi W = \mathsfbi F^{-1} \widehat{\mathsfbi D}_{t, F} \mathsfbi F = \mathsfbi D_t$, where $\widehat{\mathsfbi D}_{t,F}$ is the diagonal matrix with diagonal terms $(\widehat{\mathsfbi D}_{t, F})_{\ell \ell } = -\mathrm{i} \omega_\ell = \mathrm{i} (2\pi \ell)/T$ for $\ell = -{N_t}/{2}, \cdots, {N_t}/{2}-1$, and $\mathsfbi F$ is the discrete Fourier transform in time.
The Shannon wavelet transform is unitary, and since the SVD is unique up to multiplication by a unitary matrix, we expect the singular values of $\widetilde{\mathsfbi H}^{(k_1, k_3)}$ to be the same as those for
}
\begin{equation}\label{eq:harmonicSys}
\widehat{\mathsfbi{H}}^{(k_1,k_3)} = 
\left (
\begin{array}{cccc}
\widehat{\mathsfbi H}^{(k_1, k_3, \omega_1)} & & & \\
& \widehat{\mathsfbi H}^{(k_1, k_3, \omega_2)} & & \\
& & \widehat{\mathsfbi H}^{(k_1, k_3, \omega_3)} & \\
& & & \ddots \\
\end{array} \right ),
\end{equation}
\reviewerthree{where each $\widehat{\mathsfbi H}^{(k_1, k_2, \omega_\ell)}$ is defined according to equation \eqref{eq:def_fourier_based_resolvent}.} Moreover, we expect the response and forcing modes of both systems to be related by the unitary transform given by the Fourier and inverse-wavelet transform in time, $\mathsfbi{FW}^{-1}$. 

The results for the Fourier-based cases were computed by applying traditional resolvent analysis at each $\omega_i$ captured by our temporal grid, while the wavelet-based resolvent modes were computed by solving the full space-time system at once. \reviewertwo{We consider the results from wavelet-based resolvent analysis to be converged: the singular values obtained for $(N_t, N_2)= (128, 128)$, and for $(N_t, N_2)= (64, 128)$ produced the same leading singular values. These are not shown in this work.}
As implied by equation \eqref{eq:harmonicSys}, a single wavelet-based resolvent analysis will yield the modes corresponding to all time scales captured by the temporal grid. In order to associate each singular value from the wavelet-based resolvent analysis to a frequency, we Fourier-transform the corresponding response mode in time, and identify the index of the non-zero Fourier component. 

In figure~\ref{fig:compare}(a), we show the 10 leading singular values of the wavelet-based resolvent operator, along with the first singular value for the Fourier-based operator for different temporal Fourier parameters $\omega_\ell$. In order to associate the singular values obtained from the wavelet-based method with their frequencies, we Fourier-transform each wavelet-based response mode in time, and identify the index of the non-zero component.
\reviewertwo{The first 10 singular values obtained differ by at most $3\%$, with the largest differing by $1.3\%$, which matches our expectation}. The discrepancy can be explained by numerical and truncation errors. Though Shannon wavelet transforms are unitary in the continuous setting, Shannon wavelets do not have compact support in time. The discrete Shannon transform is thus not a unitary matrix due to the truncation of the wavelet in time, and exhibits a condition number of approximately $1.6$ in this case. 
Using wavelets that are compactly supported in time, such as the Daubechies wavelets, reduces the discrepancy, \reviewerthree{as the resulting wavelet transform matrix $\mathsfbi W$ is a unitary operator with a condition number of 1 that better preserves the singular values of $\widehat{\mathsfbi H}^{(k_1, k_3)}.$} 
Increasing the time resolution also reduces the gap between the singular values.
We note that due to the symmetry of the problem about the centreline, the singular values appear in equal pairs \citep{mckeon2010critical}. This is \reviewerthree{only} visible in figure~\ref{fig:compare}(a) for $\omega  = 20.56$, where the pair of singular values deviate slightly for each other due to numerical error. The modes corresponding to the pair of equivalent singular values are reflections of each other about the channel centreline. 

In figure ~\ref{fig:compare}(b), we compare the two methods further by plotting the streamwise component of the most amplified resolvent response mode that they each produce. For the Fourier-based method, this corresponds to the frequency $\omega \approx 17.14$. For the wavelet-based method, we first Fourier-transform the principal mode in time, and \reviewerthree{though not shown,} observe that the Fourier modes associated with $\omega \approx 17.14$ is the only non-zero component. \reviewerthree{This matches our expectation that the wavelet-based resolvent analysis in this case is equivalent to performing the traditional Fourier-based resolvent analysis for each frequency $\omega_\ell$ individually.} Figure~\ref{fig:compare}(b) shows that the modes from the two methods match. Despite the slight discrepancy in the singular values, both methods yield the same resolvent modes associated with the maximum singular value, indicating that both methods are equivalent for this stationary case. 

Although not shown, the streamwise component of the modes dominate for the principal modes computed with the two methods, and the modes form alternating low- and high-speed streamwise streaks. 
The principal forcing mode is in the form of streamwise rolls, with a negligible streamwise component. The shape of the modes is thus in line with the previous analysis of the self-sustaining process of wall turbulence \citep{jimenez1991,hamilton1995regeneration,jimenez1999,waleffe1997,schoppa2002,farrell2017,bae2021nonlinear}. \reviewerthree{We do however note that the response modes peak at $x_2 \approx 0.21$, corresponding to $x_2^+ \approx 40$, higher than the location of $x_2^+ \approx 15$ preferred by near-wall streaks. 
This is also in line with previous literature on Fourier-based resolvent analysis \citep{schmid2000stability, mckeon2010critical, mckeon2017engine, mckeon2019self}: resolvent response modes, both formulated traditionally and using the wavelet basis in time, peak at the critical layer located where $U(x_2) = \omega / k_1$. Using a version of the linearised Navier-Stokes that includes eddy viscosity} \citep{symon2023eddy} \reviewerthree{causes the modes to peak closer to $x_2^+ = 15$ so that they match observations of the near-wall cycle.}

\subsection{Transient growth mechanism of turbulent channel flow} \label{sec:transient}

\begin{figure}
    \centering
    \vspace{0.4cm}
    \includegraphics[width=0.45\textwidth]{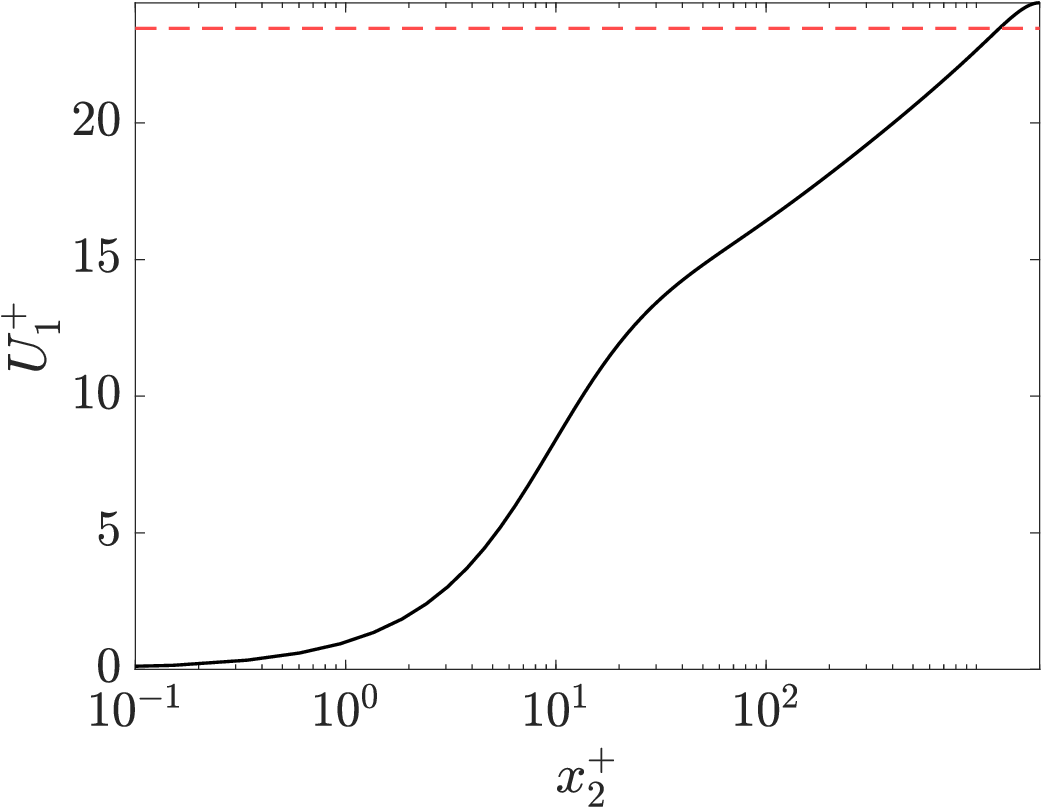}
    \caption{Mean streamwise velocity profile for channel flow at $\Rey_\tau = 2000$. The dashed line corresponds to the upper bound of the frequency band covered by the chosen wavelet, mapped to streamwise velocities using $U_1 = \omega/k_1$.}
    \label{fig:buffer}
\end{figure}

\begin{figure}
\begin{center} 
\subfloat[]{\includegraphics[width=0.474\linewidth]{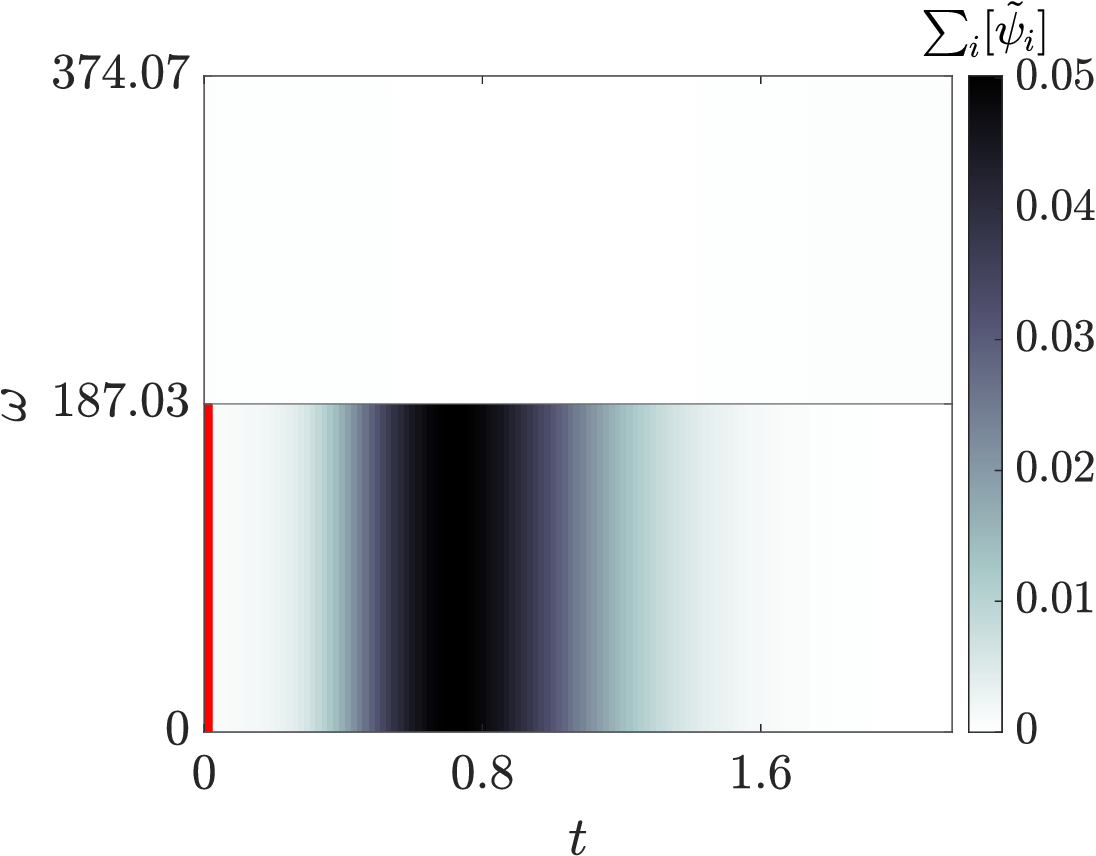}}
\hspace{0.2cm}
\subfloat[]{\includegraphics[width=0.44\linewidth]{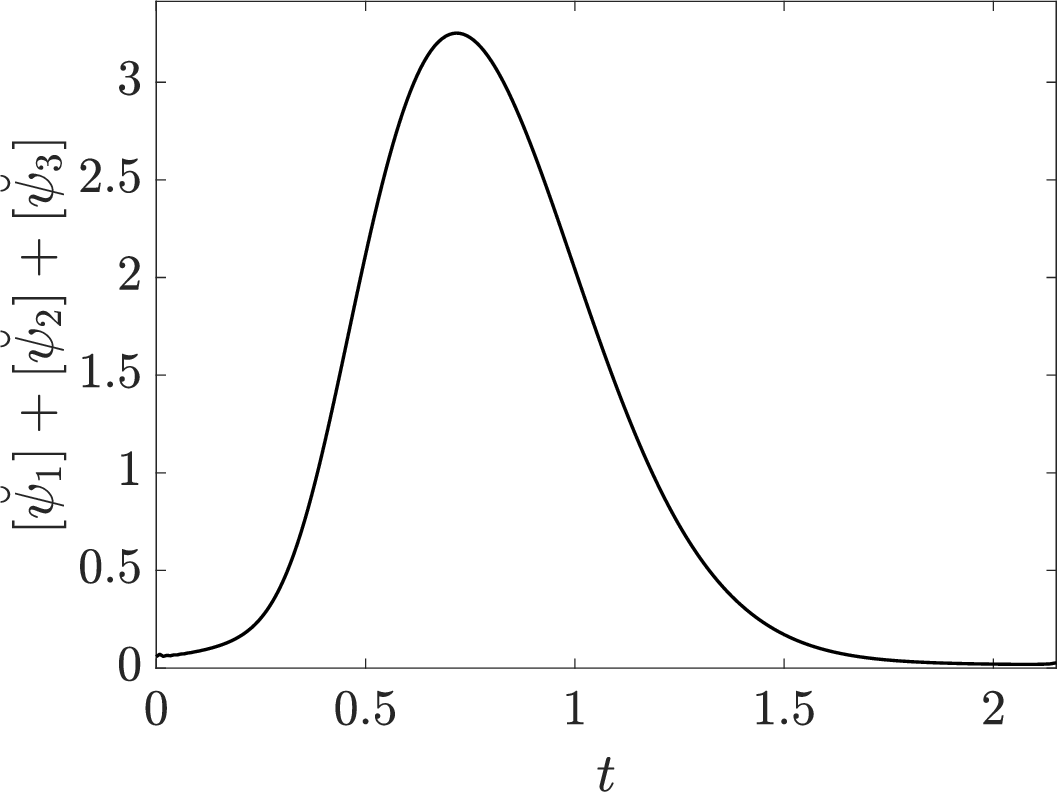}}
\end{center}
\caption{(a) Inverse wavelet-transformed principal response mode for channel flow at $\Rey_\tau = 2000$ and $k_1 = k_3 \approx 8.98$ under windowed forcing, in the frequency--time plane. The forcing window is highlighted in red. (b) $x_2$-integrated total energy of the inverse wavelet-transformed principal response mode.}
\label{fig:results}
\end{figure}

\begin{figure}
\begin{center}
\subfloat[]{\includegraphics[width=0.7\linewidth]{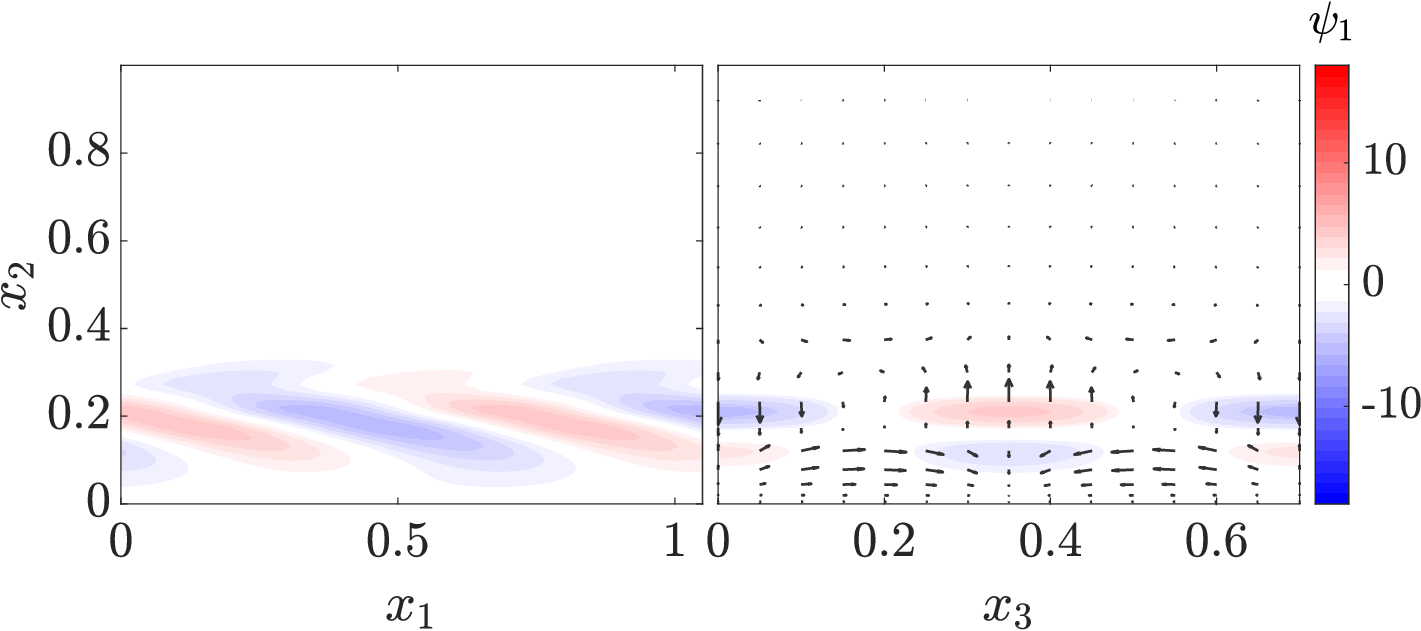}} 
\\
\subfloat[]{\includegraphics[width=0.7\linewidth]{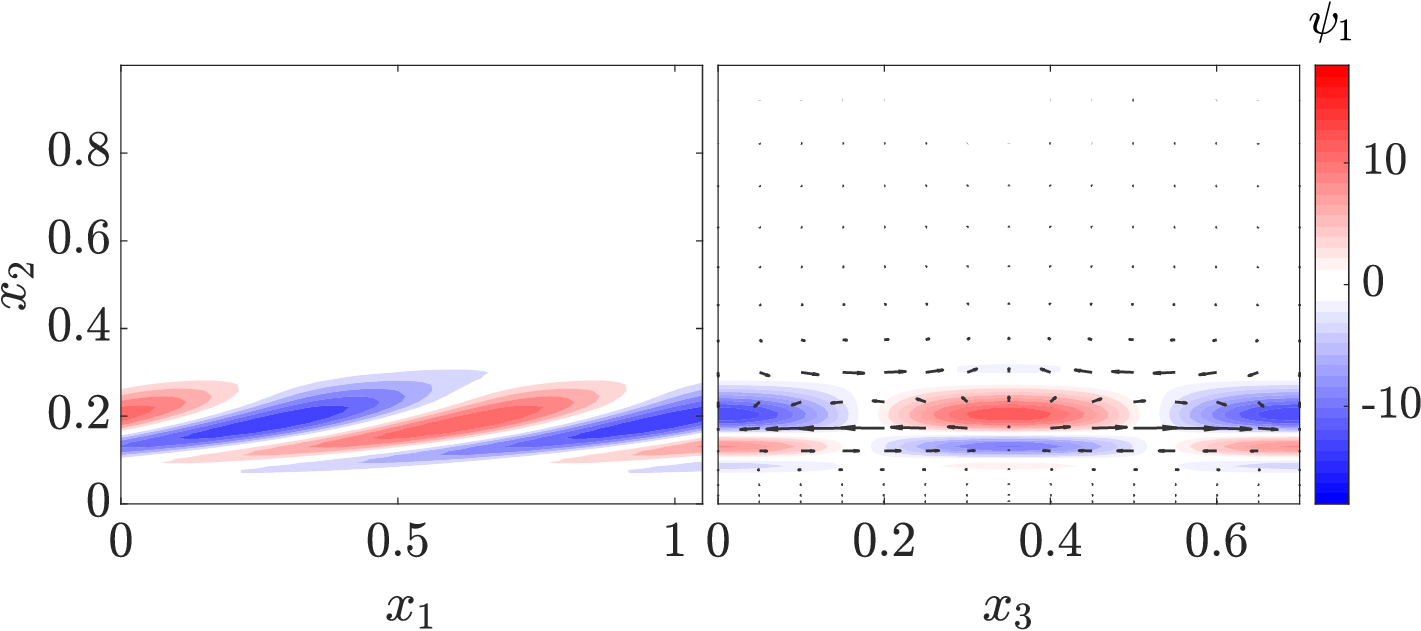}} 
\\
\subfloat[]{\includegraphics[width=0.7\linewidth]{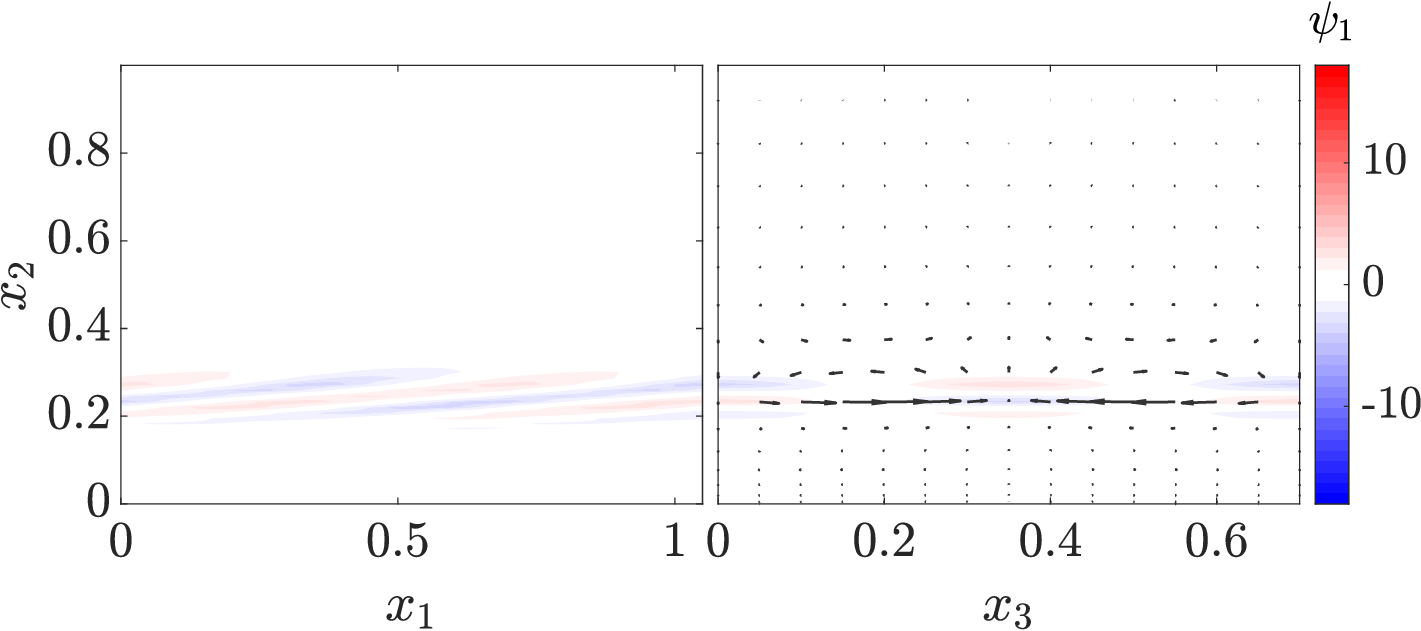}} 
\end{center}
\caption{Principal response mode for the channel flow at $\Rey_\tau=2000$ and $k_1 = k_3 \approx 8.98$ under transient forcing. The modes are shown at 
(a) $t = 0.25$, (b) $t = 0.71$, and (c) $t = 1.56$.
The left and right panels respectively correspond to half-wavelength locations of $x_3/\delta = \pi/k_3 \delta$ and $x_1/\delta = \pi/k_1 \delta$. 
}

\label{fig:streaks}

\end{figure}

\begin{figure}
    \begin{center}
    \vspace{0.4cm}
    \includegraphics[width=0.5\linewidth]{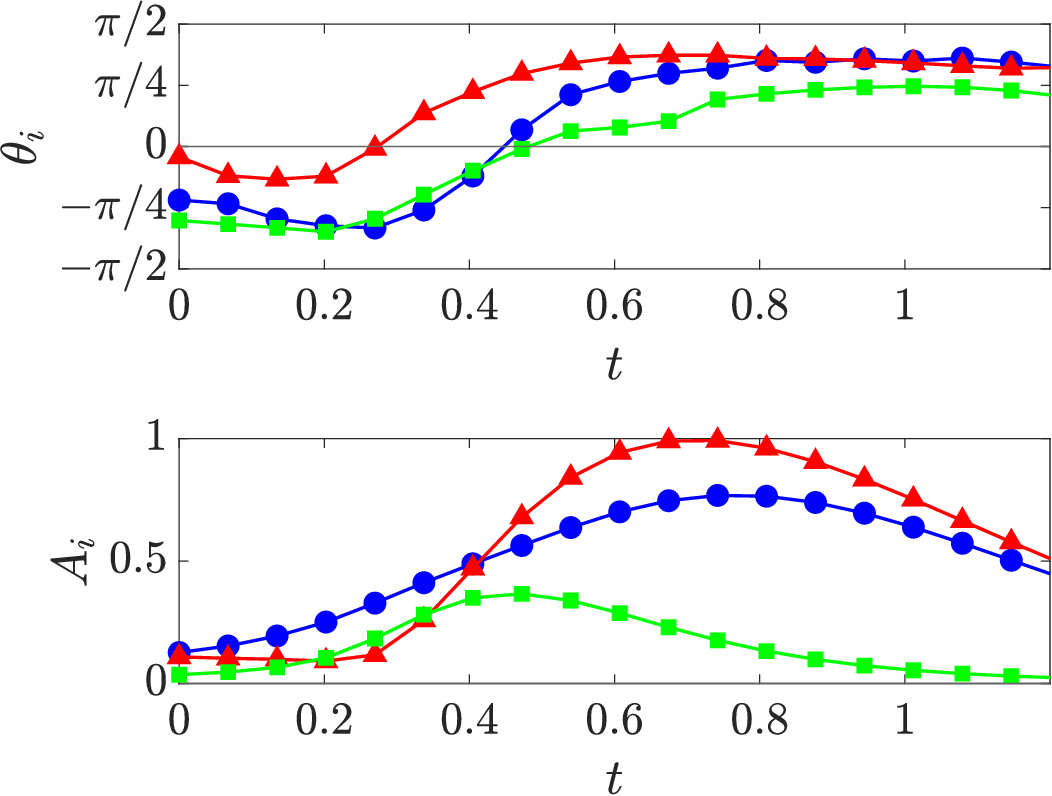}
    \end{center}
    \caption{Mode angle (top) and magnitude (bottom) of the principal response mode as a function of time.
    The streamwise component of the mode is represented by $-\bullet$ (blue), the wall-normal component by $-\blacksquare$ (green) and the spanwise component by $- \blacktriangle$ (red)}
    \label{fig:Orr}
\end{figure}

The added advantage of the wavelet-based method lies in its ability to preserve temporal localisation. The states in equation \eqref{eq:windowed_res} encode time and frequency information, which allows us to study transient problems even when the mean profile is statistically stationary. One such transient phenomenon is the Orr mechanism, a linear mechanism first described by \citet{orr1907} that has been proposed to explain transient energy amplification in shear flows \citep{jimenez2013linear, jimenez2015bursts, landahl1975, jimenez2018coherent, encinar2020momentum}.
A two-dimensional physical description is given in \citet{jimenez2013linear, jimenez2018coherent}: the mean shear profile rotates backward-tilting velocity structures forward (in the positive $x_1-$ direction), effectively extending the wall-normal distances between structures; to compensate, continuity will impose larger wall-normal fluxes, \emph{i.e.} larger wall-normal velocity perturbations. This effect amplifies the velocity perturbations until the velocity structures are tilted past the normal to the wall, after which the mechanism is reversed and the perturbations are attenuated. 
The tilting angle of the velocity perturbations does not affect the streamwise or spanwise velocities directly, and after those components are amplified by the growth of the wall-normal component through lift-up, they decay slowly through viscous effects \citep{jimenez2013linear}.
Thus, due to the long-lasting effects on the streamwise component of a brief amplification of the wall-normal velocity perturbation, it has been proposed that these Orr bursts are involved in the generation of persistent near-wall streaks \citep{jimenez2013linear, jimenez2015bursts, encinar2020momentum}.

Because of its possible role in regenerating near-wall streaks, the Orr mechanism in linearised wall-bounded flows has been examined in \citet{jimenez2013linear,jimenez2015bursts, jimenez2018coherent} and \citet{encinar2020momentum}. These studies rely on computing optimal growth trajectories for the linearised equations with viscosity as well as solutions to the inviscid linearised equations The optimal trajectories are defined as those emanating from the optimal initial condition which maximises the growth of kinetic energy under the linearised dynamics \citep{butler1993optimal, schmid2000stability}. The linear trajectories exhibit the characteristic forward tilting of velocity structures in conjunction with the transient amplification of velocity perturbations, suggesting that the Orr-mechanism is a dominant energy amplification mechanism in the linearised system. 

Optimal growth trajectories compute the singular modes for the linearised flow map between an initial condition and velocity perturbations at a later time; the question we wish to answer is whether optimal external forcing upon the linearised system, which could originate from nonlinear interactions, also exploits the Orr mechanism. For this, we use the wavelet-based resolvent analysis formulation. We note that traditional resolvent analysis has been used to reveal some evidence of the Orr mechanism in turbulent jets, where it is identified by the tilt of the optimal forcing structures against the jet shear \citep{pickering2020orr, tissot2017sensitivity, schmidt2018spectral}. In an attempt to capture the Orr mechanism in channel flow at $Re_\tau = 2000$ as in \citet{encinar2020momentum}, we use the mean profile for channel flow at $Re_\tau = 2000$ \citep{hoyas2008reynolds}, the same grid in the wall-normal direction as \S\ref{sec:app:channel} \reviewerthree{with $N_2 = 128$ collocation points,} and a uniform temporal grid with $N_t = 256$. As argued in \citet{encinar2020momentum}, we choose spatial wavelengths $\lambda_1 = \lambda_3 = 0.7$, and we choose a total time of  $T = 2.15$.

\reviewerthree{The choice of $T$ depends on the critical layer dynamics that we wish to highlight. From traditional resolvent analysis applied to turbulent stationary channel flow, we know that the resolvent Fourier modes tend to peak in magnitude at the critical layer, \emph{i.e.} where $U(x_2) = \omega / k_1$ \citep{schmid2000stability, mckeon2010critical, mckeon2017engine, mckeon2019self}. 
In this section, we study the effect of a time-localised forcing on a region $x_2^+ \in [0, 500]$ which contains the inner region of the boundary layer \citep{hoyasRe2000}.
This maps to a frequency interval $[0 ,U(x_2^+ = 500)] k_1 = [0, \omega_{\max}] \approx [0, 185]$ according to the mean profile for turbulent channel flow (figure \ref{fig:buffer}).
Thus, using a total time $T = 2.15$ and a Shannon wavelet transform with  $L = 1$, the Shannon scaling functions $\zeta(t/2^L - k)$ would cover the frequency interval $N_t/(2T)[-\pi, \pi] \approx [-187, 187]$. To restrict our forcing term to this frequency band while localising it time time,} we use the windowed wavelet-based resolvent analysis framework from \S\ref{sec:windowing}.
We set $\mathsfbi{C}$ to the identity matrix, allowing the response modes to cover the entire time and frequency range. 
We choose $\mathsfbi B$ to select the forcing terms corresponding to the relevant the Shannon scaling function. Without loss of generality, we select the shift parameter $\beta = 0$ so that the forcing term is concentrated at a time interval centred  at $t=0$. 

The principal resolvent response mode obtained from the SVD of $\tilde{\mathsfbi H}\mathsfbi B$ represents the maximally amplified response to a transient forcing term aligned with the selected wavelet, under the dynamics of the linearised Navier-Stokes.
The resulting principal response mode is confined to the frequency band determined by the forcing, as shown in figure \ref{fig:results}(a), which is expected since the time scales are decoupled in resolvent analysis for statistically stationary flows (equation \eqref{eq:harmonicSys}). \reviewerthree{The velocity fluctuations can only extract energy from the mean flow in the linearised formulation of resolvent analysis, and since the mean profile is solely constituted by a steady-state component, the response mode can only contain the frequencies injected by the forcing mode.}
We observe that the spatially integrated energy of the response mode first grows transiently, peaks at $t = 0.72$, then decays, as shown in figure \ref{fig:results}(b). 
This transient growth can be explained by the non-normality of the linearised system \citep{schmid2000stability}. The response modes at three different times are shown in figure \ref{fig:streaks}. 
\reviewerone{We note that the modes are concentrated in a region $x_2 < 0.4$ with a corresponding Corrsin shear parameter of approximately 10, which justifies the use of a linearised method to analyse the full system \citep{jimenez2013linear}.}

The transient behaviour of the modes displays characteristics of the Orr mechanism, mainly a synchronisation between the amplification of the wall-normal component of the response mode and its forward tilting.
To study the forward tilting of velocity structures in the response mode, we define the tilt angles as in \citet{jimenez2015bursts}, \emph{i.e.}
\begin{equation}
    \theta_i(x_2, t) = -\mathrm{tan}^{-1} \left( \frac{\partial_{x_2} \angle \boldsymbol{\breve \psi}_i (x_2, t)}{k_1} \right),
\end{equation}
where $\angle (\cdot)$ represents the complex angle.
In \citet{jimenez2015bursts}, the angle defined above is averaged over a region of interest. We define the energy-weighted average tilt as
\begin{equation}\label{eq:avg_tilt}
    \theta_i^{[y^+_a,y^+_b]}(t) = \frac{\int_{y_a}^{y_b} \Vert \boldsymbol{\breve \psi}_i \Vert^2 \theta_i dx_2}{ \int_{y_a}^{y_b} \Vert \boldsymbol{\breve \psi}_i \Vert^2 dx_2},
\end{equation}
and the amplitude as
\begin{equation}
     A_i^{[y^+_a,y^+_b]} = \left(\int_{y_a}^{y_b} \Vert \boldsymbol{\breve \psi}_i \Vert^2 dx_2 \right)^{\frac{1}{2}},
\end{equation}
and pick $y_a^+ = 0$ and $y_b^+ = 2000$ to capture the half-channel. The results (figure \ref{fig:Orr}) show that the amplitude of the wall-normal velocity component of $\boldsymbol{\breve \psi}_2$ indeed peaks roughly when $\theta^{[0, 2000]}_2 \approx 0$ at $t \approx 0.45$, and decays as $\theta_2^{[0,2000]}$ tilts past zero until it vanishes for $\theta_2^{[0,2000]} = \pi/4$.

\reviewerone{
We note that the spanwise component tilts forward much faster than the streamwise and wall-normal components (figure \ref{fig:Orr}). This faster tilt has been attributed to the spanwise component's placement closer to the wall where the shear is stronger \citep{encinar2020momentum}, and we do indeed find that $\boldsymbol{\breve \psi_3}$ is closer to the wall compared with $\boldsymbol{\breve \psi_2}$. However, we also find that $\boldsymbol{\breve \psi_1}$ is located at the same wall-normal height as $\boldsymbol{\breve \psi_3}$, and hypothesise 
that it tilts forward at the same rate as $\boldsymbol{\breve \psi_2}$ because of the more direct coupling between the two components via the $u_2 dU_1/dx_2$ term in the linearised momentum equations.

Moreover, both the streamwise and spanwise components of the mode peak 
at $t \approx 0.75$, \emph{i.e.} after $\theta_1^{[0,2000]}$ and $ \theta_3^{[0,2000]}$ cross the zero threshold and approximately $t \approx 0.35$ later than the amplitude peak for the wall-normal component.}
\reviewerone{The streamwise and spanwise components also decay more slowly than the wall-normal component, even as their tilt is relatively constant at $\theta_1^{[0, 2000]} = \theta_3^{[0,2000]} = \pi/4$ and after the wall-normal component vanishes. \cite{jimenez2013linear} explains this using the Squire equation, the linearised advection--diffusion equation for the wall-normal vorticity in which the wall-normal velocity perturbation acts as a forcing term. Even if this forcing disappeared, the streamwise and spanwise velocity components would continue advecting downstream and decay only due to viscous dissipation, which acts at a slower time scale than the mean shear in the near-wall region for the length scales considered. 
The non-normality of the Squire system further explains the delayed growth of the spanwise component after being forced by the wall-normal component. Note that the amplitude streamwise component rises in tandem with the wall-normal component, possibly due their tighter coupling via the presence of the $u_2 dU_1/dx_2$ term in the linearised streamwise momentum equation.
The tilt of the streamwise and spanwise components does not directly drive or suppress their amplitudes, since forward-tilting only affects their wall-normal gradients which do not appear in the equations of motion. This allows them to grow even after they attain their maximum positive tilt.
}

\reviewers{The lifetime of the wall-normal component $\boldsymbol{\breve \psi}_2$ of the principal resolvent response mode differs significantly from the bursting time scales in  \citet{jimenez2013linear, jimenez2015bursts} and \citet{encinar2020momentum}, which argue that the Orr-mechanism is a linear inviscid process whose period scales with the local mean shear. 
Defining the location of the wall-normal velocity fluctuation as its energy-weighted center of gravity
\begin{equation}\label{eq:center_of_gravity}
    x_{2, g} := \frac{\int_{0}^{1} \Vert u_2 (t_{\mathrm{max}})\Vert^2 x_2 dx_2 }{ \int_0^{1} \Vert u_2 (t_{\mathrm{max}}) \Vert^2 dx_2},
\end{equation}
where $t_{\mathrm{max}}$ is the time of maximum amplitude for $u_2$, we find that $x_{2, g} \approx 0.21$, further away from the wall than the inviscid and optimal growth solutions in \citet{jimenez2013linear} and \citet{encinar2020momentum} located at $x_{2, g} \approx 0.16 \lambda_1 \approx 0.11 $. 
Defining a shear time scale $S^{-1}$ as
\begin{equation}\label{eq:shear_timescale_1}
    S := \frac{dU_1}{dx_2}(x_2 = x_{2, g}),
\end{equation}
we find that the mode grows and decays in $tS \approx 13$ local shear units, much more slowly than the inviscid and optimal growth solutions, which grow and decay in $tS \approx 1$ and $tS \approx 3.5$ respectively. 
Moreover, the Orr mechanism studied in \citet{encinar2020momentum} is linked more broadly to energetic bursting events, which have been considered in \citet{jimenez2015bursts}, and which exhibit a bursting period of $t = 0.2$ for the wall-normal mode corresponding to $(\lambda_x, \lambda_z) = (0.7, 0.7)$, again faster than the growth time scale for $\boldsymbol{\breve \psi}_2$. }
\reviewertwo{In an attempt to explain the discrepancies between the optimal growth solution in \citep{encinar2020momentum} and the principal resolvent mode, we note that the measures of optimality differ between the optimal growth and wavelet-based resolvent analysis frameworks: while the optimal growth framework measures energy amplification as a ratio between the initial condition and the solution at a given time, wavelet-based resolvent analysis maximises the intergrated kinetic energy of the entire time interval considered. This may better capture energetic structures that persist in time. Moreover, resolvent analysis additionally provides the optimal forcing that produces the response in the velocity field, which is helpful in studying mechanisms that drive linear transient growth in turbulent flows.}

\reviewers{However, the bursting time scales of the principal resolvent mode agree with those found using turbulent channel flow data in \citet{encinar2020momentum}. Using an average shear time scale defined as
\begin{equation}\label{eq:shear_timescale_2}
    S_\Lambda := \frac{(U_1(x_2 = y_b) - U_1(x_2 = y_a))}{y_b - y_a},\; y_a = 0.7, \; y_b = 0.25,
\end{equation}
where $\Lambda = [y_a, y_b]$ captures a section of the logarithmic region, we find that $\boldsymbol{\breve \psi_2}$ takes $t S_\Lambda \approx 1.87$ average shear units to grow to half its maximum amplitude, compared with $t S_\Lambda \approx 1$ in \citet{encinar2020momentum}.
The amplitude and timescale of the process captured by $\boldsymbol{\breve \psi}$ can be further compared with other studies of bursting in the near-wall region of turbulent channel flow at moderate $\Rey_\tau$. }
\reviewers{In \citet{flores2010hierarchy}, turbulent energy is found to peak at time intervals of approximately $6 x_2$, and given that our resolvent modes are centred  at $x_2 \approx 0.2$, the trend would predict a bursting period of $t = 1.2$. This roughly matches the timescale $t \approx 1.5$ for the growth and decay of $\boldsymbol{\breve \psi_2}$. Likewise, in \citet{hwang2016self}, which studies bursting in the logarithmic region of the channel, the relation $t = 2 \lambda_3$ is found to describe the bursting period. For our case where $\lambda_3 = 0.7$, this would correspond to a bursting period of $t = 1.4$, again matching the lifetime of $\boldsymbol{\breve \psi_2}$. Thus, the Orr bursting period predicted by $\boldsymbol{\breve \psi_2}$ roughly agrees with the bursting time scale detected in turbulent channel flow, though this time scale tends to be significantly longer than the shear time scale found to govern the Orr mechanism in linear inviscid and linear optimal growth analyses.}

\section{Application to non-stationary flow} \label{sec:app_time}

We now apply wavelet-based resolvent analysis to problems with a time-varying mean flow. In particular, we study the turbulent Stokes boundary layer and a turbulent channel flow with a sudden lateral pressure gradient. The Stokes boundary layer is a purely oscillatory flow in time, and thus, Fourier-based resolvent analysis \citep{Padovan2020} still may be used. However, in the case of the temporally-changing channel flow, the flow is truly unsteady, and a Fourier transform in time is not applicable. 

\subsection{Turbulent Stokes boundary layer}\label{sec:stokes}

The Stokes boundary layer is simulated through a channel flow with the lower and upper walls oscillating in tandem at a velocity of $U^*_w(t) = U^*_{\max}$ cos$(\Omega^* t^*)$ with no imposed pressure gradient.
We non-dimensionalise velocities by $U^*_{\max}$ and lengths by $\delta^*_\Omega := \sqrt{2\nu/\Omega^*}$, which denotes the laminar Stokes boundary layer thickness. Though time and frequency are both non-dimensionalised with $ U^*_{\max}$ and $\delta^*_\Omega$, we will use $t\Omega$ as our preferred time variable for a clearer comparison with the period, and $\omega/\Omega$ as our preferred frequency variable as it represents temporal wavenumber in this case.
The relevant non-dimensional number is $\Rey_\Omega = U^*_{\max} \delta_\Omega^* / \nu$. For the current case, we consider $\Rey_\Omega = 1500$, which lies within the intermittently turbulent regime \citep{hino1976,akhavan1991,verzicco1996,vittori1998,costamagna2003}.  
This problem has been well-studied numerically and experimentally in the literature \citep{hino1976,spalart1989direct,jensen1989turbulent,akhavan1991,verzicco1996,vittori1998,costamagna2003,kerczekDavis,sarpkaya1993coherent,blondeauxVittori,carstensen,ozdemir}. 

\begin{figure}
    \begin{center}
    \vspace{0.1cm}
    \subfloat[]{\includegraphics[width=0.45\linewidth]{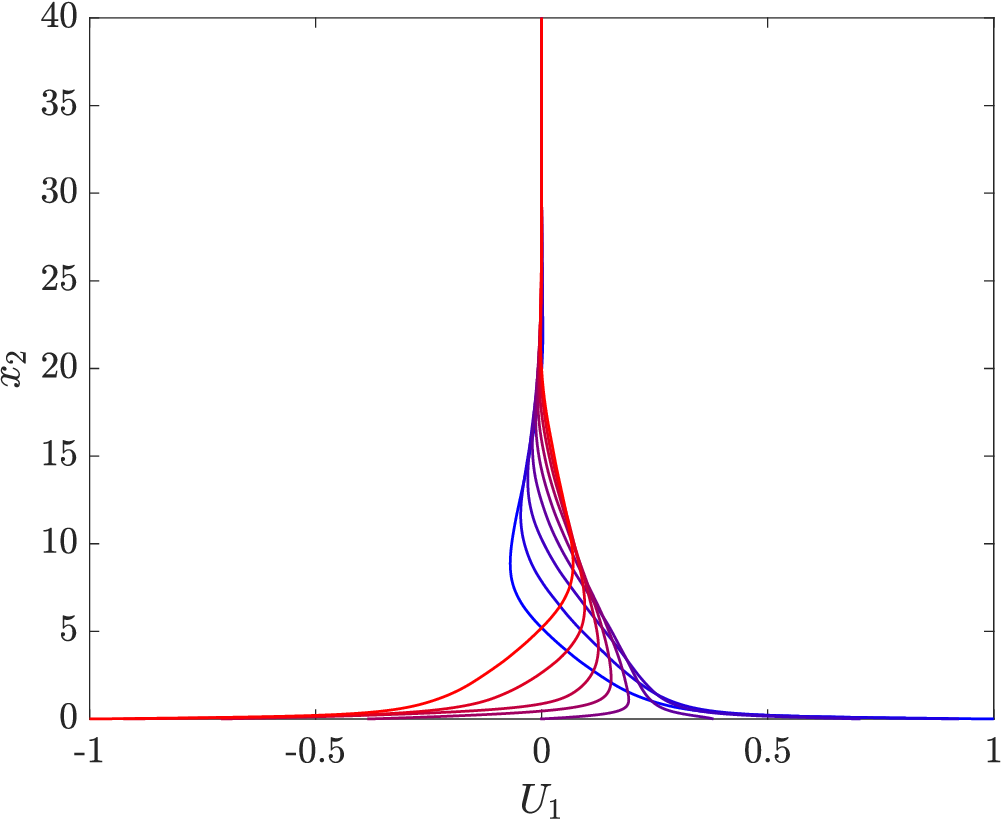}}
    \hspace{0.35cm}
    \subfloat[]{\includegraphics[width=0.45\linewidth]{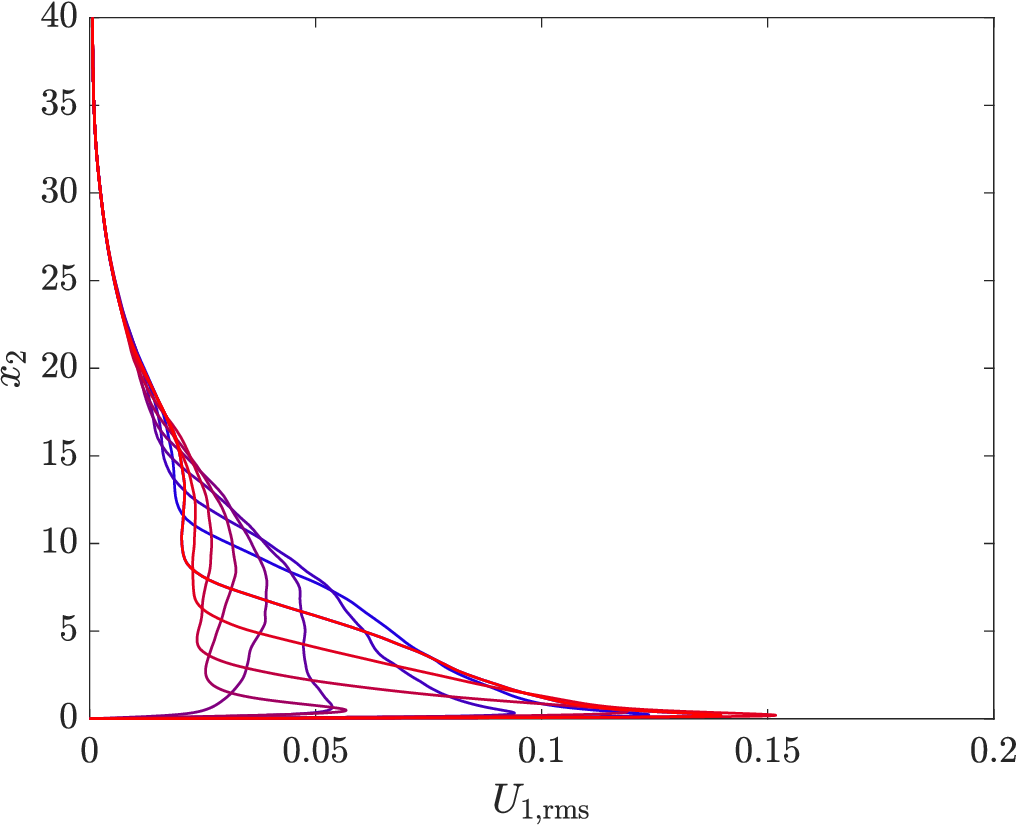}}
    \end{center}
    \caption{(a) Mean streamwise velocity profile and (b) streamwise r.m.s. velocity from $t\Omega = 0$ (blue) to $t\Omega = \pi$ (red). The profiles shown are at $t\Omega = n \pi/8$, $n = 0, 1, ... 8$.}
    \label{fig:StokesMean}
\end{figure}

To generate the mean profile and second-order statistics, we run a direct numerical simulation (DNS) using a second-order staggered finite-difference \citep{orlandi2000} and a fractional-step method \citep{kim1985} with a third-order Runge-Kutta time-advancing scheme \citep{wray1990}. Periodic boundary conditions are imposed in the streamwise and spanwise directions and the no-slip and no-penetration boundary conditions are used at the top and bottom walls. The code has been validated in previous studies in turbulent channel flows \citep{bae2018,bae2019,lozano-duran2019} and flat-plate boundary layers \citep{lozano-duran2018}, though we note that, for this problem, we modify the boundary conditions to accommodate the oscillating walls. 
The domain size of the channel for the DNS is given by $6\pi \times 80 \times 3\pi$. The domain is discretised uniformly in the $x_1-$ and $x_3-$directions using $64$ points, which corresponds to non-dimensionalised spacings of $\Delta x \approx 0.29$ and $\Delta x_3 \approx  0.15$. For the $x_2$-direction, a hyperbolic tangent grid with $385$ points is used, resulting in $\min(\Delta x_2) \approx 0.01$ and $\max(\Delta x_2) \approx  0.91$. We compute the mean velocity profiles by averaging in homogeneous directions and phase. Figure~\ref{fig:StokesMean} shows the mean and the streamwise root-mean-square (rms) velocity profiles at different times. 
Using $\Omega$ to denote the non-dimensionalised wall oscillation frequency, we note that $U_1(t\Omega + \pi) = -U(t\Omega)$ and $U_{i,rms}(t\Omega + \pi) = U_{i,rms}(t\Omega)$.
We observe that the turbulent energy peak occurs near the wall at $x_2 = 1.43$ and $t\Omega = 2.65$, and propagates away from the wall thereafter. 

To construct the resolvent operator, we first choose the spatial scales for the homogeneous directions. Using the DNS data, we calculate the streamwise energy spectrum at $x_2= 1.43$ and $t\Omega = 2.65$, the wall-normal location and phase of the peak $U_{1,rms}$. The most energetic streamwise and spanwise scales at that location are $k_1 = 0.67$ and $k_3 = 4.22$, which we choose as the streamwise and spanwise scales for the resolvent operator. To solve the discrete system, we use a Chebyshev grid in the wall-normal direction, with $N_2 = 80$, and a uniform temporal discretisation over one period $T\Omega = 2\pi$, with $N_t = 1600$. As in the previous section, $\mathsfbi{U}_{3}$, $\mathsfbi{U}_{2}$, $\mathsfbi{dU}_{2,2}$, and $\mathsfbi{dU}_{3,2}$ are zero.

We choose the first-order derivative matrix in time $\mathsfbi D_t$ and the wall-normal spatial derivative matrices to be second-order-accurate centred  finite difference matrices. We additionally choose $\mathsfbi D_t$ to be circulant to enforce periodicity in time. We compute the modes for the half-channel and enforce a no-slip and no-penetration boundary condition at the wall, and a free-slip and no-penetration boundary condition at the centreline. Because $\mathsfbi D_t$ is a finite difference matrix rather than a Fourier differentiation matrix, we must implement a filtering step, detailed in \S\ref{sec:timediff}, to exclude the high temporal wavenumbers. To apply this filtering step, we must assume that the high-frequency waves are not physically significant for the turbulent Stokes boundary layer problem. We use a two-stage Daubechies-16 wavelet transform, which is a sparse unitary operator. We note that the Daubechies-16 operator is not a perfect bandpass filter, and the numerical filtering operation simply attenuates the high-frequency waves that produce spurious SVD modes instead of excluding them outright. Nevertheless, due to the high dimensionality of the problem, it remains advantageous to use sparse transforms. We choose to constrain the forcing and response modes to the scaling functions and their shifts, which roughly cover the first quarter of all temporal wavenumbers $k_t = 0..., N_t/8$.

We compare the results obtained with the wavelet-based resolvent modes with the results from harmonic resolvent analysis \citep{Padovan2020}. The latter computes a Fourier-based resolvent analysis simultaneously for multiple temporal wavenumbers and includes the interactions between them as they are coupled by the temporally evolving mean profile. For the harmonic resolvent analysis, we use the same Chebyshev grid as in the wavelet-based method, with $N_2 = 80$. For the sake of comparing with the wavelet-based method and to account for the filtering step, we choose a frequency resolution of $N_t = 1600/4 = 400$. 
We expect the two methods to produce similar singular values and modes. The singular values and modes would be equivalent in both cases if we use a Fourier differentiation operator for the wavelet-based method as in \S\ref{sec:app:channel}. 

\begin{figure}
\begin{center}
\vspace{0.1cm}
    \subfloat[]{\includegraphics[width=0.45\linewidth]{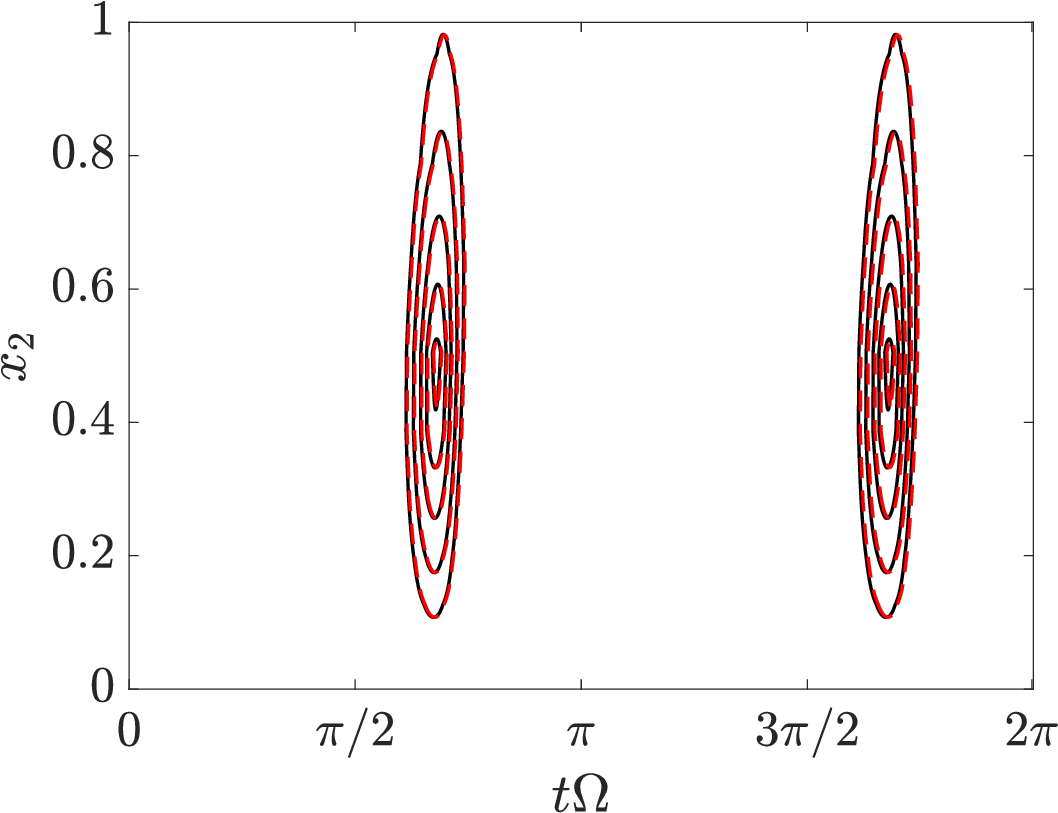}}
    \hspace{0.25cm}
    \subfloat[]{\includegraphics[width=0.45\linewidth]{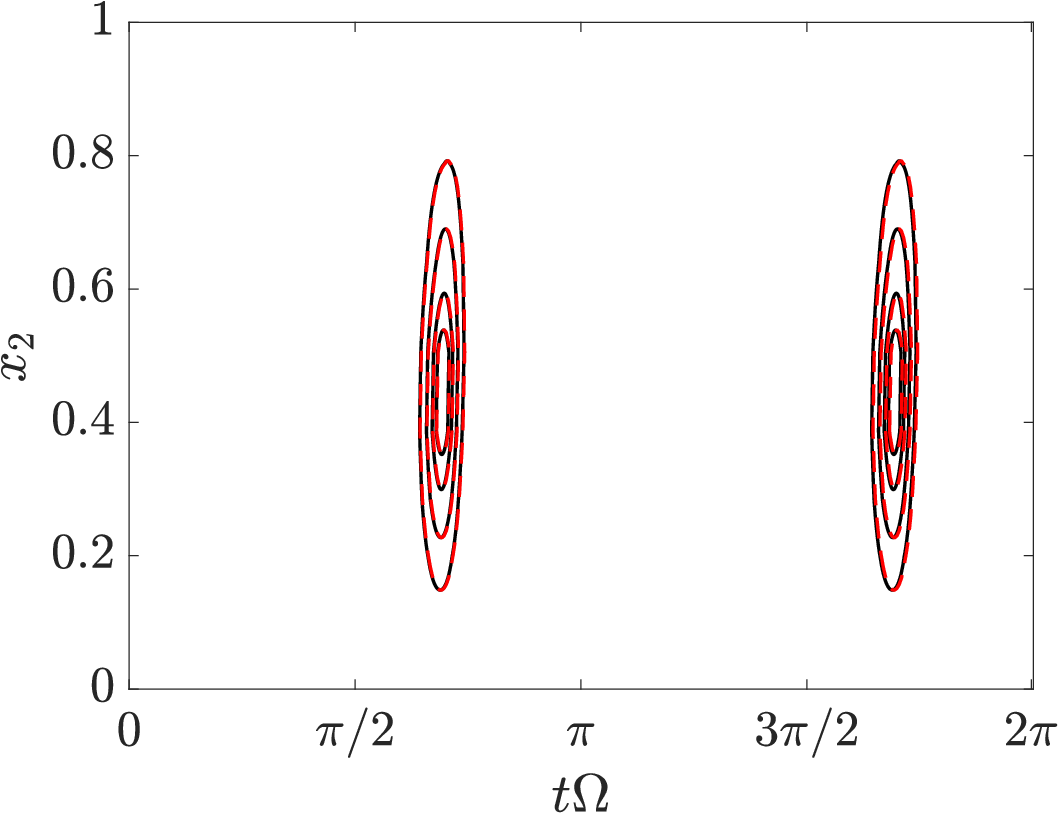}}
    \\
    \vspace{0.3cm}
    \hspace{-0.15cm}
    \subfloat[]
    {\includegraphics[width=0.49\linewidth]{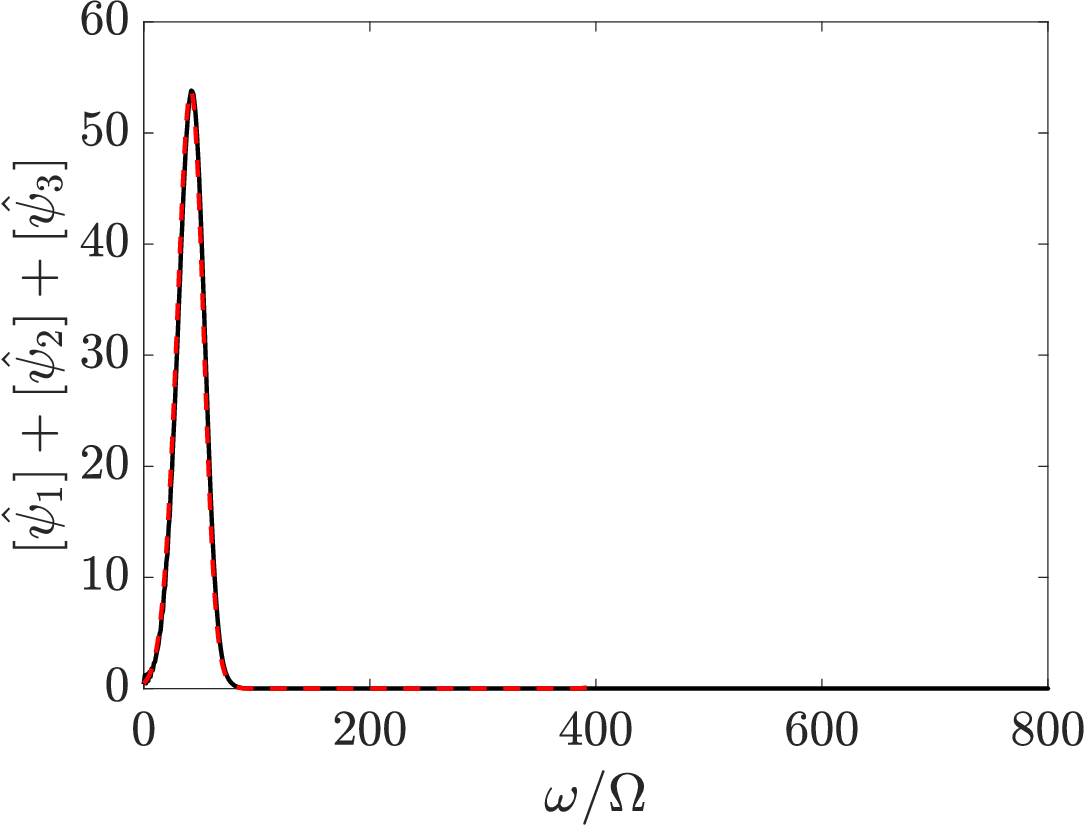}}
    \hspace{-0.05cm}
    \subfloat[]{ \includegraphics[width=0.48\linewidth]{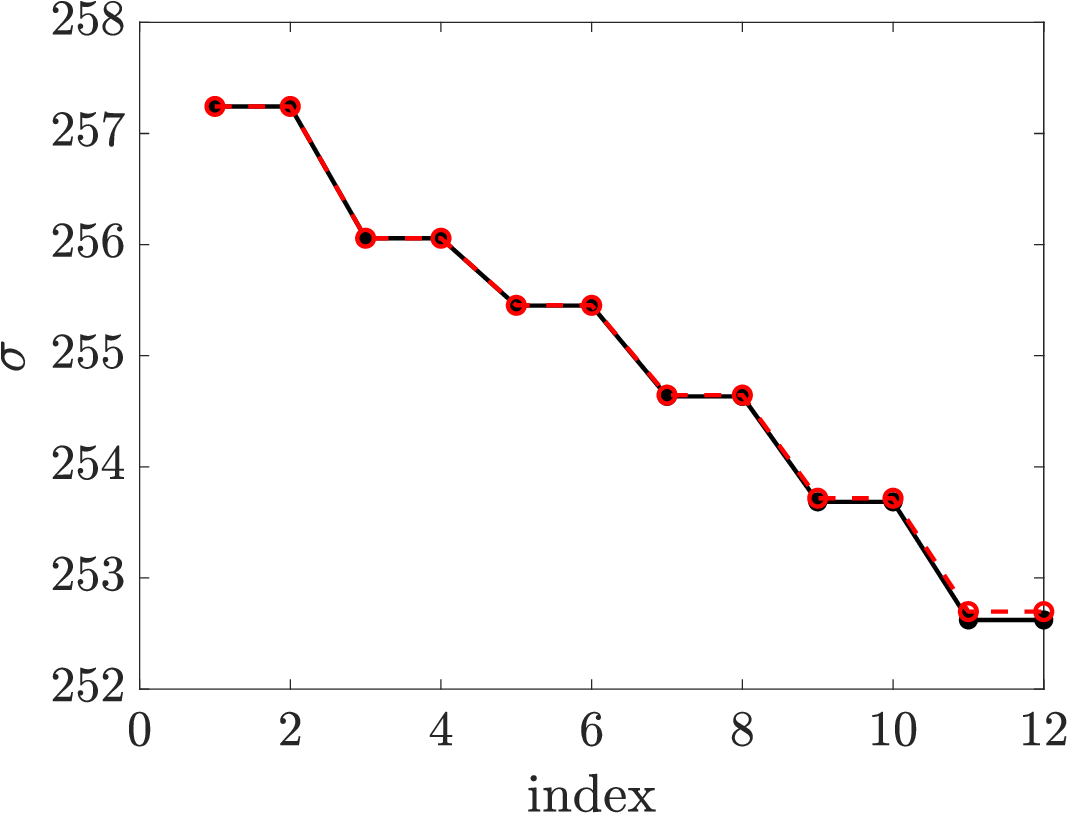}}
\end{center}
\caption{Magnitude contours ($25\%$, $50\%$, $75\%$ and $90\%$ of the maximum value) of (a) the wall-normal component of the principal resolvent forcing mode and (b) the streamwise component of the principal resolvent response mode for the turbulent Stokes boundary layer; (c) $x_2$--integrated Fourier spectrum in time for the principal response modes; (d) singular values from the SVD of the resolvent operators. Results from harmonic resolvent analysis are shown in red, and those from wavelet-based resolvent analysis in black.}
    \label{fig:compare_harmonic_wavelet}
\end{figure}

\begin{figure}
    \begin{center}
    \vspace{0.65cm}
    \subfloat[]{\includegraphics[width=0.65\linewidth]{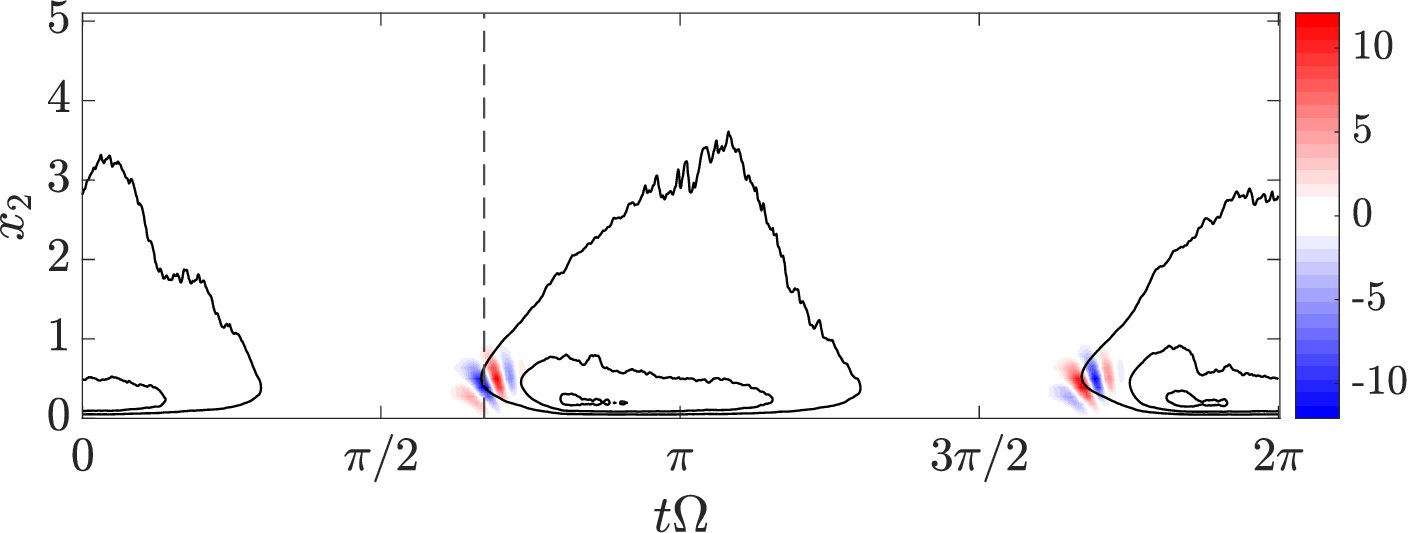}}
    \\
    \vspace{0.8cm}
    \hspace{0.05cm}
    \subfloat[]{\includegraphics[width=0.66\linewidth]{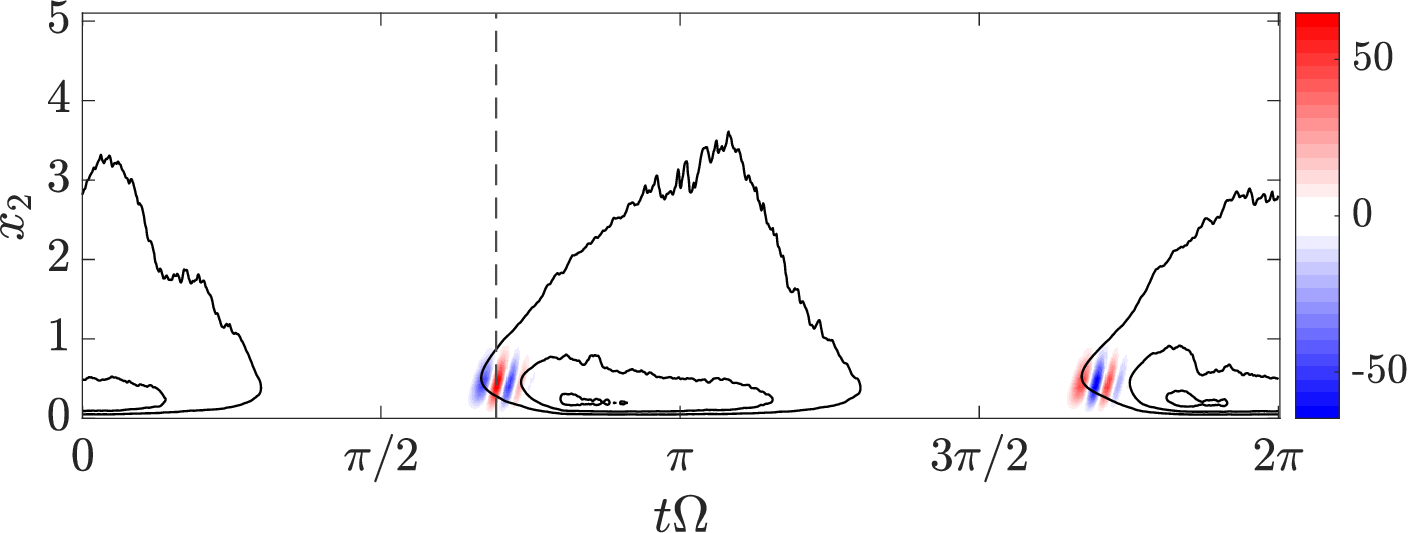}}
    \end{center}
    \caption{Real part of (a) $\breve \phi_1$, and (b) $\breve \psi_1$ for the turbulent Stokes boundary layer. The black contour lines are $U_{1, rms}$ with the levels indicating $50\%, 75\%, 95\%$ of its maximum value. The vertical dashed lines show the times of the amplitude peak for the input mode ($t\Omega =  2.11$) and output mode ($t\Omega = 2.18$).}
    \label{fig:stokesModes}
\end{figure}

\begin{figure}
    \begin{center}
        \subfloat[]{\includegraphics[width=0.49\linewidth]{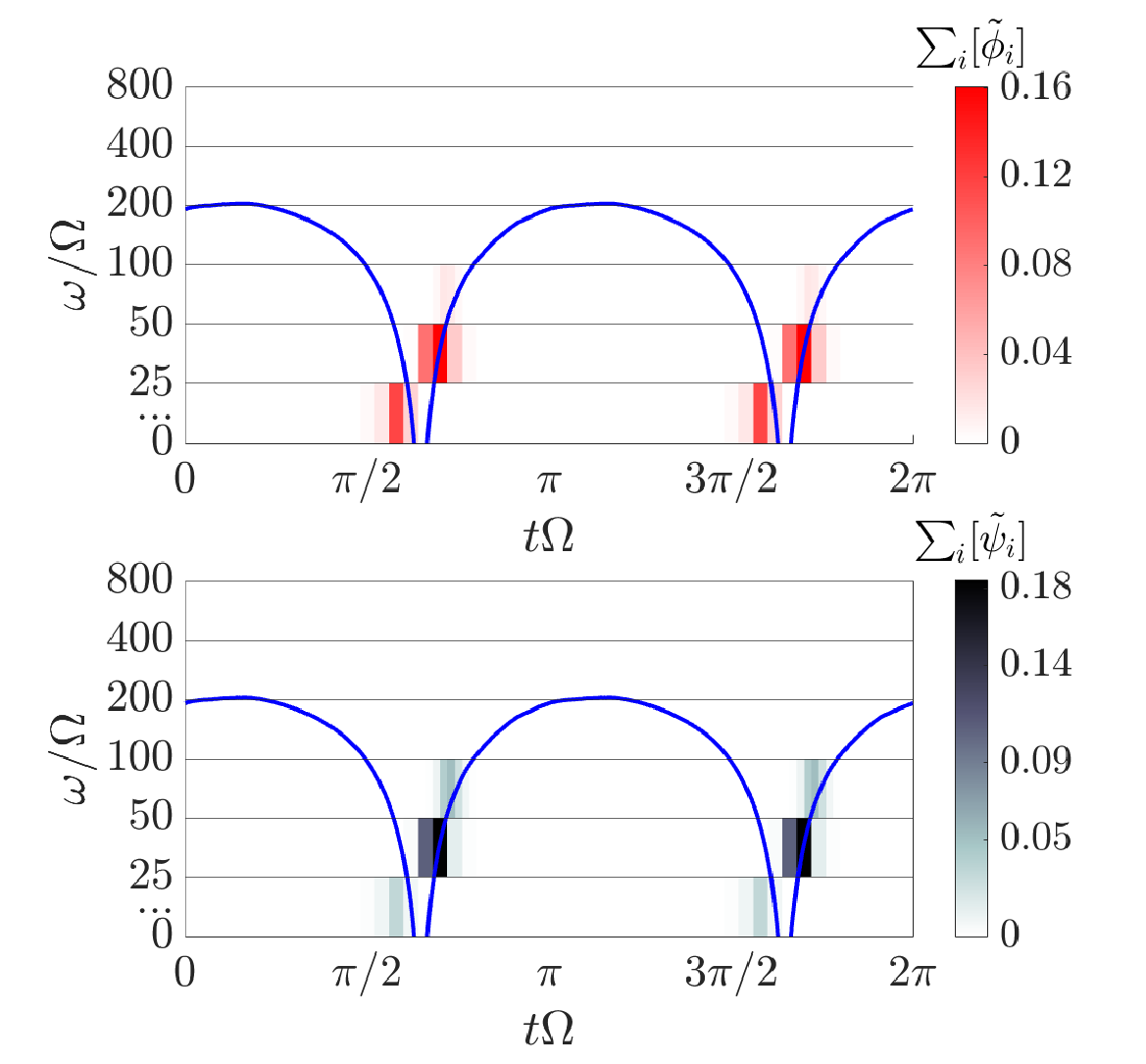}}
        \subfloat[]{\includegraphics[width=0.435\linewidth]{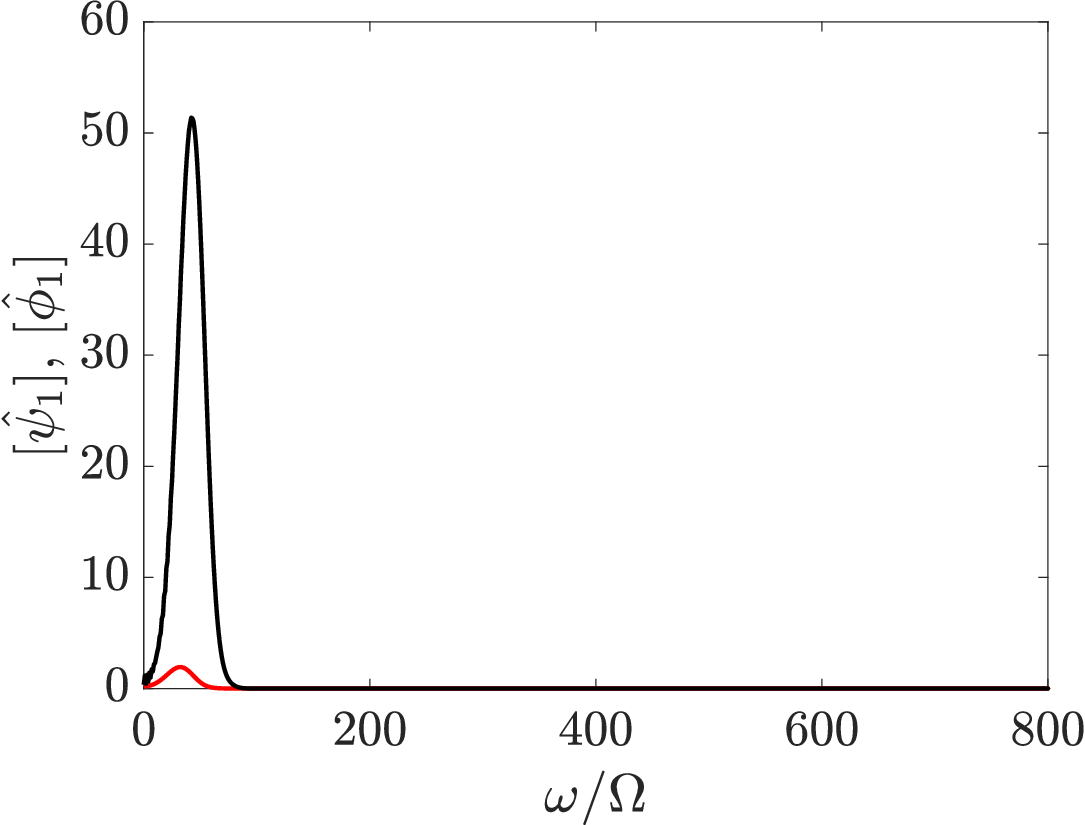}}
    \end{center}
    \caption{ (a) Principal forcing (red, top panel) and  response (black, bottom panel) modes in the time-frequency plane; the blue line indicates $U_1k_1/\Omega$ at $x_{2, \mathrm{avg}}^\mathrm{resp.} = x_{2, \mathrm{avg}}^\mathrm{forc.} \approx 0.46 $; (b) $x_2$--integrated frequency content in the streamwise component of the forcing (red) and response (black) modes. } 
    \label{fig:scalogram_in_out}
\end{figure}

The modes obtained from harmonic resolvent analysis agree well with those obtained from wavelet-based resolvent analysis. They occur at the same $x_2$ location, 
and time
(figures \ref{fig:compare_harmonic_wavelet}(a-b)), and exhibit roughly the same frequency content (figure \ref{fig:compare_harmonic_wavelet}(c)). Moreover, the SVD of the wavelet-based and harmonic resolvent operators yield very similar singular values. The first twenty singular values are shown in figure \ref{fig:compare_harmonic_wavelet}(d). 
Despite Daubechies-16 wavelets being imperfect bandpass filters, filtering-out high-frequency waves using the sparse wavelet transform succeeds in producing resolvent modes that match the leading modes from harmonic resolvent analysis. 
\reviewertwo{We consider that the results for harmonic resolvent analysis shown in figure \ref{fig:compare_harmonic_wavelet} are converged: the ones that use a coarser grid with $N_2 = 80$ and $N_t = 300$ produced the same 10 leading singular values up to $10^{-4}$.}
Moreover, the windowed wavelet-based resolvent operator exhibits significant sparsity and can be analysed efficiently, despite the larger dimension of the system. \reviewertwo{Indeed, the harmonic and wavelet-based resolvent analyses run in similar wall times.}

The principal input and output modes corresponding to the chosen spatial scales and boundary conditions in time are shown in figure \ref{fig:stokesModes}(a). \reviewertwo{We observe that the modes are located at roughly the same wall normal height as the peaks in $U_{1, rms}$, \emph{i.e.} $x_2 \approx 0.5$.}
We also observe that the principal input and output modes are synchronised with the peaks in $U_{1, rms}$, \reviewertwo{though the modes tend to peak slightly earlier than $U_{1, rms}$. This suggests that linear amplification might provoke the transition to turbulence}. 
\reviewertwo{The preliminary results presented in figure \ref{fig:stokesModes}} suggest that the study of energy amplification in the Stokes boundary layer is a good candidate for the use of linearised methods, similarly to the turbulent channel flow \citep{jimenez2013linear}.

We also observe that the principal input mode precedes the principal output mode in time, with the peak of the former occurring  $\Delta t\Omega \approx 0.063$ before the peak of the latter. Wavelet-based resolvent analysis is able to capture the natural response time between forcing and response terms under the dynamics of the linearised Navier-Stokes. This time delay is also in line with a physical interpretation of the modes in which the input modes cause the output modes and must thus occur earlier. 
The extent to which this captures important causal mechanisms within the full nonlinear system is yet to be determined. In future works, it would be interesting to project flow fields onto these time-separated resolvent forcing and response modes to test whether better correlations can be obtained between them in the transformed bases.

Additionally, we notice that the only non-zero wavelet coefficients of both the principal forcing and response modes are those corresponding to the bottom quarter of the set of resolved frequencies \emph{i.e.} the lowest four bands in the scalograms shown in figures \ref{fig:scalogram_in_out}(a). This validates the windowing step described above. 
We also see in figure \ref{fig:scalogram_in_out}(a) that the frequency content of the principal modes varies with time. The principal forcing mode is initially composed of lower-frequency waves, whose frequencies are centred  in a band $[0, 25\Omega]$; these waves are gradually shifted up to frequencies centred  in $[25\Omega, 50\Omega]$. 
Likewise, the waves composing the principal response mode, initially at frequencies centred  in $[25\Omega, 50\Omega]$ are also shifted up to higher frequencies. We propose that this frequency shift is due to the time-varying mean streamwise velocity $U_1$, which acts as a convection velocity and accelerates the resolvent forcing and response waves. \reviewerthree{Thus, we expect the frequency content of the forcing and response modes to vary in tandem with the changing mean streamwise velocity profile.}
We define the average location of the streamwise modes as 
\begin{equation}
 x^\mathrm{resp.}_{2, \mathrm{avg}} := \frac{\int_0^T \int_0^1 x_2 |\breve \psi_1|^2 dx_2 dt }{  \int_0^T \int_0^1 |\breve{\psi}_1|^2 dx_2 dt},
\end{equation}
and
\begin{equation}
 x^\mathrm{forc.}_{2, \mathrm{avg}} := \frac{\int_0^T \int_0^1 x_2 |\breve \phi_1|^2 dx_2 dt }{  \int_0^T \int_0^1 |\breve{\phi}_1|^2 dx_2 dt},
\end{equation}
and plot the frequency shift due to the mean convection $U_1k_1/\Omega$ at the average mode locations in figure \ref{fig:scalogram_in_out}(a). We observe a good correlation between the shift in the frequency content of the forcing and response modes and the change in the mean velocity.

We can also use the changing mean velocity profile to explain the difference in the frequency content between the forcing and response modes. We propose that this difference in frequency content is due to the different peaking times of the forcing and response modes: since the modes occur at difference phases of the oscillating mean profile, they will be convected at different velocities.
To verify this, we first Fourier-transform $\boldsymbol{\breve \phi}$ and $\boldsymbol{\breve \psi}$ in time to extract their frequency content with better precision, 
and observe in figure \ref{fig:scalogram_in_out}(b) that the average frequency shift between the forcing and response modes is
\begin{equation}
   \Delta \omega := \frac{\int_0^{\Omega Nt/2} \omega[\hat \psi_1] d\omega}{\int_0^{\Omega Nt/2}  [\hat \psi_1] d\omega} - \frac{\int_0^{\Omega Nt/2} \omega  [\hat \phi_1]d\omega}{\int_0^{\Omega Nt/2}  [\hat \phi_1]d\omega} \approx 10.2 \; \Omega.
\end{equation}
We then define the average temporal location of the modes as
\begin{equation}
t^\mathrm{resp.}_{ \mathrm{avg}} := \frac{\int_0^T  t [\breve \psi_1]  dt }{  \int_0^T [\breve{\psi}_1] dt},
\end{equation}
and
\begin{equation}
t^\mathrm{forc.}_{ \mathrm{avg}} := \frac{\int_0^T t [\breve \phi_1] dt }{  \int_0^T \int_0^1 [\breve{\phi}_1] dt}.
\end{equation}
Assuming that both the optimal forcing and response prefer the same natural frequency $\omega_0$ with a corresponding streamwise wave speed $c_0 = \omega_0/k_1$, we estimate the shift with
\begin{equation}\label{eq:doppler}
    \left ( |U(x^\mathrm{resp.}_{2, \mathrm{avg}}, t^\mathrm{resp.}_{\mathrm{avg}})|  - |U(x^\mathrm{forc.}_{2, \mathrm{avg}}, t^\mathrm{forc.}_{\mathrm{avg}}|) \right ) k_1/\Omega \approx 10.5,
\end{equation}
which roughly matches the observed shift. We repeat this analysis for two other sets of spatial parameters,  $(k_1, k_3) = (0.67, 2.67)$ and  $(k_1, k_3) = (1.33, 2.67)$, though the plots are not shown. For the first set, the estimated frequency shift is found to be $\Delta \omega / \Omega \approx 27$ using equation \eqref{eq:doppler}, roughly matching the measured mean frequency shift of $\Delta \omega / \Omega \approx 26$ computed using a Fourier transform of the response and forcing modes. Similarly, for the second set of these length scales, the estimated frequency shift due to convection by the mean flow is $\Delta \omega / \Omega \approx 14.6$ and the measured mean frequency shift is $\Delta \omega / \Omega \approx 13.4$. 

Wavelet-based resolvent analysis for this non-stationary flow thus reveals how a time-varying mean profile affects the linear amplification of perturbations. The mean velocity profile not only determines the spatial structure of the modes like in \S\ref{sec:app:channel}, but  also their transient behaviour, and in this case, acts as a convection velocity that modulates their frequency content and wave speeds.

\subsection{Channel flow with sudden lateral pressure gradient} \label{sec:3DChannel}

\begin{figure}
    \begin{center}
    \vspace{0.1cm}
    \subfloat[]{\includegraphics[width=0.45\linewidth]{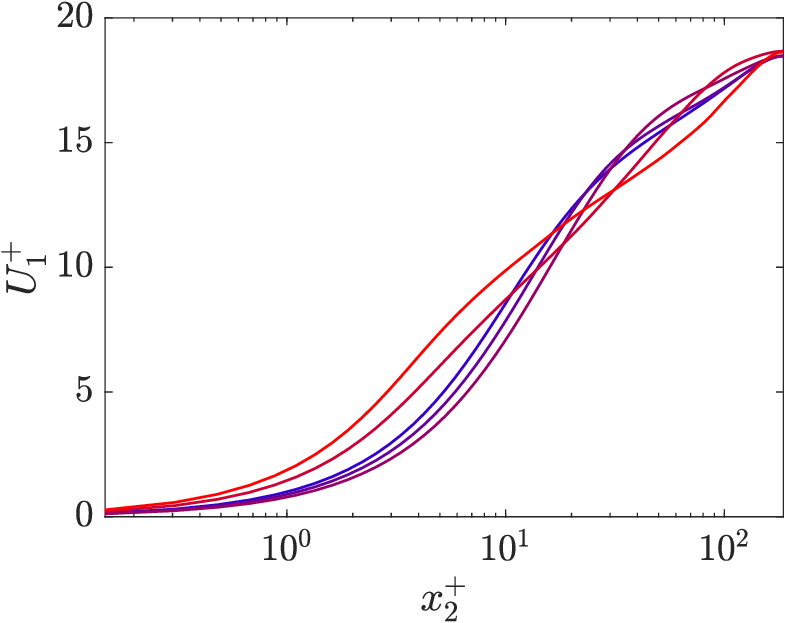}}
    \hspace{0.2cm}
    \subfloat[]{\includegraphics[width=0.45\linewidth]{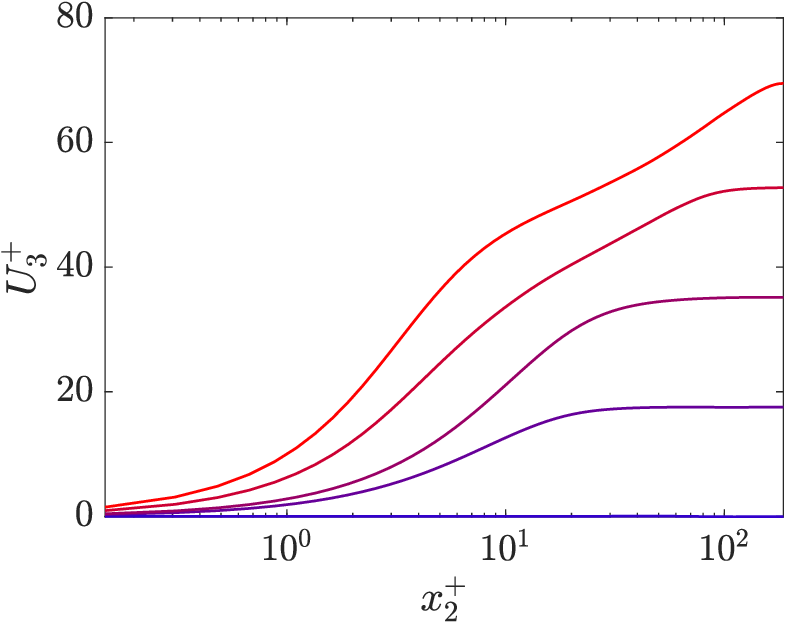}}
    \end{center}
    \caption{Mean (a) streamwise and (b) spanwise velocity profile from $t = 0$ (blue) to $t  = 2.34$ (red). The times shown are $t = 0, \, 0.58, \, 1.17,\, 1.76,\, 2.34$. Data taken from \citet{noneq}. Time $t$ is non-dimensionalised with $u_{\tau, 0}/\delta$}
    \label{fig:3DChannelProf}
\end{figure}

\begin{figure}
    \begin{center}
    \subfloat[]{\includegraphics[width=0.45\linewidth]{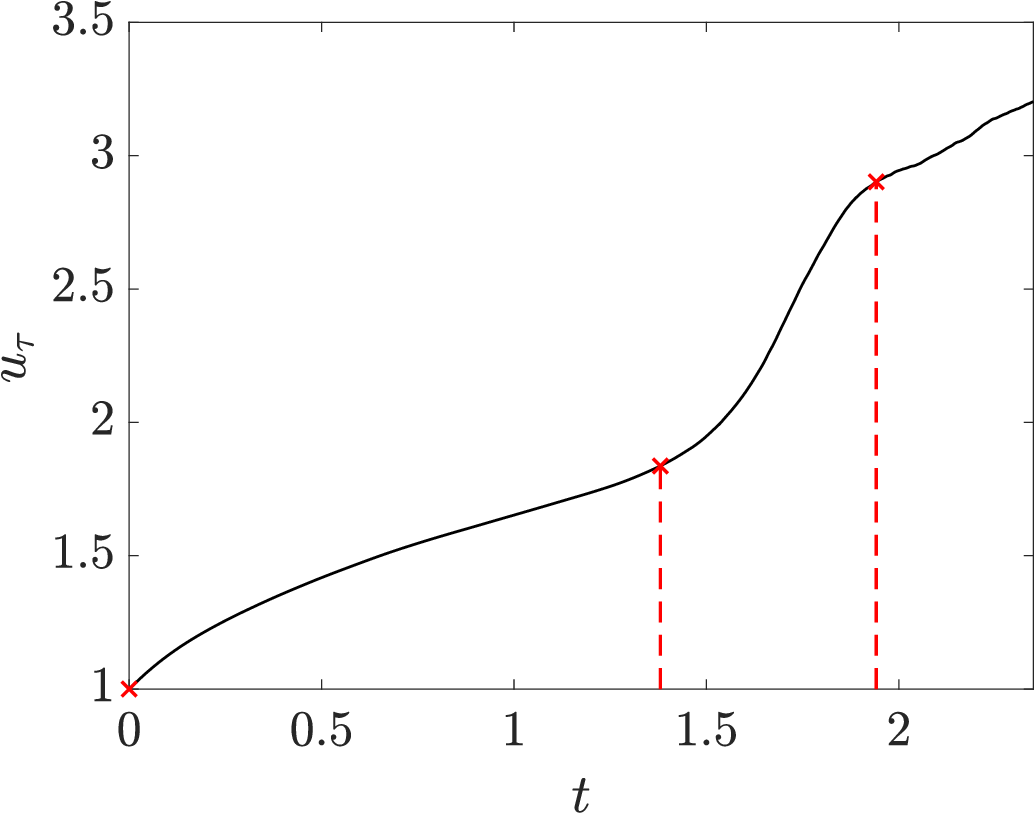}}
    \hspace{0.2cm}
    \subfloat[]{\includegraphics[width=0.45\linewidth]{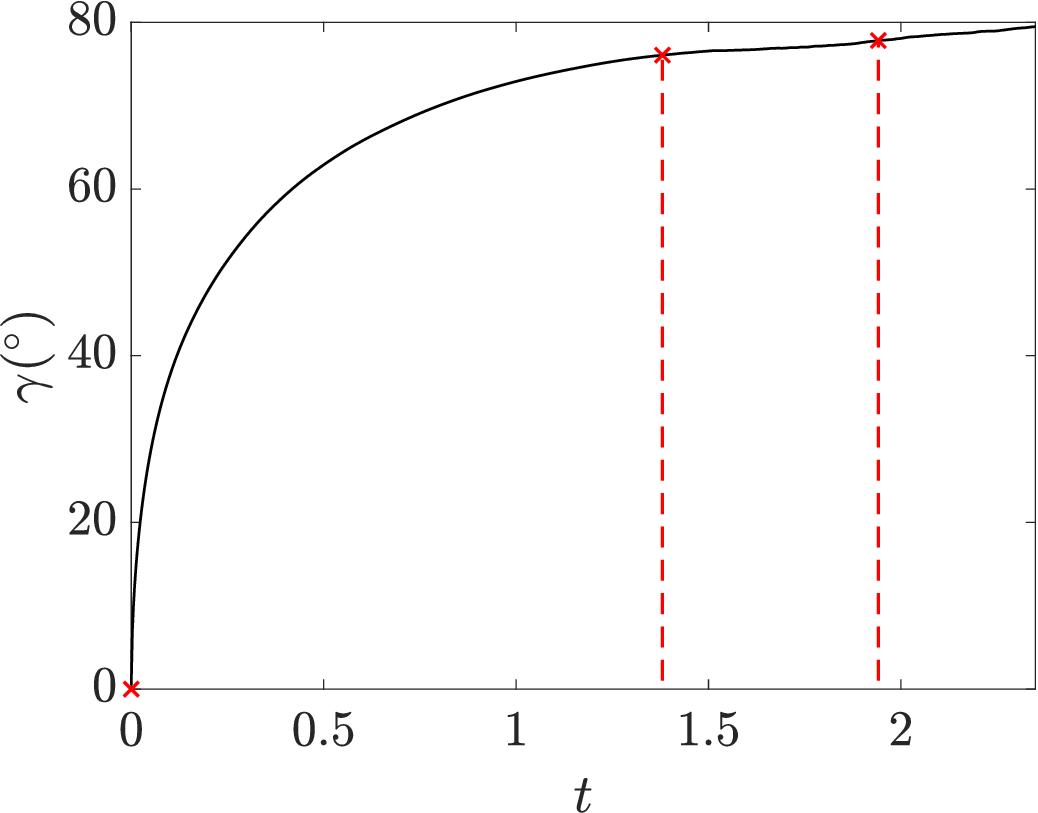}}
    \end{center}
    \caption{(a) Friction velocity $u_{\tau}$ and (b) wall shear-stress angle $\gamma = \tan^{-1}(\tau_3 / \tau_1)$ as a function of time. Data taken from \citet{noneq}. The vertical dashed lines are at $t = 0, 1.38, 1.94$, and correspond to the choice of $\lambda_1^+$ and $\lambda_3^+$ for the modes plotted in figure \ref{fig:3DChannel}}
    \label{fig:utau_angle}
\end{figure}

Finally, we study a fully developed turbulent channel flow at $\Rey_\tau = 186$ that is subjected to a sudden lateral pressure gradient $dP/dx_3 = \Pi dP/dx_1$ at $t=0$ with $\Pi = 30$ \citep{Moin1990, noneq}. This flow, commonly referred to as a three-dimensional (3-D) channel flow, has an initial transient period dominated by 3-D non-equilibrium effects. Eventually, the flow will reach a new statistically steady state with the mean flow in the $(dP/dx_1, dP/dx_3)$ direction parallel to the wall. In the transient period, the tangential Reynolds stress initially decreases before increasing linearly, with depletion and increase rate that scales as $\Pi x_2/\delta$ \citep{noneq}. 

The mean flow profiles are obtained from \citet{noneq} and have non-zero streamwise and spanwise components $U_1$ and $U_3$ (figure \ref{fig:3DChannelProf}) as well as non-zero wall-normal gradients of streamwise and spanwise components $\mathsfbi{dU}_{1, 2}$ and $\mathsfbi{dU}_{3, 2}$. In this section, we non-dimensionalise velocity by initial friction velocity $u^*_{\tau,0}$, lengths by the channel half-height $\delta^*$, and time by $\delta^* / u^*_{\tau,0}$. The Reynolds number for this problem is $\Rey = \Rey_{\tau, 0} := u^*_{\tau, 0} \delta^* / \nu$. The time domain of the simulation is $T = 2.34$. 
To construct the discrete resolvent operator, we use a Chebyshev grid of size $N_2 = 65$ in the $x_2-$ direction extending from $x_2 = 0$ to $x_2 = 1$. For the spatial derivatives in the $x_2-$ direction, we choose second-order-accurate finite difference matrices. We enforce a no-slip and no-penetration boundary condition at the wall and a free-slip and no-penetration condition at the centreline. The boundary condition for the temporal finite difference operator $\mathsfbi D_t$ is chosen to enforce a Neumann-type condition, $\partial_t(\cdot)|_{t=0} = \partial_t(\cdot)|_{t=T} = 0$. To reduce the impact of the boundary condition on the modes at $t=0$ we extend $U_1$ and $U_3$ to the time interval $t \in [-0.58, 2.34]$ and assume $U_i(t \leq 0, y) = U_i(t = 0, y)$ and $dP/dx_3(t < 0) = 0$. When the modes are plotted, we only show the original time domain $t \in [0, 2.34]$ and exclude the contribution from negative times. We use a temporal resolution of $N_t = 1000$ for the extended time frame. In this case, we note that we do not obtain spurious modes due to the distortion of high frequency waves, and that filtering-out those waves as in \S\ref{sec:stokes} has little effect on the results.

Regarding the spatial scales for the homogeneous directions,
we choose them to capture near-wall streaks at three different times: $t = 0$, $1.3$ and $1.94$. We thus tune them to represent the aspect ratio characteristic of near wall streaks, \emph{i.e.} $\lambda_1^+ \approx 10 \lambda_3^+$ for a mean flow with a dominant streamwise component, and $\lambda_3^+ \approx 10 \lambda_1^+$ for a mean flow with a dominant spanwise component. Here, $(\cdot)^+$ indicates the wall scaling with $\Rey_{\tau,0}$, before the lateral pressure gradient is applied. 
To capture near-wall streaks at  $t = 0$, we choose $(\lambda_1^+,\, \lambda_3^+) = (\lambda_{1,0}^+,\, \lambda_{3,0}^+) := (1000,\,100)$ as in \S\ref{sec:app:channel}, which corresponds to the spatial scales preferred by the near-wall streaks at $\Rey_\tau = 186$ prior to the lateral pressure gradient. Under the shear conditions at $t = 1.3, \; 1.94$, we must take into account the stronger mean shear in the spanwise direction 
\citep{noneq} by multiplying by a factor of $\Rey_{\tau}(t) / \Rey_{\tau, 0} = u^*_\tau(t) / u^*_{\tau, 0} = u_\tau$, plotted in figure~\ref{fig:utau_angle}(a). We also take into account the new orientation of the streaks by applying a rotation by the wall-shear stress angle $\gamma(t) = \tan^{-1}(\tau_3/\tau_1)$, where $\tau_i$ is the instantaneous wall-shear stress in the $x_i$ direction (see figure \ref{fig:utau_angle}(b)). \reviewerthree{Using the expressions 
$\lambda^+_1 \approx \left(\lambda^+_{1, 0} \mathrm{cos}(\gamma(t)) - \lambda^+_{3,0} \mathrm{sin}(\gamma(t)) \right ) u_\tau(t)$ and $\lambda^+_3 \approx \left(\lambda^+_{1, 0} \mathrm{sin}(\gamma(t)) + \lambda^+_{3,0} \mathrm{cos}(\gamma(t)) \right ) u_\tau(t)$,} 
we obtain spatial parameters $(\lambda_1^+, \lambda_3^+) = (264, 1827)$ corresponding to $t = 1.3$, and $(\lambda_1^+,\, \lambda_3^+) = (329, 2898)$ corresponding to $t = 1.94$.

\begin{figure}
\begin{center}
\subfloat[]{
\includegraphics[width=0.48\linewidth]{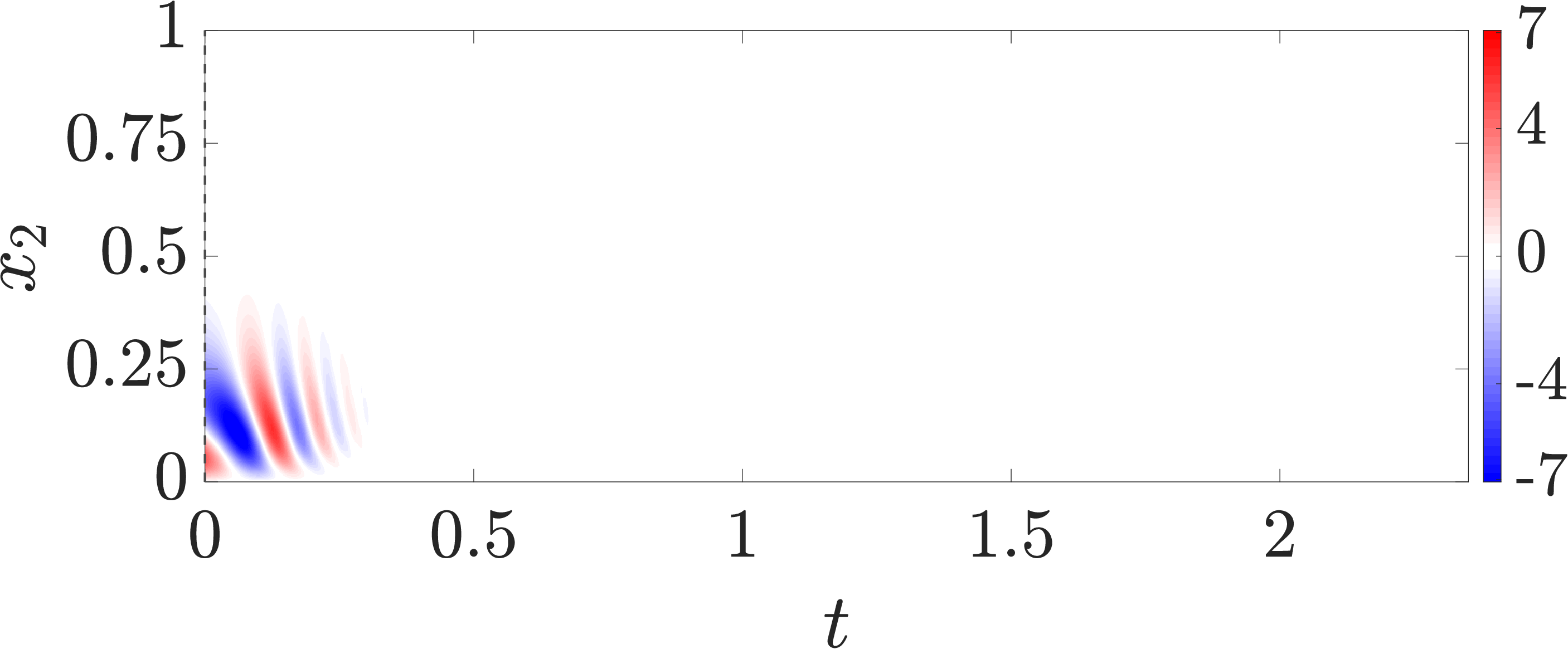}}
\hspace{0.2cm}
\subfloat[]{    \includegraphics[width=0.48\linewidth]{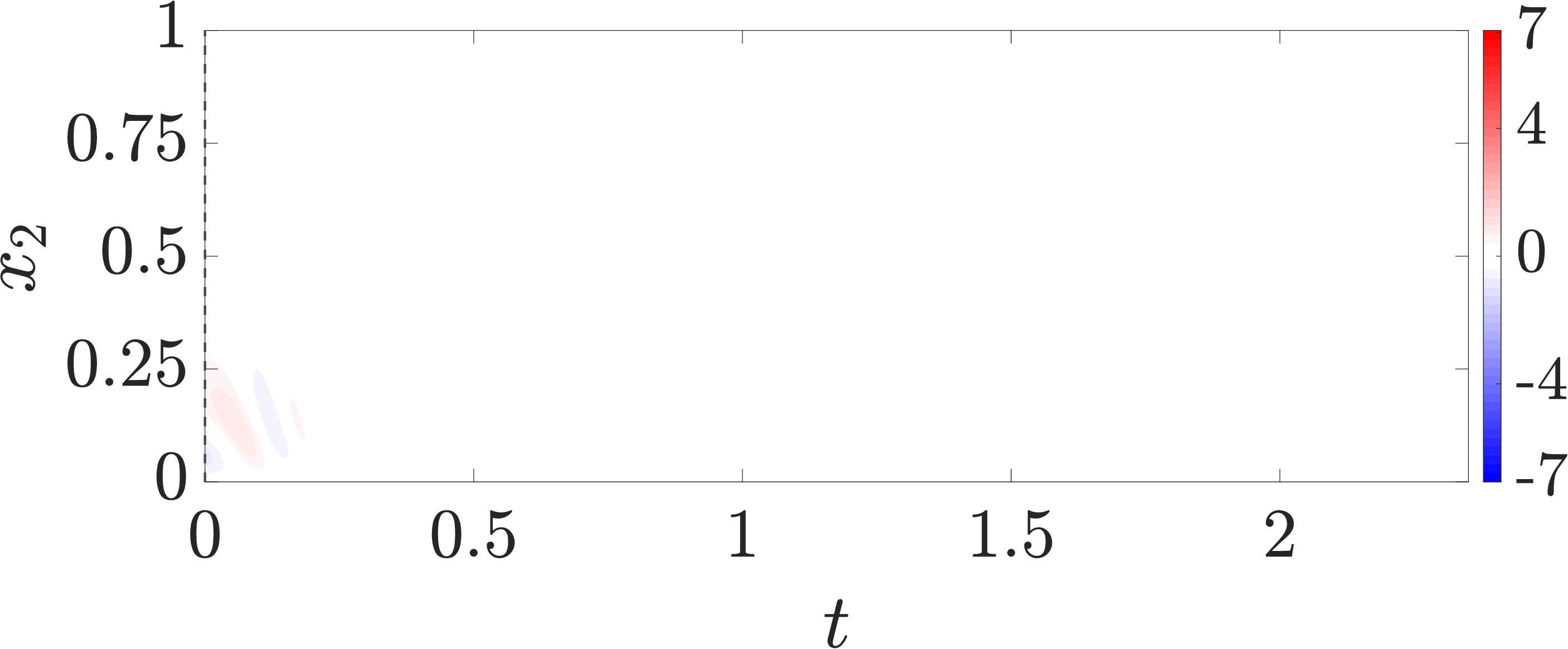}}
\\
\subfloat[]{\includegraphics[width=0.48\linewidth]{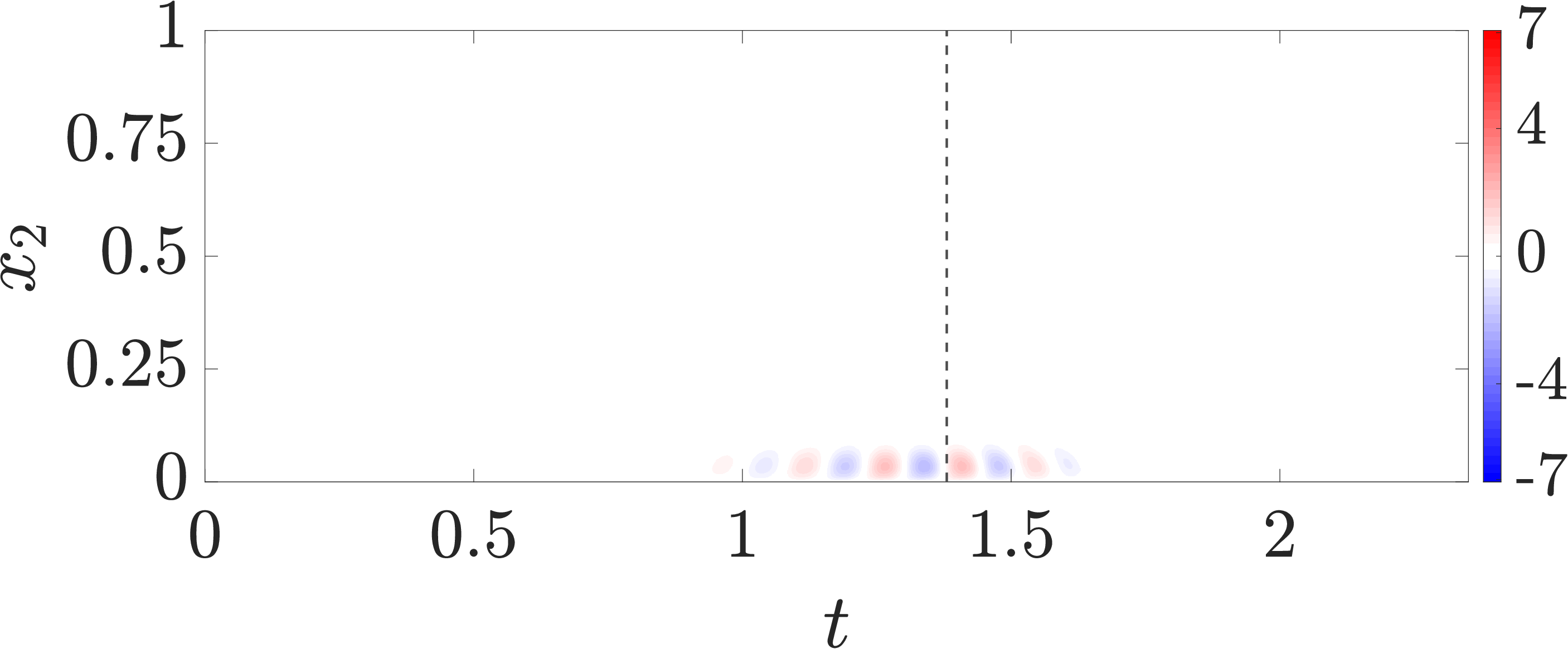}}
\hspace{0.2cm}
\subfloat[]{\includegraphics[width=0.48\linewidth]{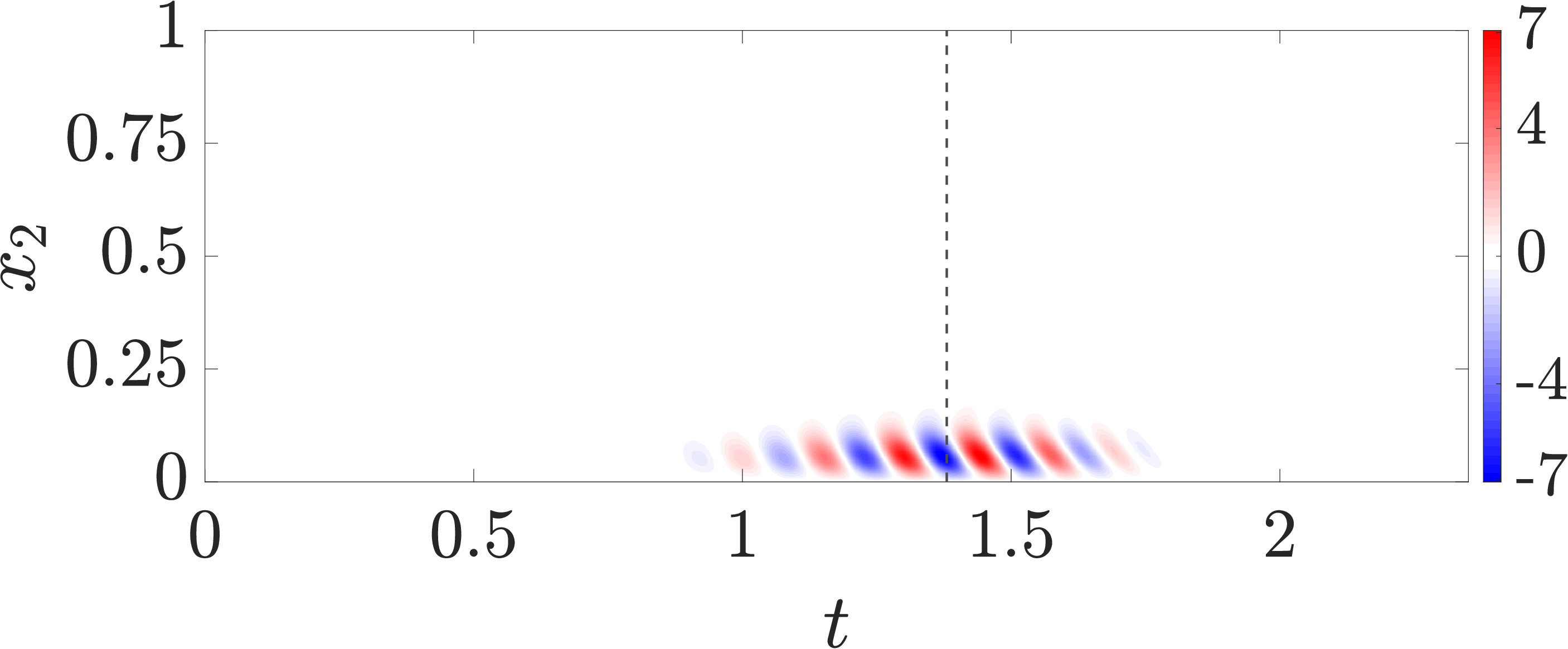}}
\\
\subfloat[]{\includegraphics[width=0.48\linewidth]{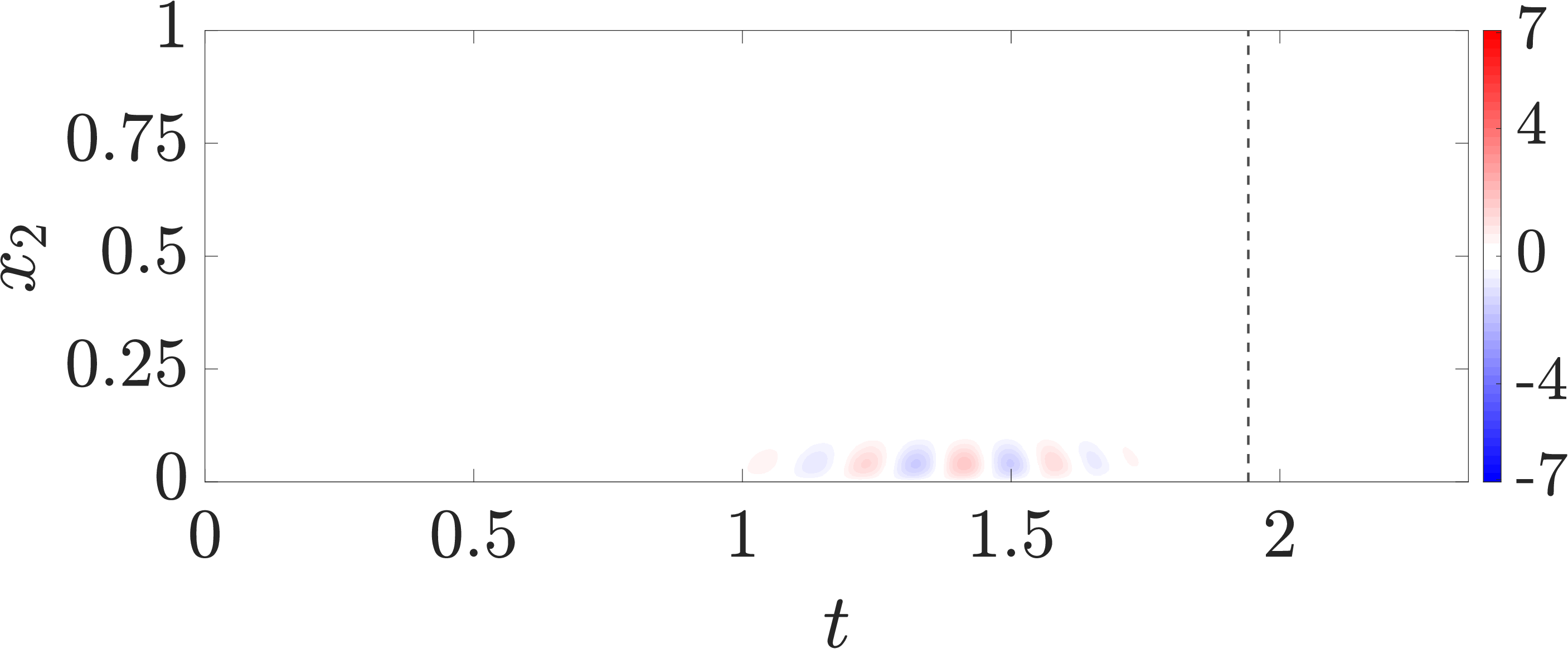}}
\hspace{0.2cm}
\subfloat[]{\includegraphics[width=0.48\linewidth]{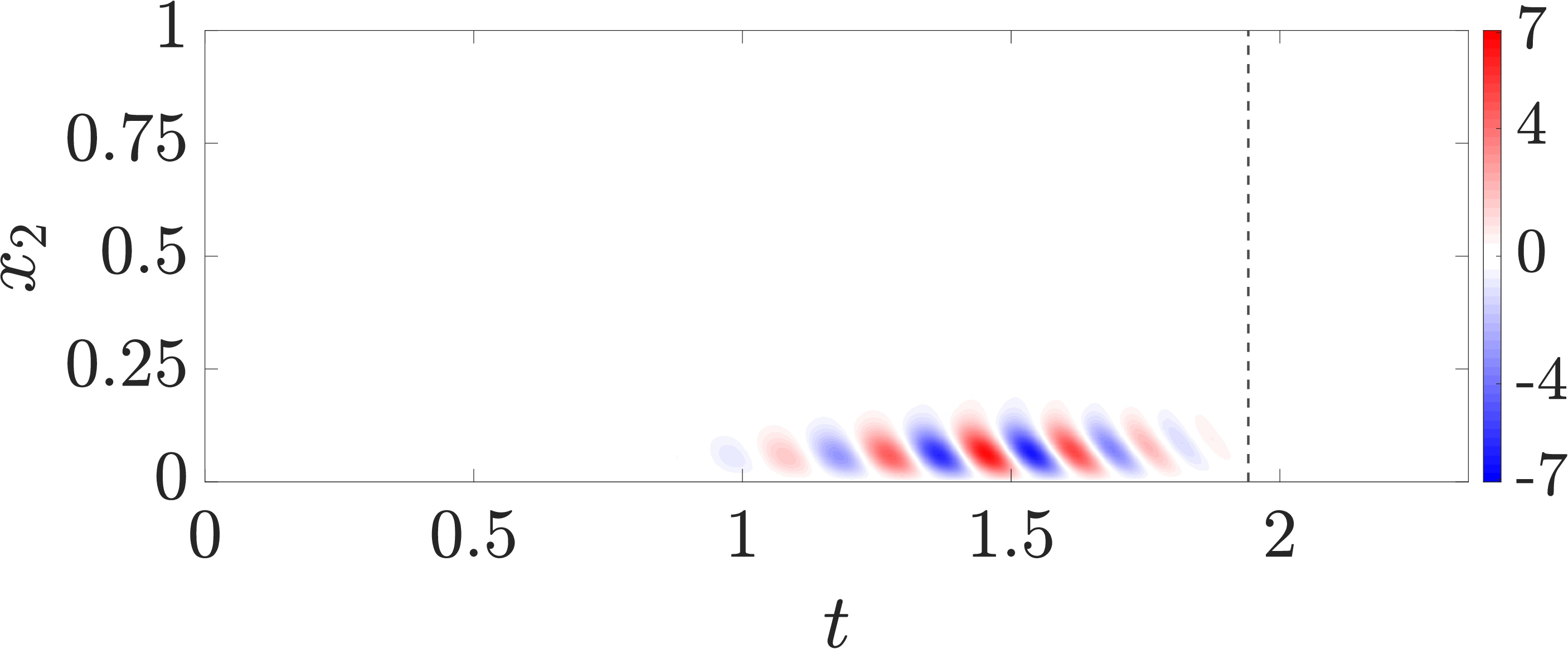}}
\end{center}
\caption{Real part of $\breve \psi_1$ (left) and $\breve \psi_3$ (right) for the turbulent channel subject to a spanwise pressure gradient. The chosen spatial scales are (a, b) $\lambda_1^+ = 1000$, $\lambda_3^+ = 100$, (c, d) $\lambda_1^+ = 264$, $\lambda_3^+ = 1827$, and (e, f) $\lambda_1^+ = 329$, $\lambda_3^+ = 2898$. The vertical dashed lines mark (a, b) $t = 0$, (c, d) $t=1.38$, and (e, f) $t=1.94$.}
\label{fig:3DChannel}
\end{figure}

\begin{figure}
    \begin{center}
    \subfloat[]{\includegraphics[width=0.45\linewidth]{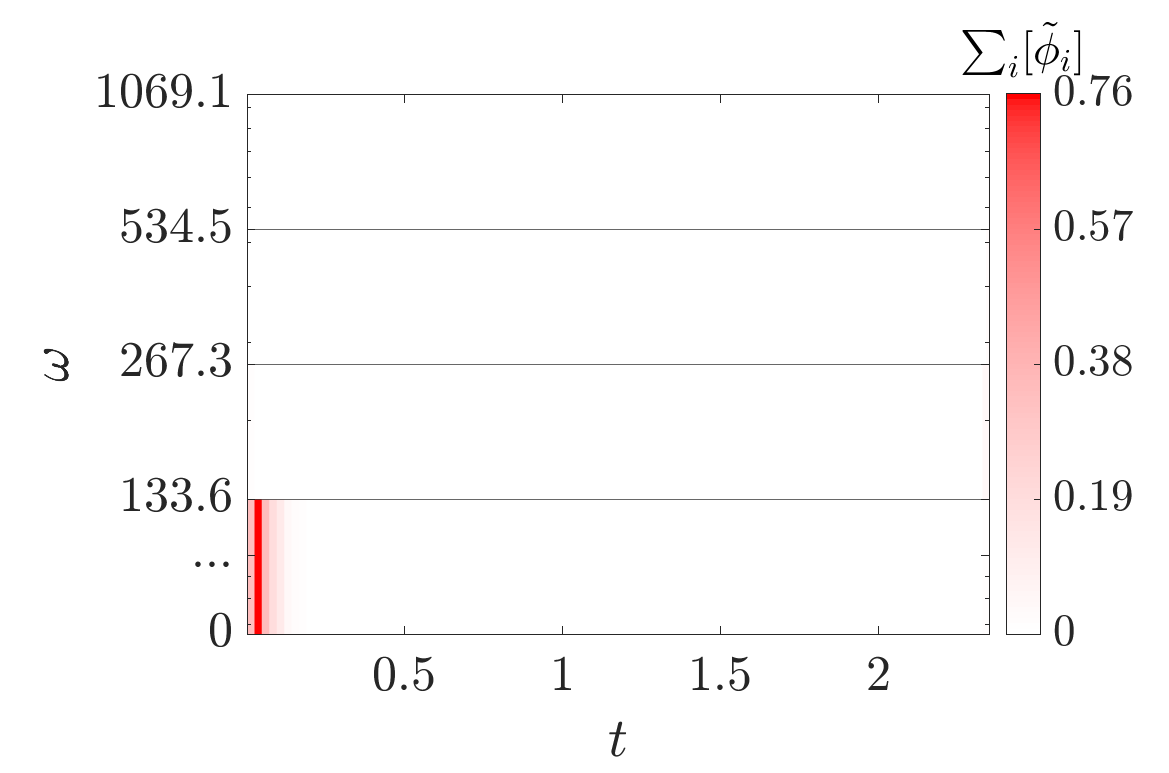}}
    \subfloat[]{\includegraphics[width=0.45\linewidth]{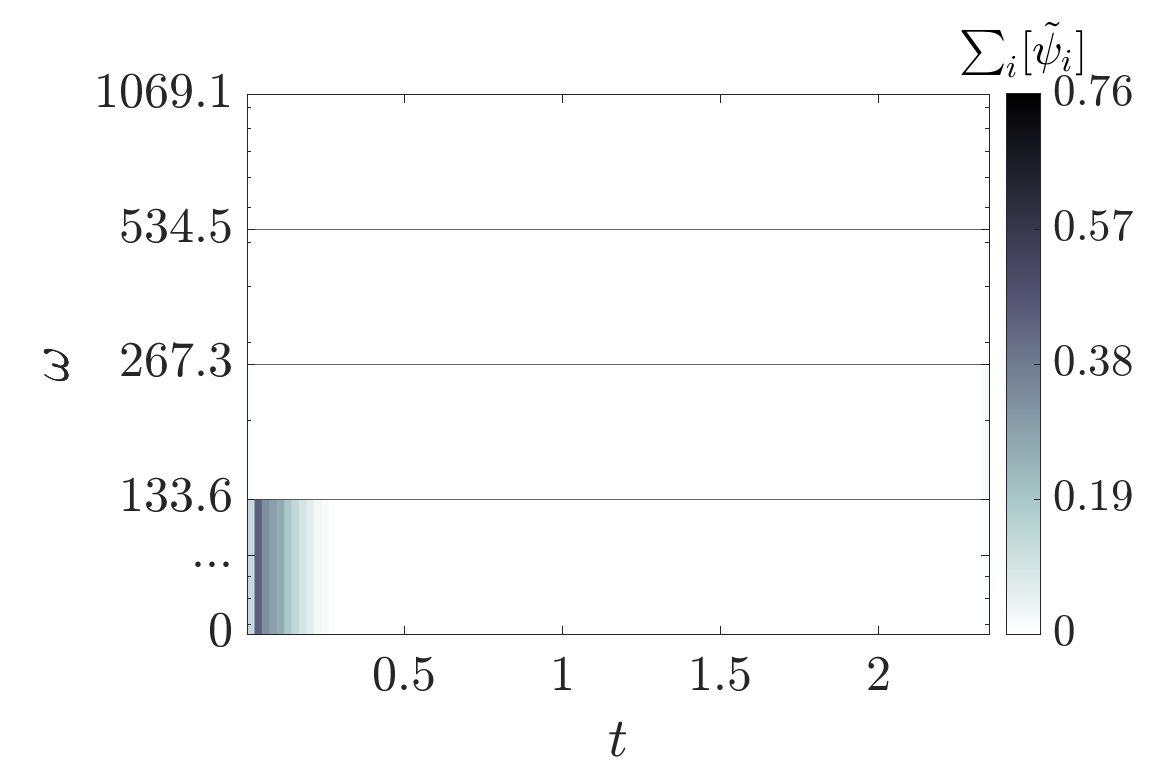}}
    \\
    \subfloat[]{\includegraphics[width=0.45\linewidth]{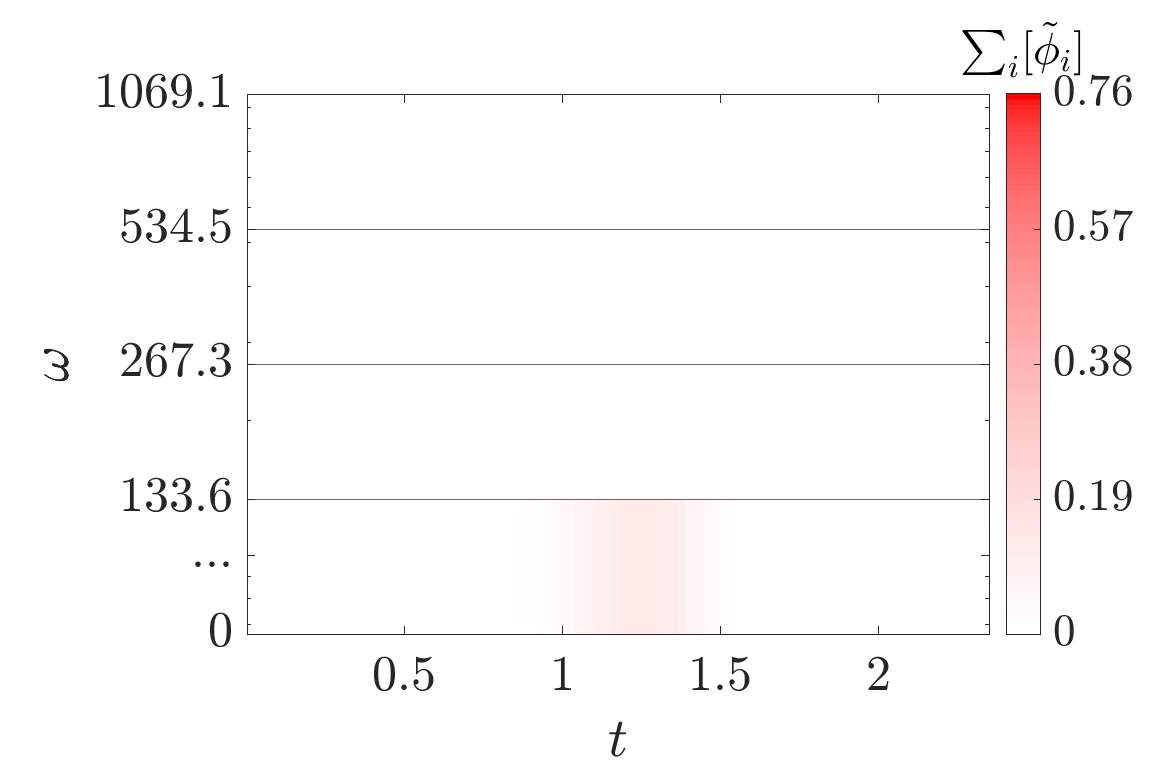}} 
    \subfloat[]{\includegraphics[width=0.45\linewidth]{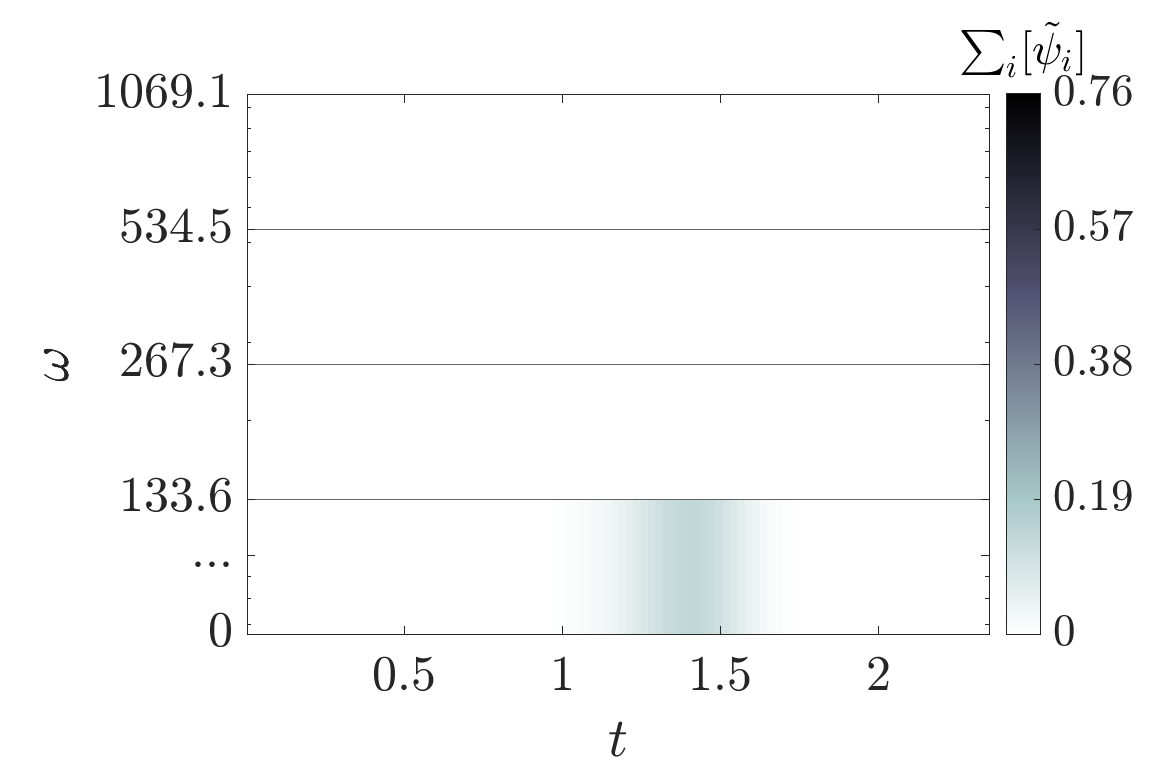}}
    \\
    \subfloat[]{\includegraphics[width=0.45\linewidth]{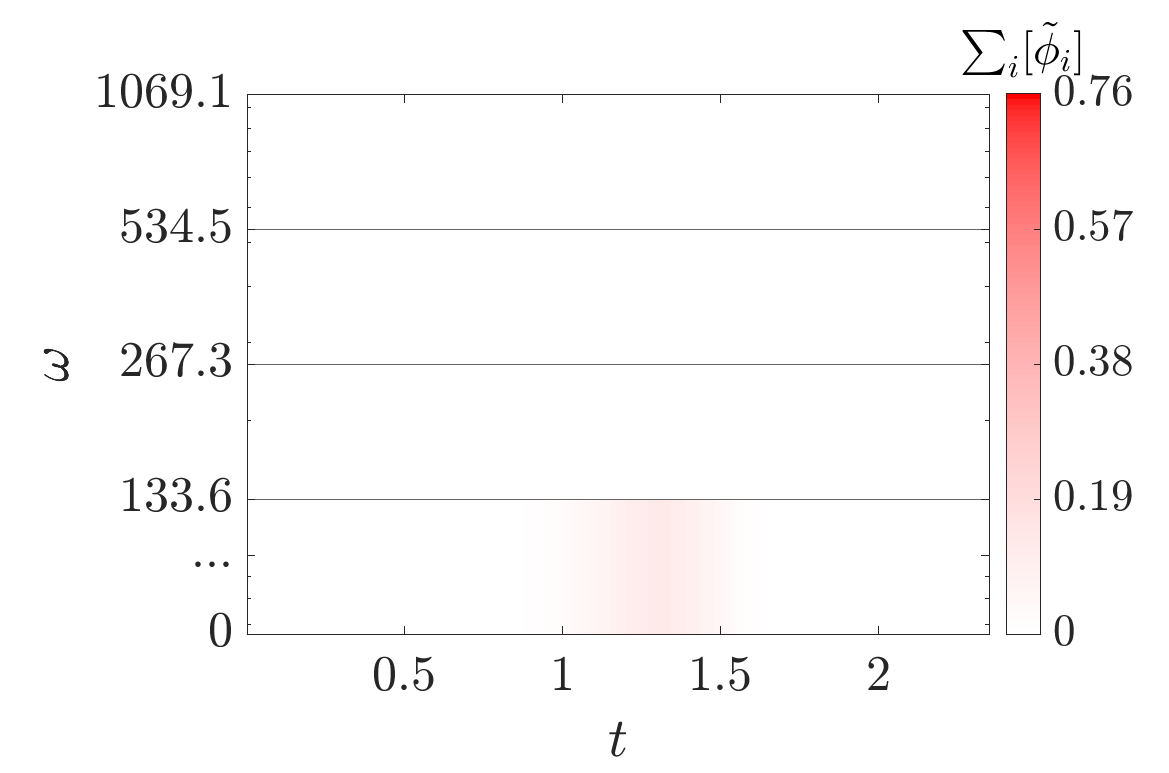}}
    \subfloat[]{\includegraphics[width=0.45\linewidth]{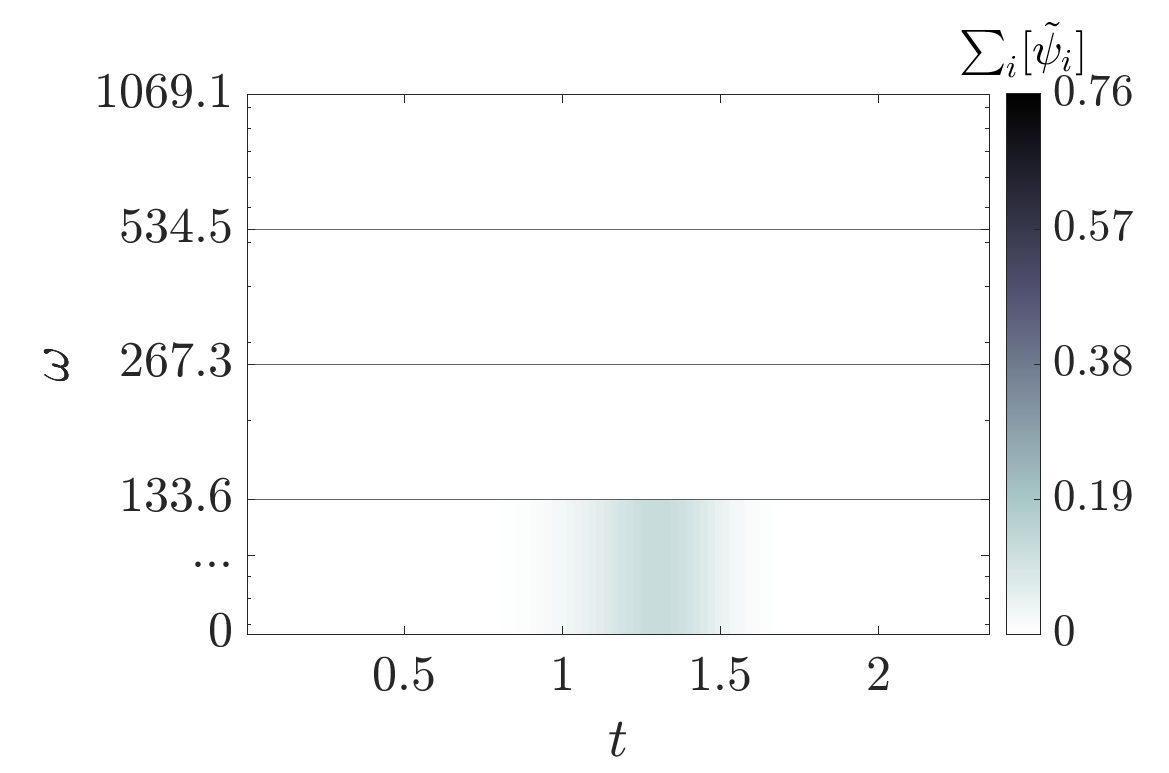}}
    \end{center}
    \caption{Principal forcing (left) response (right) modes in the frequency-time plane for the turbulent channel subject to a spanwise pressure gradient, for (a, b) $\lambda_1^+ = 1000$, $\lambda_3^+ = 100$, (c, d) $\lambda_1^+ = 264$, $\lambda_3^+ = 1827$, and (e, f) $\lambda_1^+ = 329$, $\lambda_3^+ = 2898$.}
    \label{fig:3DScalogram}
\end{figure}

\begin{figure}
    \begin{center}
    \subfloat[]{\includegraphics[width=0.32\linewidth]{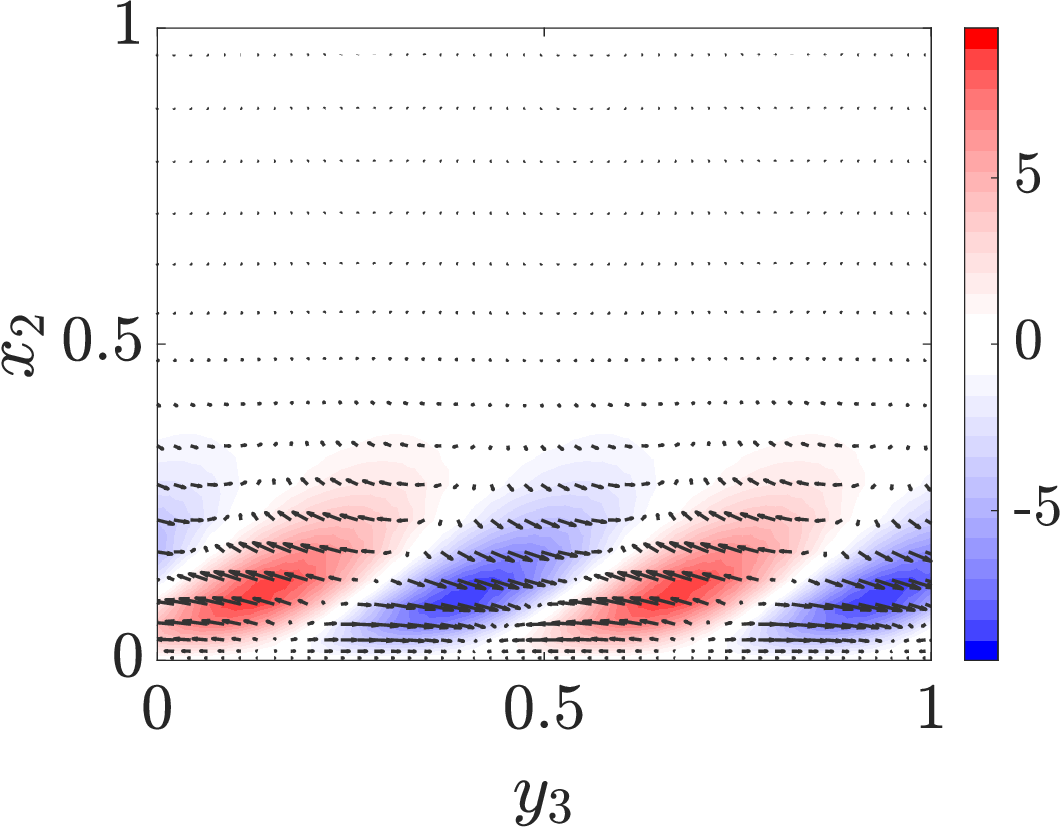}}
    \hspace{0.01cm}
    \subfloat[]{\includegraphics[width=0.32\linewidth]{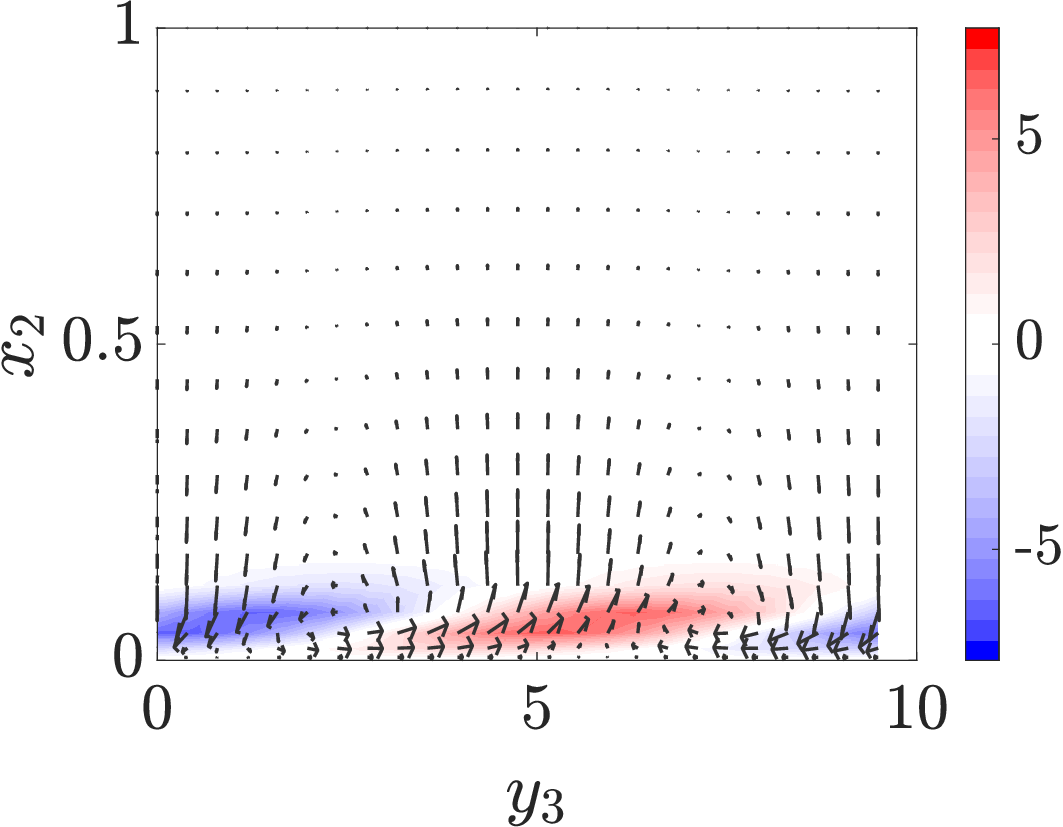}}
    \hspace{0.01cm}
    \subfloat[]{\includegraphics[width=0.32\linewidth]{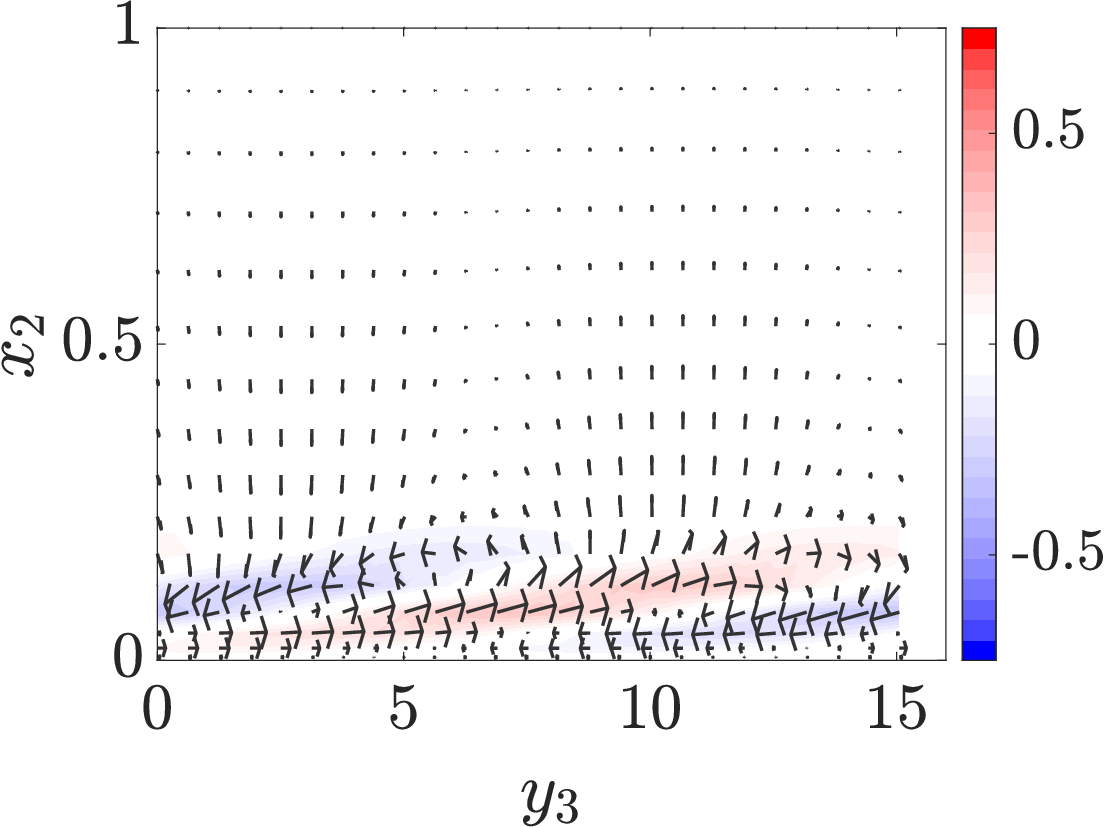}}
    \end{center}
    \caption{Principal response mode in the reference frame rotated about $x_2$ by an angle $\gamma$, at (a) $t = 0$, (b)  $t=1.38$, and (c) $t =1.94$. The contours represent the streamwise component; the arrows represent the velocity field in the wall-parallel plane.
    }
    \label{fig:mode_rotation}
\end{figure}

\begin{figure}
    \begin{center}
    \subfloat[]{\includegraphics[width=0.32\linewidth]{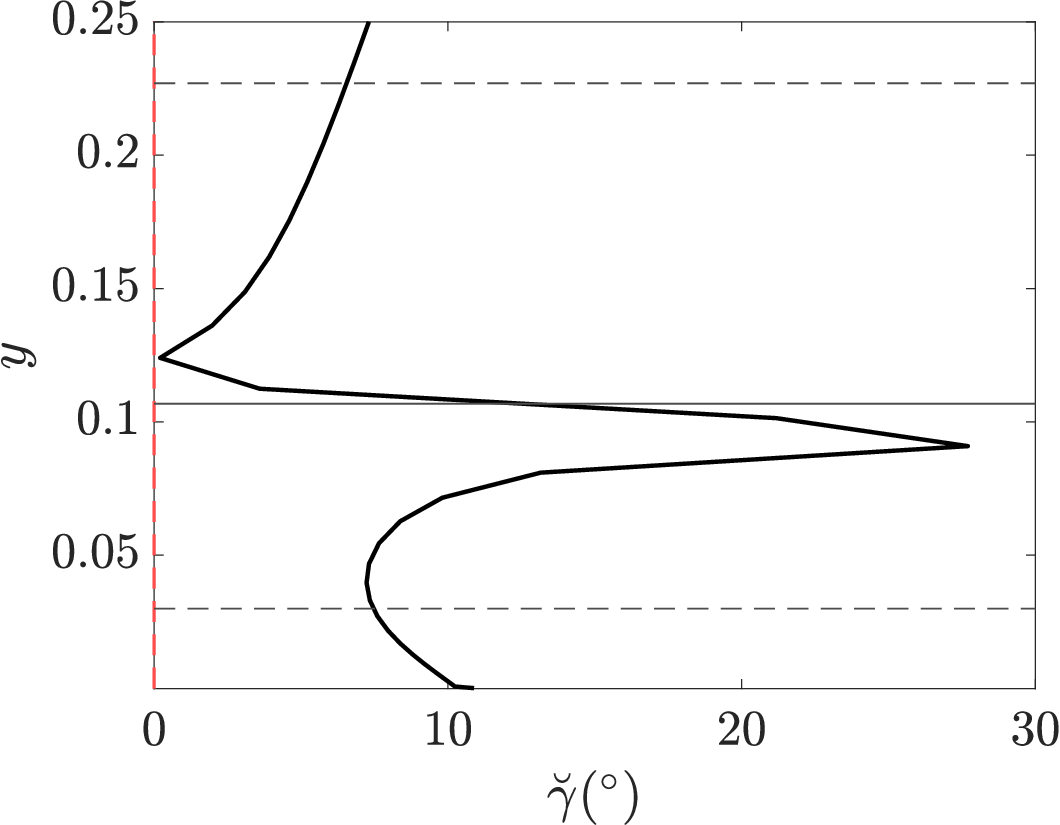}}
    \hspace{0.01cm}
    \subfloat[]{\includegraphics[width=0.32\linewidth]{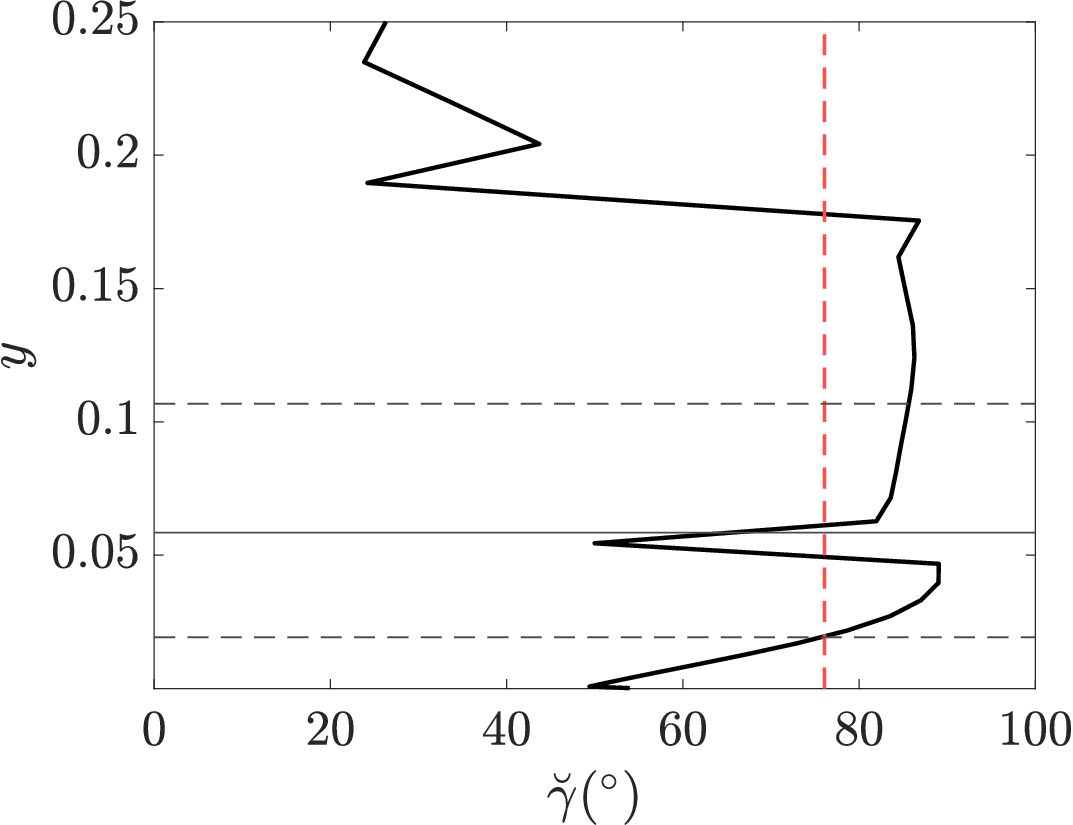}}
    \hspace{0.01cm}
    \subfloat[]{\includegraphics[width=0.32\linewidth]{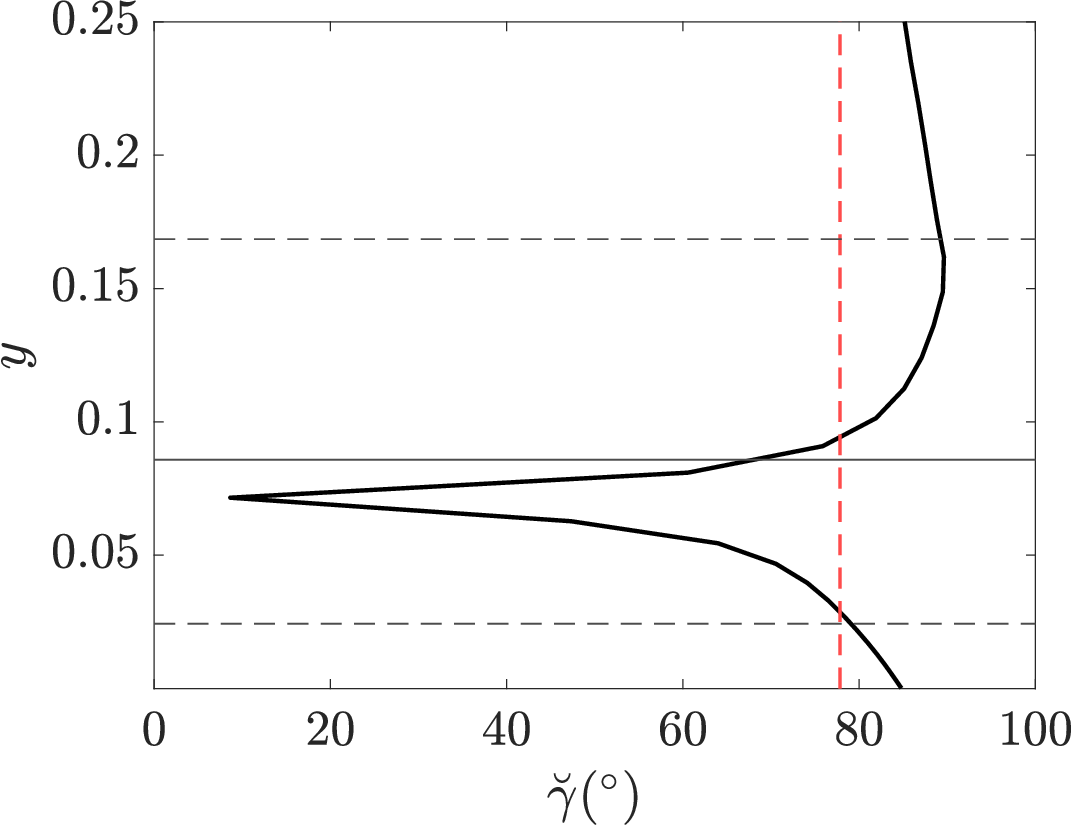}}
    \end{center}
    \caption{Shear angle of the principal response mode for (a) $(\lambda_1^+, \lambda_3^+) = (1000, 100)$ at $t = 0$, (b) $(\lambda_1^+, \lambda_3^+) = (264, 1827)$ at $t = 1.38$, and (c) $(\lambda_1^+, \lambda_3^+) = (329, 2898)$ at $t = 1.94$. The red vertical line (\color{red}- -\color{black}) indicates the wall shear-stress angle at the chosen times; the horizontal black line indicates the location of the peak of kinetic energy $|\breve \psi_1|^2 + |\breve \psi_2|^2 + |\breve \psi_3|^2$ at the target times, and the dashed horizontal lines indicate where the mode energy is at $10\%$ of the peak.}
    \label{fig:mode_angles}
\end{figure}

\begin{figure}
    \begin{center}
    \subfloat[]{\includegraphics[width=0.32\linewidth]{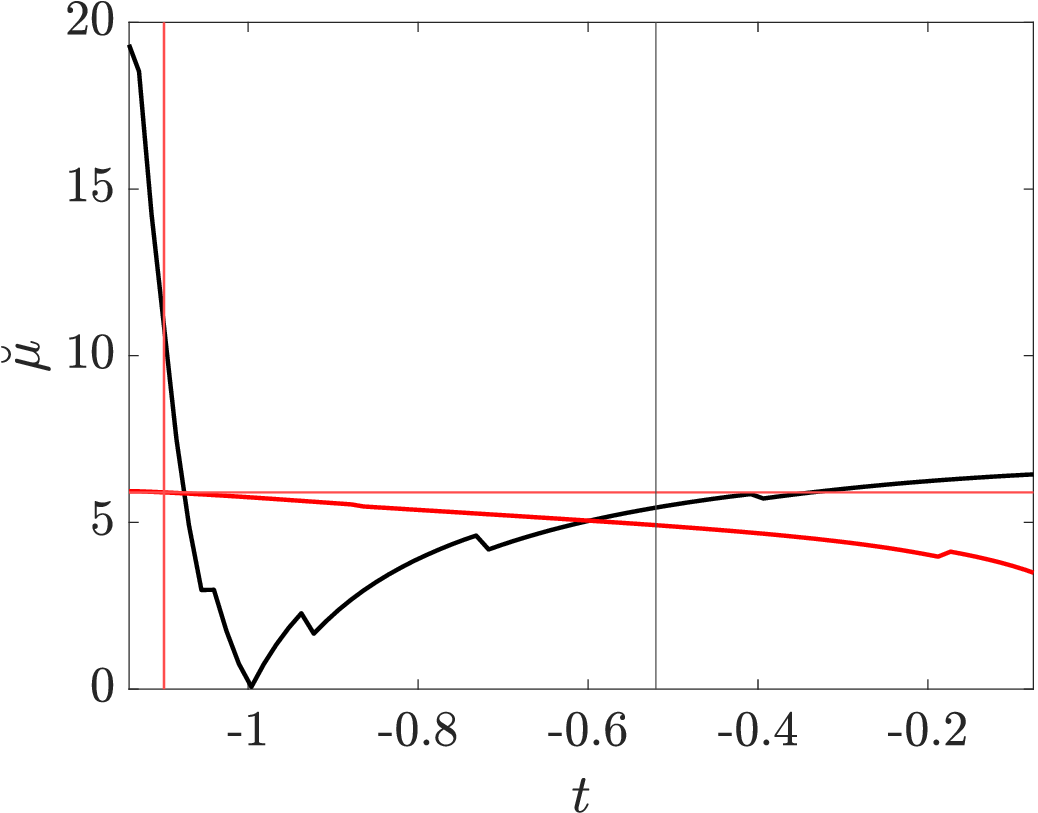}}
    \hspace{0.01cm}
    \subfloat[]{\includegraphics[width=0.32\linewidth]{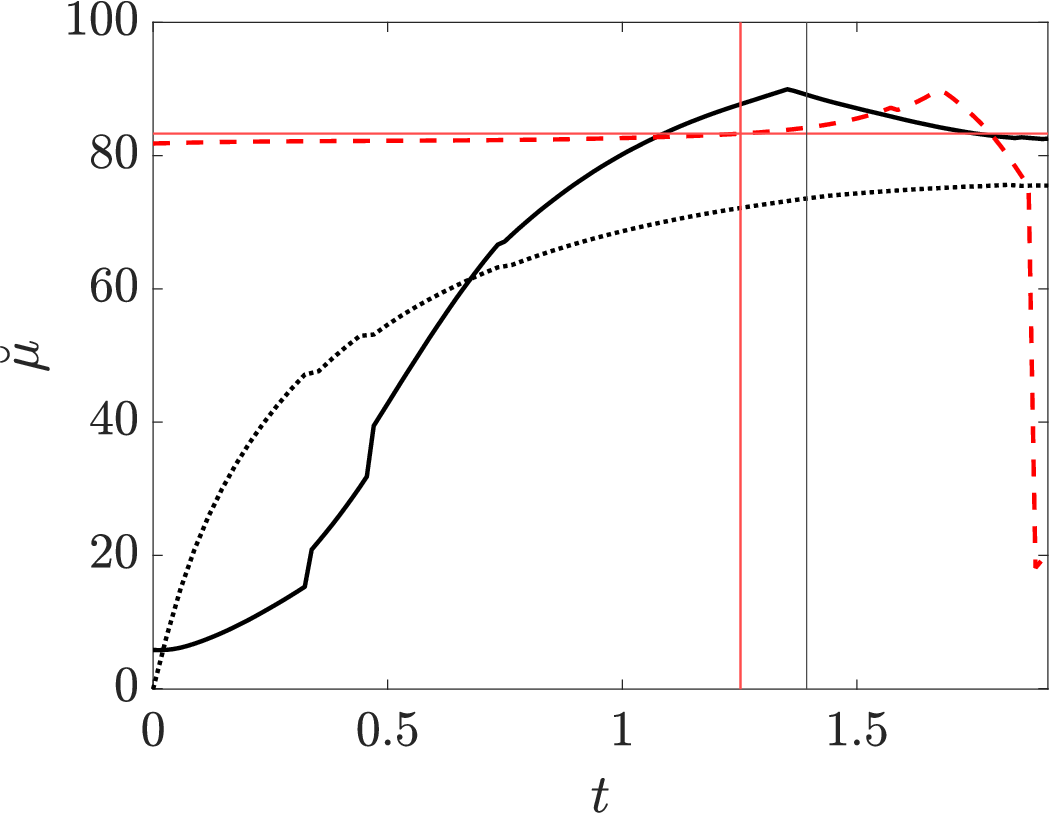}}
    \hspace{0.01cm}
    \subfloat[]{\includegraphics[width=0.32\linewidth]{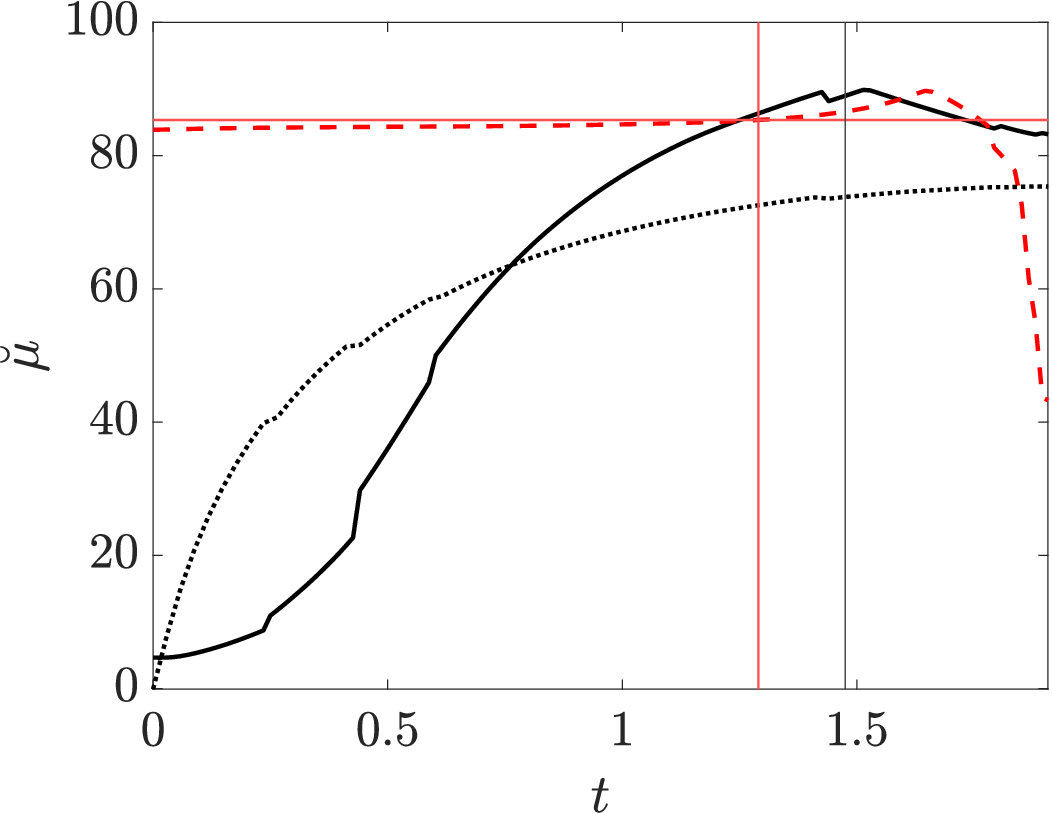}}
    \end{center}
    \caption{Flow angle of the principal forcing (red) and response (black) modes for (a) $(\lambda_1^+, \lambda_3^+) = (1000, 100)$, (b) $(\lambda_1^+, \lambda_3^+) = (264, 1827)$, and (c) $(\lambda_1^+, \lambda_3^+) = (329, 2898)$. The vertical lines indicates the amplitude peaks for the forcing (red) and response (black) modes, and the dotted black line represents $\gamma(t)$}
    \label{fig:flow_angles_vs_t}
\end{figure}

\begin{figure}
    \begin{center}
    \subfloat[]{\includegraphics[width=0.32\linewidth]{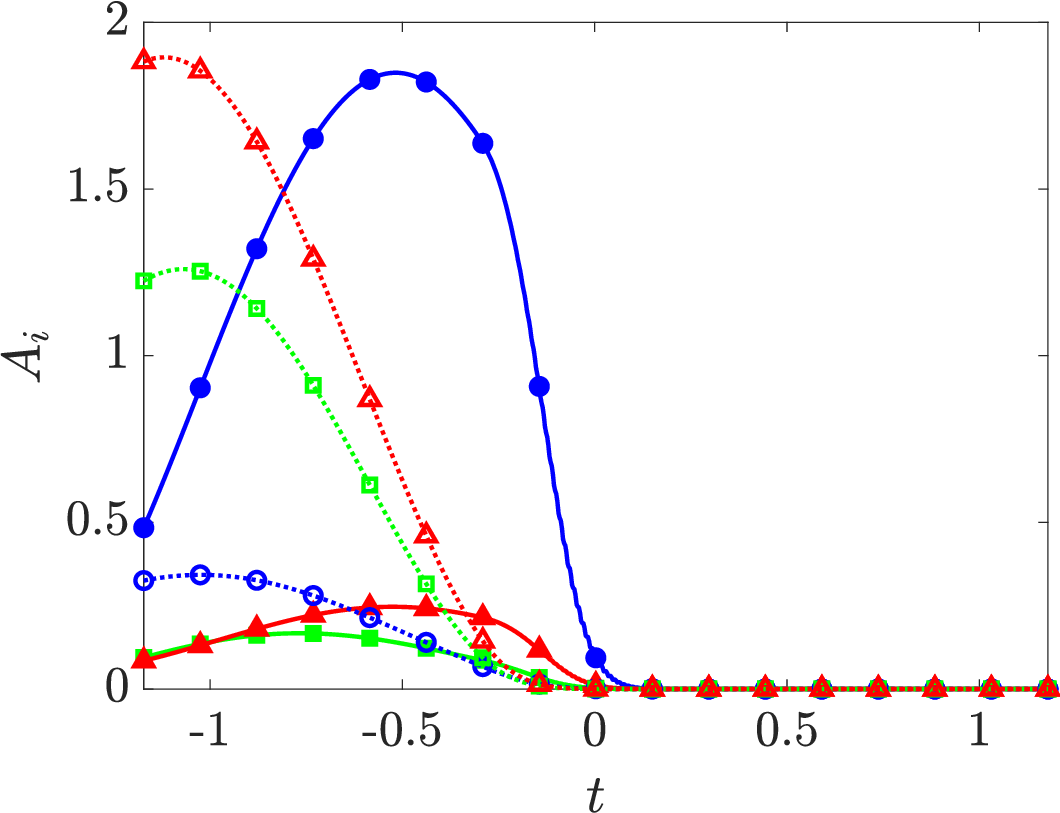}}
    \hspace{0.01cm}
    \subfloat[]{\includegraphics[width=0.32\linewidth]{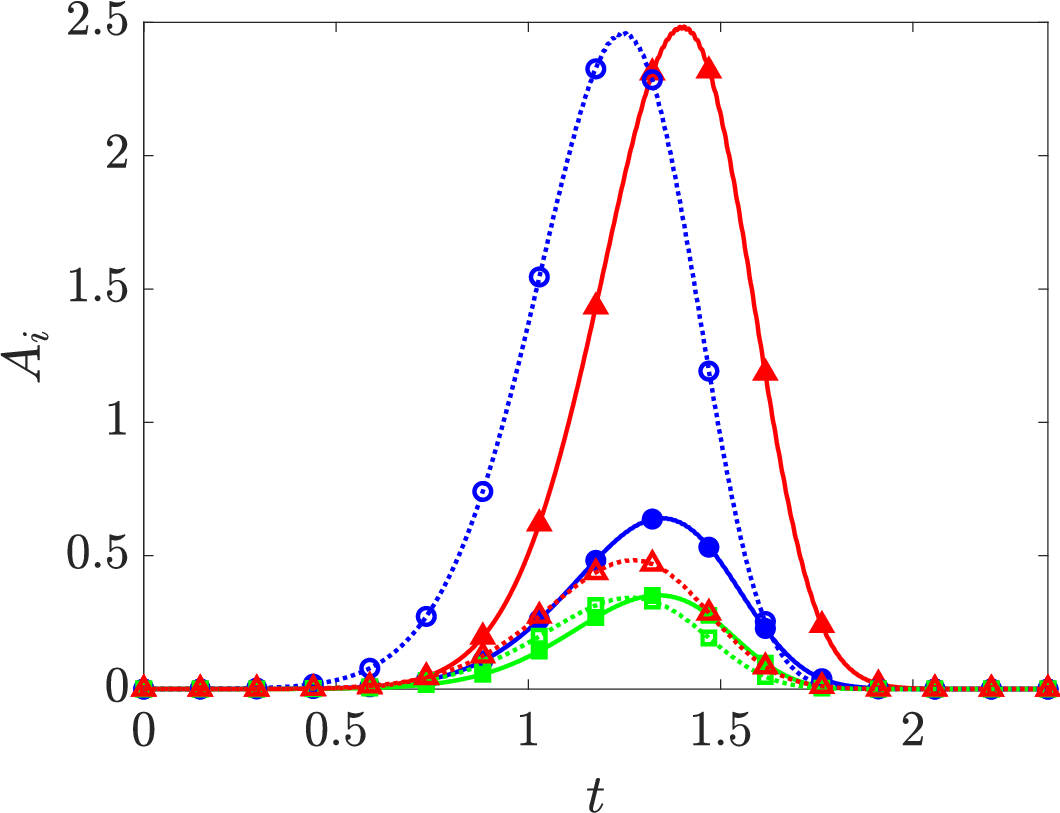}}
    \hspace{0.01cm}
    \subfloat[]  {\includegraphics[width=0.32\linewidth]{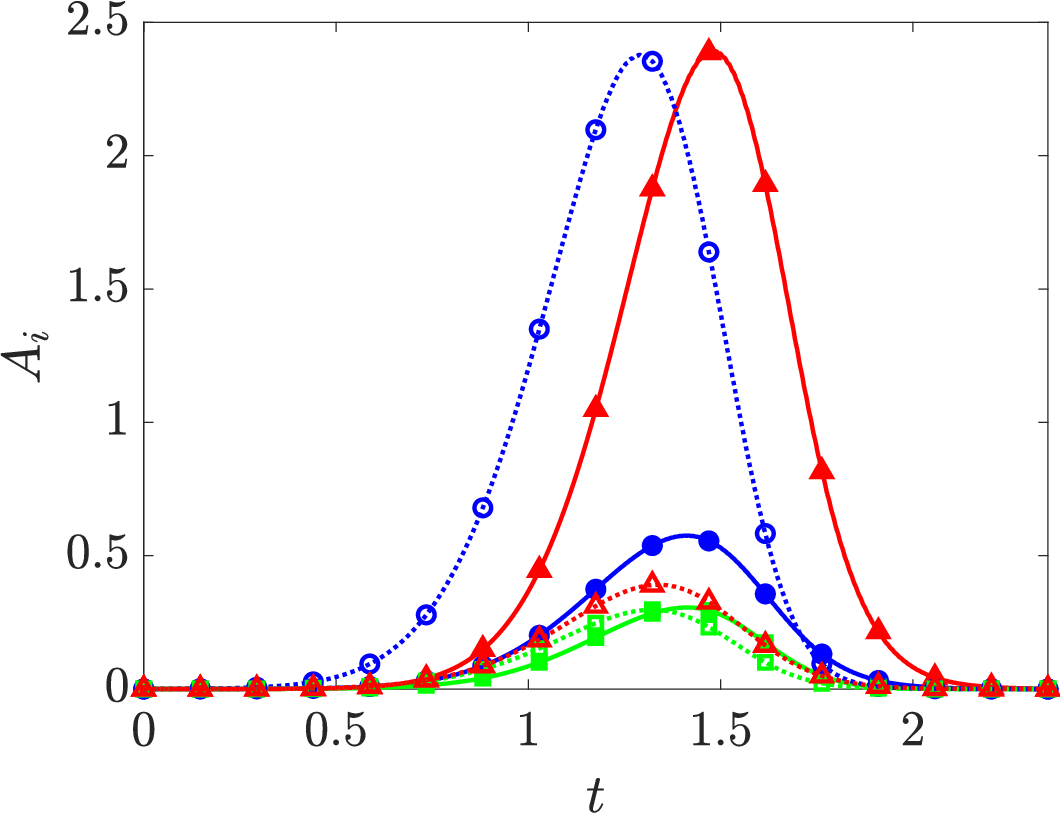}}
    \end{center}
    \caption{Amplitudes for the streamwise (blue $\bullet$), spanwise (red $\blacktriangle$), and wall-normal (green $\blacksquare$) components of the principal forcing (dotted line, hollow markers) and response (solid line, filled markers) modes, for (a) $(\lambda_1^+, \lambda_3^+) = (1000, 100)$, (b) $(\lambda_1^+, \lambda_3^+) = (264, 1827)$, and (c) $(\lambda_1^+, \lambda_3^+) = (329, 2898)$. }
    \label{fig:3D_forcing_response_amplitudes}
\end{figure}

The resolvent modes for the first wavelength pair $(\lambda_1^+,\lambda_3^+) = (1000, 100)$ are shown in figure~\ref{fig:3DChannel}(a, b). The magnitude of the modes in frequency-time space is also plotted in figure~\ref{fig:3DScalogram}(a, b). The resolvent modes are temporally centred  around $t = 0$ and exhibit a predominant streamwise component. The modes are located in a region  $x_2 < 0.25$, which corresponds to $x_2^+ < 45$, i.e., the buffer region. Thus, at $t = 0$, the modes capture the highly energetic near-wall streaks. The subsequent temporal decay of these modes can be explained by the changing flow conditions, notably the growth of the spanwise wall-shear stress $\tau_3$, and consequently $u_\tau$ (see Figure \ref{fig:utau_angle}). Under these conditions, the spatial scales preferred by the near-wall streaks stretch as $u_\tau$ increases and the wall-shear stress tensor rotates towards the $x_3$ direction.
 

The response mode for the second pair of spatial scales, $(\lambda_1^+,\lambda_3^+) = (264, 1827)$, tuned to conditions at $t = 1.38$, are plotted in \ref{fig:3DChannel}(c,d). The frequency-time map of the modes is shown in figure~\ref{fig:3DScalogram}(c, d). Similar to the first case, the modes are centred  around $t  = 1.38$, indicating that the wavelet-based resolvent analysis is able to identify the non-equilibrium effects of the non-stationary flow. We note that the spanwise component of the response mode is much more dominant than the streamwise component, which reflects the new wall-shear angle $\gamma = 75.7^\circ$. 
Finally, for the third case, $(\lambda_1^+,\lambda_3^+) = (329, 2898)$, tuned to conditions at $t = 1.94$, we observe that the modes (figure~\ref{fig:3DChannel}(e,f) and figure~\ref{fig:3DScalogram}(e, f)) are not centred  around the target time. We speculate that this is due to the temporal boundary condition at $t  = 2.34$. As the flow is not at a statistically steady state at this time, a Neumann boundary condition may not be the most suitable boundary condition. The modes cannot grow beyond the boundary due to the boundary condition and are artificially damped near the end of the temporal domain. 

Figure~\ref{fig:mode_rotation} shows the principal response mode in the physical domain for the three target length scale pairs, in the reference frame rotated by $\gamma$ about the $x_2$--axis. We denote the streamwise and spanwise directions in the rotated reference frame by $y_1$ and $y_3$, respectively, with $y_1 = x_1 \mathrm{cos}(\gamma) - x_3 \mathrm{sin}(\gamma)$ and $y_3 = x_1 \mathrm{sin}(\gamma) + x_3 \mathrm{cos}(\gamma)$
The modes resemble each other qualitatively, and capture elongated near-wall streaks in the direction of the rotated flow. The length and spanwise spacing of the streaks increases with wall shear stress, as expected.
Moreover, the response modes for $(\lambda_1^+,\lambda_3^+) = (264, 1827)$ and  $(\lambda_1^+,\lambda_3^+) = (329, 2898)$ (figure~\ref{fig:mode_rotation}(b)) are concentrated closer to the wall than for $(\lambda_1^+,\lambda_3^+) = (1000, 100)$ (figure~\ref{fig:mode_rotation}(a)), which indicates that the region of high-sensitivity to forcing moves closer to the wall as $Re_\tau$ increases. 
This is in line with the behaviour of near-wall turbulence: for higher $Re_\tau$, the buffer and logarithmic layers, which contain the bulk of turbulent energy in channel flow, are closer to the wall. \reviewertwo{We note, however, that the mode corresponding to  $(\lambda_1^+,\lambda_3^+) = (329, 2898)$ is expected to be located at an even lower wall-normal height than the mode corresponding to $(\lambda_1^+,\lambda_3^+) = (264, 1827)$, and attribute its higher location to the effect of the temporal boundary condition at the end of the temporal domain.}


\reviewertwo{We can compute flow and shear angles from the principal resolvent mode and compare them to the wall shear stress angle $\gamma$ extracted from DNS. We define the mode shear angle as 
%
%
\begin{equation}
    \breve \gamma(x_2, t) = \left | \mathrm{tan}^{-1} \left ( \widefrac{  \mathrm d \psi_3/ \mathrm d x_2}{ \mathrm d \psi_1\mathrm /dx_2} \right )   \right | _{x_1 = x_{1, \mathrm{max}}, x_3 = x_{3, \mathrm{max}}},
\end{equation}
%
where $\psi_i$ denotes the inverse Fourier and wavelet transform of the $i^{\mathrm{th}}$ mode component, and $(x_{1, \mathrm{max}}, x_{3, \mathrm{max}})$ denotes the location in the $x_1$--$x_3$ plane where the amplitude of the response mode $\boldsymbol{\psi}$ is maximal at each $x_2$. We plot the results for the three spatial parameters at their respective target times (figure \ref{fig:mode_angles}). 
For the mode corresponding to $(\lambda_1^+, \lambda_3^+) = (1000, 100)$, the mode shear angle $\breve \gamma$ matches the wall shear angle of zero from DNS at a wall-normal location slightly farther from the wall than the location of peak amplitude. 
For $(\lambda_1^+, \lambda_3^+) = (264, 1827)$ and $(\lambda_1^+, \lambda_3^+) = (329, 2898)$, we observe that $\breve \gamma$ matches $\gamma$ well at the amplitude peaks. For all three sets of spatial parameters, the mode angles at the wall itself differ significantly from the wall shear angle from DNS, though this is expected since the modes obtained are lifted from the wall and decay to match the wall boundary condition.
}

\reviewerone{The amplification of the response mode given by the leading singular value of the resolvent operator differs across the chosen spatial parameters. The energy of the principal response mode corresponding to $(\lambda_1^+, \lambda_3^+) = (1000, 100)$, which peaks for $t < 0$, is amplified by a factor of $\sigma_1^2 \approx 8$, while the energy of the modes corresponding to $(\lambda_1^+, \lambda_3^+) = (264, 1827)$ and $(\lambda_1^+, \lambda_3^+) = (329, 2898)$, respectively peaking at $t = 1.34$ and $t = 1.94$, are respectively amplified by a factor of $\sigma_1^2 \approx 0.4$ and $\sigma_1^2 \approx 0.6$, indicating an effective energy suppression. A decrease in Reynolds stresses was similarly observed in the fully turbulent system simulated during the development of the spanwise mean flow \citep{lozano2021cause}, and it was proposed that the developing spanwise mean flow generates smaller transverse structures close to the wall which disrupt the coherence of the dominant streamwise rolls and inhibit them from vertically transporting momentum upward from the near-wall region.

The behaviour of the wavelet-based resolvent modes allows us to expand this proposed explanation. 
The optimal forcing for the mean flow conditions at a time $t$ is in the form of rolls approximately pointing in the direction $\gamma(t)$, and its corresponding response mode will eventually align itself with the forcing angle after a period of transient growth as shown in figures \ref{fig:flow_angles_vs_t} (a) and (b).
Due to the nonnormality of the linearised system, the optimal forcing peaks before the target time $t$, and thus the optimal forcing may be instantaneously misaligned with the rotating mean flow during the transient growth period of the response. Indeed, the forcing mode is not computed to maximise the instantaneous kinetic energy amplification, but maximises instead the time-integrated kinetic energy of the response over the entire temporal domain.

More insight can be gained by considering the linearised equation governing the wall-normal vorticity $v_2$
\begin{equation}\label{eq:squire}
    \left ( \partial_t + U_1\partial_{x_1} + U_3 \partial_{x_3} - \frac{1}{\Rey}(\partial_{x_1}^2 + \partial_{x_2}^2 + \partial_{x_3}^2 ) \right) v_2 = \left (-\frac{\mathrm dU_1}{\mathrm dx_2} \partial_{x_3} + \frac{\mathrm dU_3}{\mathrm dx_2} \partial_{x_1} \right ) u_2 + g_2,
\end{equation}
where $g_2 := \partial_{x_3} f_1 - \partial_{x_1} f_3$ denotes the external forcing. We note that in this problem, the mean wall-normal flow $U_2$, and the wall-normal gradients $\mathrm d U_1/ \mathrm d x_2$ and $\mathrm d U_3/ \mathrm d x_2$ are zero. Introducing a flow angle $\mu := \mathrm{tan}^{-1}(U_3/U_1)$ and a shear angle $\upsilon := \mathrm{tan}^{-1} \left ((\mathrm d U_3 /\mathrm dx_2)/ (\mathrm d U_1 /\mathrm dx_2) \right )$, we can rewrite equation \eqref{eq:squire} in the reference frame locally rotated by $\mu$ as
\begin{multline}
\label{eq:squire_rotated}
    \left ( \partial_t + (U_1^2 + U_3^2)^{\frac{1}{2}} \partial_{x'_1} - \frac{1}{\Rey}(\partial_{x'_1}^2 + \partial_{x_2}^2 + \partial_{x'_3}^2 ) \right) v_2 =  
    \\ \left( \frac{\mathrm dU_1}{\mathrm dx_2}^2 + \frac{\mathrm dU_3}{\mathrm dx_2}^2 \right)^{\frac{1}{2}}
    \left ( - \cos(\mu-\upsilon) \partial_{x_3'} + \sin(\mu -\upsilon)\partial_{x_1'} \right ) u_2 + g_2,
\end{multline}
where $(x_1', x_2, x_3')$ denote the coordinates rotated anticlockwise by angle $\mu$ in the $x_1 - x_3$ plane. The left-hand side of equation \eqref{eq:squire_rotated} is identical to the classical Squire equation for perturbations about a one-dimensional streamwise mean flow, with an effective advection velocity of $(U_1^2 + U_3^2)^{1/2}$. The right-hand side includes a lift-up term modified by the misalignment between the mean shear and velocity profiles. If the mean flow is at equilibrium and $\mu = \upsilon$, we obtain an identical lift-up term to the classical Squire equation, with an effective mean shear of $\left( (\mathrm dU_1\mathrm /dx_2)^2 + (\mathrm dU_3/\mathrm dx_2)^2 \right)^{1/2}$. The lift-up term in the Squire equation constitutes a way for the wall-normal velocity perturbations to force the streamwise and spanwise components, as discussed in \S\ref{sec:transient}  and \cite{jimenez2013linear, jimenez2018coherent}. Through this coupling, the streamwise and spanwise velocity components can be efficiently forced by $f_2$ along with $f_1$ and $f_3$.
However, in a non-equilibrium rotating flow, the mean shear profile lags behind the mean velocity, \emph{i.e.} $\mu \neq \upsilon$, causing a misalignment between the two profiles which is particularly pronounced for regions farther away from the wall.
The term pertaining to the misalignment in equation \eqref{eq:squire_rotated} damps the effect of the wall-normal velocity perturbation $u_2$ on the wall-normal vorticity, and consequently, the streamwise and spanwise velocity perturbations. This is illustrated by figure \ref{fig:3D_forcing_response_amplitudes}: for the initial one-dimensional mean flow configuration in figure \ref{fig:3D_forcing_response_amplitudes}(a), the optimal forcing mode exploits the lift-up mechanism via a strong wall-normal component; for the rotated configurations in figures \ref{fig:3D_forcing_response_amplitudes}(b) and (c), the wall-normal component of the optimal forcing is significantly attenuated. 

Thus, the lag of the mean shear profile disrupts the coupling between the wall-normal, and the streamwise and spanwise velocity perturbations. In a non-equilibrium rotating mean flow, wall-normal velocity perturbations are less efficient at producing slow-decaying streaks in the flow direction (than in \S\ref{sec:transient}, for example), and the mean flow selects rolls with a much weaker wall-normal component.
}
\section{Conclusion} \label{sec:conclusion}
This work expands the resolvent analysis framework to non-stationary flow problems. The resolvent operator is traditionally constructed for flow quantities that are Fourier-transformed in the homogeneous spatial directions and in time. Such a resolvent operator cannot be used to study time-localised nonlinear forcing or a time-varying mean flow. Instead, we construct a wavelet-based resolvent operator by applying a wavelet transform in time while keeping the Fourier transform for the homogeneous spatial directions, thus trading the functional dependence on time with a dependence on $\alpha$ and $\beta$. 

This resolvent operator, provided we use an orthonormal wavelet basis, is equivalent to the Fourier-based resolvent analysis for statistically stationary flows. Even in such cases, wavelet-based resolvent analysis can be modified through windowing in order to explore the effects of transient forcing localised to time scales of interest, such as those characterising the logarithmic layer. 
In the case of transiently forced channel flow, the wavelet-based resolvent analysis with windowing reveals that 
the optimal response modes are transiently amplified rolls. The significant transient energy growth of these streaks is expected of non-normal systems. 
Moreover, the optimal forcing and response modes exhibit characteristics of the Orr mechanism, which supports the claim that this mechanism plays an important role in the linear amplification of velocity perturbations.

The wavelet-based resolvent analysis is notable in its ability to reflect the effects of a non-stationary mean flow. In the case of the turbulent Stokes boundary layer, the wavelet-based resolvent modes, which encode time, allow us to track the spatial and temporal location of the peak amplification alongside the varying mean flow. The resolvent modes reveal an increased sensitivity to forcing and perturbation amplification near the peaks of the streamwise r.m.s. velocity. This suggests that linear mechanisms may be an important source of energy amplification in this type of flow, as is believed for channel flow.
We also observe that the input modes precede the output modes, opening the possibility to study causality in turbulent flows using resolvent analysis.
Wavelet-based resolvent modes also encode frequency information. This ability sheds new light on the properties of linear amplification in the Stokes oscillating boundary layer: there exists an optimal forcing frequency to which the linearised flow is most sensitive, but the corresponding optimal response trajectory is shifted to higher frequencies by the decelerating mean flow. Wavelet-based resolvent analysis can thus be a useful tool for analyzing systems in which forcing and response prefer different frequencies.

Finally, for the 3-D channel flow, the resolvent modes are able to identify the effect of the varying flow conditions, mainly the increasing shear velocity and rotating wall shear stress, on the principal resolvent modes. 
We compute the resolvent modes using the length scales preferred by near-wall streaks for flow conditions at three different times. 
The resulting resolvent response modes peak around the chosen times, with the exception of the time close to the end of the temporal domain. The predominant velocity component for the resolvent modes progressively shifts from the streamwise component to the spanwise one, mirroring the reorientation of the mean flow. Wavelet resolvent modes reflect time-varying mean flow conditions and help locate energetic near-wall streaks in space and time, and identify their preferred spatial scales. This can shed light on the flow conditions that amplify these coherent structures, \reviewerone{or, conversely, suppress them. The wavelet-based modes reflect a damping of the effectiveness of the lift-up mechanism at energising streamwise near-wall streaks, which mirrors the reduction of the wall-normal transport of streamwise momentum observed in DNS}. Thus, the cases considered in this work showcase the versatility of the wavelet-based formulation in analyzing transient linear energy amplification in flows with either statistically stationary and non-stationary mean profiles. 

\section*{Acknowledgments}
The authors acknowledge support from the Air Force Office of Scientific Research under grant number FA9550-22-1-0109.
\section*{Declaration of Interests} The authors report no conflict of interest.

\bibliographystyle{jfm}
\bibliography{bibliography}

\end{document}